\documentclass[prc,twocolumn,secnumroman,tightenlines,amssymb,aps,nobibnotes,
               superscriptaddress,showpacs,balancelastpage,floatfix,nofootinbib]{revtex4}
\usepackage{graphicx}
\usepackage{multirow}           
\usepackage{color}	
\expandafter\ifx\csname package@font\endcsname\relax\else
\expandafter\expandafter
\expandafter\usepackage
\expandafter{\csname package@font\endcsname}
\fi
\DeclareRobustCommand\substyle{\name@idx{document substyle}}
\DeclareRobustCommand\classoption{\name@idx{document class option}}
\DeclareRobustCommand\classname{\name@idx{document class}}
\def\name@idx#1#2{{\ttfamily#2}
\index{#2\space#1=\string\ttt{#2}\space#1}\index{#1>#2=\string\ttt{#2}}}

\newcommand{\mc}[3]{\multicolumn{#1}{#2}{#3}}

\begin{document}







\title[]
      {Nuclear Jacobi and Poincar\'e Transitions at High Spins and  
       Temperatures: Account~of~Dynamic~Effects~and~Large-Amplitude Motion}
       
\author{K. Mazurek}
\email{Katarzyna.Mazurek@IFJ.edu.pl}
\affiliation{Institute of Nuclear Physics PAN, ul.\,Radzikowskiego 152,
             Pl-31342 Krak\'ow, Poland}
\author{J. Dudek}
\email{Jerzy.Dudek@IPHC.CNRS.fr}
\affiliation{IPHC/DRS and Universit\'e de Strasbourg,
             23 rue du Loess, B.P.\,28, F-67037 Strasbourg Cedex 2, France}
\author{A. Maj}
\email{Adam.Maj@IFJ.edu.pl}
\affiliation{Institute of Nuclear Physics PAN, ul.\,Radzikowskiego 152,
             Pl-31342 Krak\'ow, Poland}
\author{D. Rouvel}
\email{David.Rouvel@IPHC.CNRS.fr}
\affiliation{IPHC/DRS and Universit\'e de Strasbourg,
             23 rue du Loess, B.P.\,28, F-67037 Strasbourg Cedex 2, France}

\date{\today}


\begin{abstract}

We present a theoretical
analysis of the competition between so-called nuclear Jacobi and Poincar\'e
shape transitions in function of spin - at high temperatures. The latter
condition implies the method of choice - a realistic version of the nuclear
Liquid Drop Model (LDM), here: the Lublin-Strasbourg Drop (LSD) model. We address specifically the fact that the Jacobi and Poincar\'e shape transitions are accompanied by the flattening of total nuclear energy landscape as function of the relevant deformation parameters what enforces large amplitude oscillation
modes that need to be taken into account. For that purpose we introduce an approximate form of the collective Schr\"odinger equation whose solutions are used to calculate the most probable deformations associated with both types of
transitions and discuss the physical consequences in terms of the associated
critical-spin values and transitions themselves.

\end{abstract}


\maketitle


\section{Introduction}
\label{Sect.I}

Atomic nuclei whose properties are governed by strong interactions acting among constituent nucleons can, to 
an approximation, be considered compact since the volumes of nuclei remain close to the sums of the volumes of these 
nucleons. This fact combined with the short range of the nucleon-nucleon interactions and incompressibility of the
nuclear matter allows to introduce a classical notion of nuclear surfaces - at first a paradox, since the nucleons 
{\em and} nuclei are quantum systems. These surfaces define what is referred to as nuclear shapes.

The notion of generally non-spherical nuclear shapes remains an underlying classical element of quantum mean-field 
theories of the nucleonic motion in nuclei as described using phenomenological but realistic Woods-Saxon or Yukawa-folded nuclear potentials. These two particular realizations of the nuclear 
mean-field have been used in the past, combined with the so-called Strutinsky method, Ref.\,\cite{VMS68}, to calculate very 
successfully various nuclear properties such as masses, deformation energies, nuclear moments, angular momenta of 
excited states and, more generally, rotational properties together with their evolution with nuclear spin and 
temperature - to mention just a few elements on the much longer list.

Among important characteristics of atomic nuclei are quadrupole (or higher) charge (and/or mass) moments. Numerous 
measurements show that the majority of the atomic nuclei are deformed in their ground and excited states and strictly 
speaking, among nearly 3000 nuclei so far investigated in laboratory only about a dozen may be considered strictly 
spherical. In fact, these are the nuclei with fully occupied $j$- and/or $N$-shells such as $^{16}$O, $^{40,48}$Ca, $^{100,132}$Sn or
$^{208}$Pb and only very few others. 

It is well known that an increase of excitation energy, which within nuclear mean-field theories can be translated into an increase of nuclear temperature, leads to diminishing and possibly to a full disappearance of quantum (shell) effects. Under such conditions, nuclear energy can be described as a sum of the repulsive Coulomb interactions among the protons together with centrifugal stretching effects associated with the collective rotation and an effective nuclear attraction formally modeled with the help of the concept of the nuclear surface tension. One of the simplest but at the same time a very successful modeling of such a physical situation has been achieved within the Nuclear Liquid Drop Model whose ingredients are indeed the competing Coulomb and the surface tension mechanisms, cf., e.g., Refs.\,\cite{CWe35,HBe36,WDM66,CPS74}, possibly combined with collective rotation. 

It then follows that theoretical modeling of leading features of motion of macroscopic charged drops, of planets and stars and of atomic nuclei may, under the discussed conditions, become analogous. The shape evolution which occurs in nuclei may take a form of what is referred to as Jacobi, Ref.\,\cite{Jac84}, and Poincar\'e, Ref.\,\cite{Poi85}, transitions following suggestions in the above historical articles that such transitions may be induced by rotation of certain astronomical objects. Of course, proposing an analogy between the forms of behavior of stellar and nuclear objects, as suggested in 
\cite{RBK61}, may be of certain aesthetic interest, however, predictions at which critical spin values and in which 
nuclei the considered transitions take place is a totally different, challenging issue; for early discussion 
cf.~Refs.\,\cite{CPS74} and \cite{Sie86}. 

The issue of Jacobi and Poincar\'e shape transitions has received some attention both from the experimental as well as modeling view points also more recently as e.g.~in Ref.\,\cite{DWa02}, where a discussion of, among others, Jacobi shape 
transitions represented with the help of the moments of inertia in the form of `gigantic back-bending' can be found. However, aiming at the use of the LSD approach we would like to mention a number of results obtained by combining the LSD analysis with the experimental results within the Cracow-Strasbourg collaboration, during the last ten years or so. In particular a successful determination of the presence of the Jacobi transitions in $^{46}$Ti has been reported in Refs.\,\cite{AMa01,AMa04} and through observation of the high-energy gamma-rays and $\alpha$-particles in Ref.\,\cite{MKm07}. Results of an analogous study of the neighboring $^{42}$Ca nucleus can be found in 
Ref.\,\cite{MKm05} (cf.~also the study of the $^{47}$V case, Ref.\,\cite{DPa10}) and more recently on $^{88}$Mo nucleus in Ref.\,\cite{MCi11}. In all the cases studied a good correspondence between LSD modeling and experiments has been 
reported.  Theoretical predictions for $^{132}$Ce based on an extension of the LSD approach to include the modeling of the Giant Dipole Resonance width, Ref.\,\cite{KMa07,KMa08}, show a good correspondence with measurement as well. Signals of 
the presence of very large deformations at high spins have been obtained in several nuclei in the mass range $A\sim 120$ in an attempt combining the search for hyper-deformed nuclear configurations in conjunction with the Jacobi 
transitions using triple-gamma-coincidence measurements, Ref.\,\cite{BHe03}. Some preliminary results and discussion concerning the predictions of the Poincar\'e shape transitions in a few Barium nuclei can be found in 
Ref.\,\cite{KMa11} [cf.~also Ref.\,\cite{AMa10}]. 

Before becoming more precise about the exact subject of the present work let us recall some earlier efforts in the context of the shape transitions. Early calculations with the Liquid Drop Model suffered from inaccuracies in reproducing the experimental fission barriers, cf.~introduction section in~\cite{Sie86} and references therein. Publications which followed focused first of all on including a description of diffuseness properties of the nuclear surface, Refs.\,\cite{Kra79,Mol81}, reducing in this way discrepancies between the model and experimental data. Many calculations of the nuclear energies on the way to fission discussed first of all the static properties of the nuclear potential energies such as energy-positions of the minima and saddle points in function of both the deformation and increasing angular momentum. Extensive calculations in~\cite{Sie86}, in addition to reviewing the model predictions based on the techniques of those times, addressed in particular the issue of nuclear potential 
behavior in the vicinity of the characteristic (minima, saddle) points by calculating the second order Taylor expansion around those points and allowing for local description of the stiffness properties.  

In the present article we address first of all the rotation-induced shape transitions in hot rotating nuclei with the help of the Liquid Drop Model in its so-called Lublin-Strasbourg Drop (LSD) realization of Refs.\,\cite{KPD03,JDu04}, 
cf.~also Ref.\,\cite{KPo04,KP13a,FAI13}. This approach is combined with that of the collective model, allowing to go beyond traditional static description of the nuclear shape changes and calculate, among others, the most probable deformations or the most probable fission-fragment mass-asymmetry with the help of the nuclear collective wave functions. 

The presentation is organized as follows. The next Section contains the discussion of the position of our physics problem together with a short description of the macroscopic energy algorithm (LSD) chosen for this article.  In Section III we derive what appears to us as the only possible algorithm of treating and analyzing the multi-dimensional deformation spaces in which total collective nuclear energies will be calculated and possibly competing Jacobi and Poincar\'e transitions analyses. At the same time we introduce a short description of the collective model and related Schr\"odinger equation in curvilinear spaces which will be applied, within approximations, in Section VIII. Section IV describes the  technical aspects of the extension of the original LSD model with particular accent on possible inadequacies of the macroscopic modeling of the nuclear neck area. Section V contains the discussion of the technical aspects related to the classical subject of deformation dependence of the fission barriers 
whereas Section VI is focused on some selected aspects of the spherical-harmonics basis cut off properties.
In Section VII we address the issue of the spin dependence of the fission barriers, followed by the discussion of the problem of the large amplitude motion accompanying the shape transitions of the Jacobi and Poincar\'e type in Section VIII. Summary and conclusions are contained in Section IX.   


\section{Present Realization of the Liquid Drop Model}
\label{Sect.II}


In this article we aim at a {\em possibly realistic} description of the shape
transitions in hot nuclei in function of spin. By the very definition, the 
shape transitions of interest are those taking place {\em before the fission 
limit}, i.e.~at spins lower than the critical-spin values for fission,
$L<L_{\rm fiss.}$. 

Let us remark in passing that the precise values of the latter reference quantity may be difficult to determine uniquely. What we are interested in, in the present context, are the limiting spin values at which physical system looses totally its stability. In other words, the system's measured life-times become too short to be able to prove the presence of such systems in nature so that it makes no sense to dwell upon the associated shape transitions. Within the barrier-penetration model the corresponding {\em limiting barrier heights are necessarily finite}, whereas, on the other hand, one possibility would be to define $L^{\rm th}_{\rm fiss.}$ as the closest integer (half-integer) value at which calculated fission barrier totally vanishes -- in contrast to the argument just presented. As it turns out certain arbitrariness in this respect will have no impact on conclusions of the present article, whereas the notion itself will be occasionally convenient in the discussion. 
 
As a matter of convention, in what follows, we apply the term {\em transition} while speaking about the sequences of shapes in function of {\em increasing} spin - the way of speaking which has no impact on the conclusions.


\subsection{Goal's Impact on the Chosen Strategy}
\label{Sect.II-A}

One of the goals of this article is to investigate the tools that can be used in studying nuclear symmetry- and symmetry-breaking phenomena within 3D geometry [interested reader may consult Ref.\,\cite{JDu10} for the principles and an overview] -- in particular at high temperatures. One of the most successful tools capable of producing the results close to experiment within the large scale calculations is the so-called Macroscopic-Microscopic method of Strutinsky in which the {\em macroscopic} (read: Liquid Drop Model) and the {\em microscopic} (read: Phenomenological Nuclear Mean-Field Theory) combine to produce a joint scheme. However, it is important to examine through comparison with experiment not only both of these tools combined -- but when possible -- to be able to extract the information on one of these tools alone. Studying Jacobi and Poincar\'e shape transitions at high temperatures offers a {\em unique possibility of controlling  -- vs.~experiment -- the performance of the macroscopic tool 
separated from the impact of shell effects} (cf.~examples of the studies cited in the previous Section, obtained with the use of the LSD model) because of the disappearance of the shell effects at sufficiently high temperatures. 

Following this line, in the present article we focus on the macroscopic part alone employing the LSD model with which a number of rather successful pilot-projects, Refs.\,(\cite{AMa01}-\cite{KMa11}) has been performed. In the rest of this Section we discuss briefly the arguments related to precision in the description of the shape-transition aspect. A related, important but slightly more technical aspect of choosing the mathematical approach to study the properties of the nuclear potential in multi-dimensional spaces will be presented in Sect.\,\ref{Sect.III} in relation to the quantum theory of nuclear collective motion.

Any macroscopic energy expression will be capable of providing the total
potential energy maps for predefined spin sequences and thus will be able to
predict \underline{a certain} evolution of the family of shapes with spin. 
However, the model which predicts, for instance, fission barriers which are
systematically too high (too low) compared with experiment is very likely to 
provide not only the incorrect/inexact overall shape evolution with spin,
but also incorrect critical spin values at which the competing shape transitions
occur -- possibly leading to confusion. A possible undesired result would be the 
prediction of certain shape transitions at spins at which a considered nucleus 
does not exist anymore because of fission -- but other undesired mechanisms can 
also be envisaged. 

In particular: One of those unwanted effects, yet likely to occur, is related to the competition between Jacobi and Poincar\'e shape transitions involving shapes 
of distinct classes and leading, within their respective classes to different types of symmetries below and above the critical transition-spin values, say, $L^{\rm crit.}_{\rm J}$ and $L^{\rm crit.}_{\rm P}$, respectively. The Poincar\'e transitions lead to the left-right shape-asymmetry with predicted asymmetric fission fragment mass-distributions. Should Poincar\'e transitions {\em follow} ($L^{\rm crit.}_{\rm J} < L^{\rm crit.}_{\rm P}$) the one of the Jacobi type, the overall elongation of the nuclei undergoing the Jacobi transition will be larger as compared to the opposite case, the corresponding shapes very different and the predictions of the mass asymmetry drastically influenced. Should the model fail to obtain the right order - the model predictions will never be verifiable against the experiment implying a possibly very limited usefulness of the resulting macroscopic description with possibly misleading or erroneous conclusions and conflicting interpretations of the experimental data.

From the above remarks it becomes clear that through an optimization of the
description of the nuclear energies on the way to fission one may avoid  
possible undesirable effects just mentioned. In this context there are at 
least two mechanisms which seem to us obvious to be taken into account. One 
of them is related to the so-called {\em congruence energy effect} introduced 
and discussed by other authors (for details cf.~the following Sections, in 
which the corresponding mechanism will be studied in detail). The other one is
related to critical phenomena which accompany the strong shape fluctuations 
present at the Jacobi and Poincar\'e shape transitions - a mechanism 
deserving a special comment which follows.

Jacobi and Poincar\'e shape transitions represent not only {\em just a certain} shape evolution - but first of all - the characteristic symmetry breaking transformations. In the case of Jacobi transitions in function of increasing spin these are the axially-symmetric shapes which evolve fast\footnote{The phrase `fast evolution' should be understood as a relatively important change in shape at the equilibrium deformation,  accompanying relatively small increase in spin, measured e.g.~in steps of $\Delta L=2\hbar$.} into a family of tri-planar symmetry ones. Similarly, the inversion-invariant (`left-right symmetric') shapes before Poincar\'e transitions are replaced by the inversion-breaking shapes after the transition. In what follows it will be practical to introduce symbols $\mathcal{S}_{<}$ and $\mathcal{S}_{>}$ to denote the symmetries `before' and `after' one of the two types of the shape transitions. Although it does not necessarily always need to be so -- calculations 
show, that the realistic macroscopic energy expressions lead to the significant flattening of the energy landscapes at spins long below the critical spin values at which the static equilibrium (static energy minimum) shapes change their 
symmetry: $\mathcal{S}_{<} \to \mathcal{S}_{>}$. In several domains of physics these conditions give rise to critical phenomena with strong fluctuations of related observables, possibly accompanied by phase transitions, which often require a special attention.

When such a symmetry-change occurs, the flatness of the energy landscape implies that it costs very little energy for the nucleus to go from one deformation area to a neighboring one, the corresponding collective  wave-function varies 
little, the probabilities remain comparable and the nucleus undergoes a {\em large amplitude vibrational motion}. Under these conditions the {\em static} shapes i.e.~the ones corresponding to the static energy minimum and the {\em most probable} deformations, which we refer to as dynamical, may (and often do) differ considerably. So do the theory predictions e.g.~the ones related to nuclear moments and electromagnetic transitions or, possibly, fission-fragment 
mass-distributions and/or other observables -- based on the static deformations as compared to the dynamical ones. Since one of our goals is to develop the modeling which, remaining simple, is as realistic as possible, a description of the large amplitude shape fluctuations in the considered transitions will be 
taken explicitly into account. This will be done by solving the appropriate approximate form of the collective Schr\"odinger equations, as discussed in Section \ref{Sect.VIII}. 


\subsection{Comments about the Earlier Lublin-Strasbourg Drop Realization 
            of the Model}
\label{Sect.II-B}

One of the relatively recent realizations of the Liquid Drop Model, the 
Lublin-Strasbourg Drop (so-called LSD) model, c.f.~Refs.\,\cite{KPD03,JDu04}, 
introduces some extra degrees of freedom associated with the {\em curvature} of 
the nuclear surfaces. More precisely, as noticed in the cited references: One 
may introduce infinitely many geometrically distinct surfaces, which may have the same {\em surface-area}, but differing in their forms. They give rise to distinct conditions of the nucleonic motion inside of the considered nuclei and yet, within traditional Liquid Drop Models they contribute the same surface energy. The surface-curvature term as introduced within the LSD model allows to improve the description of measurable quantities such as fission barriers and masses, noticeably.

Within a classical modeling of attractive short-range interactions among 
particles surrounding a given one inside a nucleus all their contributions
will be mutually compensated, unless we approach nuclear surface. There, there
are no interaction partners {\em outside} of the delimiting surface, and an
uncompensated effective attraction pulling the particles towards nuclear 
interior will be created. This attraction will depend on the number of 
considered particles per unit volume which in turn will be different for 
e.g.~locally concave vs.~locally convex surface areas - where from the need 
of introducing the surface-curvature considerations. It then follows that for 
any varying surface with fixed surface area, the nuclear surface-energy contributions will be constant whereas the curvature contribution will vary depending on the variation of the local curvature. 

Using the concept of the curvature of nuclear surface, the parameters of the LSD
model have been adjusted to the nuclear masses in \cite{KPD03} and as it turned out, the description of nuclear fission barriers has been considerably improved with respect to the best-performing preceding versions of the Liquid Drop Model. This has been achieved without introducing any fit conditions related to the
experimental barrier heights. Such an improvement can be considered as a
demonstration of an  intrinsic, physical consistency of the LSD model which
contains one more (and known in classical physics) an effective-interaction
mechanism - the surface-curvature. 

An improved description of certain nuclear properties ultimately encourages
exploration of this new version of the Liquid Drop Model to describe nuclear
mechanisms which so far were considered to correspond to a higher level of
sensitivity and/or of realistic prediction difficulty for the models in question. Our choice here is to investigate the rotation-induced shape transitions whose critical spins depend in a sensitive way on the details of the energy expression. 

The macroscopic Nuclear Liquid Drop Model energy in its LSD form can be expressed using a number of terms representing the nuclear energy as function of the proton and neutron numbers, $Z$ and $N$, respectively, as well as nuclear 
deformation in the form:
\begin{eqnarray}
   E_{\rm total}(N,Z;\alpha;L)
   =
   E(N,Z)
   &+&
   E_{\rm Coul.}(N,Z;\alpha)                                     \nonumber\\
   &+&
   E_{\rm surf.\,}(N,Z;\alpha)                                   \nonumber\\
   &+&
   E_{\rm curv.}(N,Z;\alpha)                                     \nonumber\\                                        
   &+&
   E_{\rm rotat.}(N,Z;\alpha;L),\;\;\;\;
                                                                 \label{eqn.01}
\end{eqnarray}
where $L$ denotes nuclear collective angular momentum, and where all the deformation parameters have been abbreviated to $\alpha$. Above we find deformation-dependent Coulomb electrostatic energy term, 
$E_{\rm Coul.}(N,Z;\alpha)$, the surface, $E_{\rm surf.}(N,Z;\alpha)$, and curvature, $E_{\rm curv.}(N,Z;\alpha)$ terms (the latter characteristic of the LSD realization of the model) and the rotational energy,
$E_{\rm rotat.}(N,Z;\alpha;L)$, respectively. The first term on the right-hand side in Eq.\,(\ref{eqn.01}) denotes by definition the combined deformation-independent terms -- possibly including the original, deformation-independent 
version of the so-called congruence energy expression -- see below. 

The above expression can be standardized to represent the atonic mass. In such a case the deformation independent term is given by (for a complete atomic mass formula used here cf.~also~Ref.~\cite{KPD03} and 
references therein):
\begin{eqnarray}
      E(N,Z) 
      &=&  
      Z M_{\rm H} 
      + 
      N M_{\rm n} 
      - 
      0.00001433 \, Z^{2.39}                                     \nonumber \\
      &+&
      E_{\rm vol.}(N,Z)
      +
      E_{\rm cong.}(N,Z),
                                                                 \label{eqn.02} 
\end{eqnarray}
where the term proportional to $Z^{2.39}$ parametrizes the binding energy of 
the electrons whereas the other two terms represent $Z$ masses of the Hydrogen 
atom and $N$ masses of the neutron, respectively. Deformation-independent congruence energy term as used in the literature in the past, will be replaced
by a deformation-dependent one, c.f.~Eq.\,(\ref{eqn.09}) and the surrounding text. 

The volume energy above is parametrized as:
\begin{equation}
      E_{\rm vol.}(Z,N)
      =
      b_{\rm vol.}\,(1 - \kappa_{\rm vol.} \, I^2\,)\,A  ,      
                                                                 \label{eqn.03} 
\end{equation} 
where $I=(N-Z)/(N+Z)$. [All the parameters appearing implicitly in Eq.\,(\ref{eqn.01}), such as $b_{\rm vol.}$ and $\kappa_{\rm vol.}$ and the ones which appear below, are collected in TABLE \ref{tab.01}.] 

The Coulomb LDM term in its `traditional form' reads:
\begin{equation}
      E_{\rm Coul.}(N,Z;\alpha)
      =
      \frac{3}{5} \, 
      e^2 \frac{Z^2}{r_0^{\rm ch} A^{1/3}}\, B_{\rm Coul.}(\alpha) 
      - 
      C_{4}\frac{Z^2}{A}, 
                                                                 \label{eqn.04} 
\end{equation} 
with the mass number $A=Z+N$, electric charge unit denoted $e$, and the 
so-called charge radius parameter $r_0^{\rm ch}$. The term proportional to
$Z^2/A$ represents the nuclear charge-density diffuseness-correction whereas the deformation dependent term, $B_{\rm Coul.}(\alpha)$, denotes the Coulomb energy of a deformed nucleus normalized to that of the sphere with the same volume. 

The surface energy in its `traditional' LDM form reads:  
\begin{equation}
      E_{\rm surf.}(N,Z;\alpha)
      =
      b_{\rm surf.}\,(1 - \kappa_{\rm surf.} I^2\,)\,A^{2/3} 
      B_{\rm surf.}(\alpha) .
                                                                \label{eqn.05} 
\end{equation}
Above, the deformation dependent term is defined as the surface energy of a deformed nucleus normalized to that of the sphere of the same volume. 

The curvature term is given by:
\begin{equation}
      E_{\rm curv.}(N,Z;\alpha)
      =
      b_{\rm curv.}\,(1 - \kappa_{\rm curv.} \, I^2\,)\,A^{1/3} 
      B_{\rm curv.}(\alpha)
                                                                \label{eqn.06} 
\end{equation}
with
\begin{equation}
     B_{\rm curv.}(\alpha) 
     =
     \int_0^\pi \hspace{-2mm} d\vartheta \int_0^{2\pi} \hspace{-3mm} d\varphi
     \left[
          \frac{1}{R_1(\vartheta,\varphi;\alpha)}
          +
          \frac{1}{R_2(\vartheta,\varphi;\alpha)}
     \right], \qquad
                                                                \label{eqn.07} 
\end{equation}
where $R_1$ and $R_2$ are deformation-dependent principal radii of the nuclear surface at the point-position defined by spherical angles $\vartheta$ and $\varphi$.

Finally, the rotation-energy term is defined as usual by
\begin{equation}
      E_{\rm rot.}(Z,N;\alpha)
      =
      \frac{\hbar^2}{2\mathcal{J}(Z,N;\alpha)} L(L+1),
                                                                \label{eqn.08} 
\end{equation}
with the classical moment of inertia $\mathcal{J}$ calculated at the given deformation $\alpha$. In the following applications we assume, without loosing generality, that rotation takes place about the $\mathcal{O}_y$-axis. 

The parameters entering all the above expressions are given in TABLE \ref{tab.01}.
\begin{table}[h]
\caption[TT]{The parameters of the LSD model fitted to the
             measured atomic masses only (from Ref.\,\cite{KPD03}).
                                                                \label{tab.01}}
\begin{center}
\begin{tabular}{ccr}
        Term         &\quad Units & \quad LSD$\;$  \\
                                                      \hline\hline
$b_{\rm vol.}$        &\quad  MeV  &\quad -15.4920  \\
$\kappa_{\rm vol.}$   &\quad   1   &   1.8601       \\
                                                      \hline
    $b_{\rm surf.}$   &\quad  MeV  &  16.9707       \\
 $\kappa_{\rm surf.}$ &\quad   1   &   2.2938       \\
                                                      \hline
    $b_{\rm cur.}$    &\quad  MeV  &   3.8602       \\
 $\kappa_{\rm cur.}$  &\quad   1   &  -2.3764       \\
                                                      \hline
        $r_0^{ch}$    &\quad  fm   &   1.21725      \\
        $C_{4}$       &\quad  MeV  &   0.9181       \\
                                                      \hline\hline
\end{tabular} 
\end{center}
\end{table}

In the present article the parameters given above are kept without modification, whereas the deformation-dependent congruence energy term will introduce a new parametric freedom as discussed in the following Section.


\subsection{Comments about Deformation-Dependent Congruence Energy Term and
            Critical Spins}
\label{Sect.II-C}

The congruence energy contribution to the nuclear Liquid Drop Model has been
originally introduced in a purely phenomenological manner without taking into
account its possible deformation dependence in \cite{WDM66}. It has been
modified next by introducing a multiplicative shape-dependent factor,
\cite{WDM97,WDM99}, aiming at the improvement of the description of the fission
process and, in particular, of the transformation of an original parent nucleus
into two separated fission fragments. The shape-dependent factor in
Ref.\,\cite{WDM97} has been defined in terms of the ratio between the radius of the neck and the mean value of the radii of the nascent fragments, 
cf.~Eqs.\,(5) and (7) in the quoted reference, whereas in Ref.~\cite{MNS89} in terms of the cross-sections through the neck and the maximum cross-section through the smallest nascent fragment. Such phenomenological definitions are based on the intuition that for geometrically-compact shapes, i.e.~relatively far from the neck formation, the congruence energy should not be sensitive to small deformation changes. To the contrary, this factor is expected to have an increasing impact on the total nuclear energy -- caused by the congruence 
mechanism for the more and more necked-in shapes. It should be emphasized at this point that the intuitive argumentation quoted can by no means be treated as a replacement for a better founded microscopic one [which however, to our knowledge, does not exist in the literature so far].

The arguments in favor of introducing  deformation dependence in the congruence
energy term bring us to the necessity of modification of the structure of the
original LSD expression in (\ref{eqn.01}) in that
\begin{equation}
   E(N,Z) \to E_0(N,Z) + E_{\rm Cong.}(N,Z;\alpha),
                                                                 \label{eqn.09}
\end{equation}
where the new deformation-independent term becomes 
$
  E_0(N,Z) = Z M_{\rm H} 
           + 
           N M_{\rm n} 
           - 
           0.00001433 \, Z^{2.39}
           +
           E_{\rm vol.}
$, 
whereas deformation-dependent congruence-contribution is denoted 
$E_{\rm Cong.}(N,Z;\alpha)$ from now on.

As already mentioned, our goal is, among others, to investigate high-temperature competition between the nuclear Jacobi- and Poincar\'e-type transitions in function of increasing spin. Their experimental detection is relatively indirect but can be achieved for instance by investigating the shape of the Giant Dipole Resonance in function of nuclear spin, 
Refs.\,\cite{AMa01,AMa04,MKm07,MKm05,DPa10,MCi11,KMa07,KMa08}. Poincar\'e shape transitions in turn consist in shape transformations that break the left-right symmetry thus leading to the asymmetric fission-fragment 
mass-distributions. Even though experimental tests in question may require very distinct instrumental conditions to address each of the discussed mechanisms separately, their combination focusing on the same nuclei in independent measurements may be necessary for testing the theoretical model predictions and arriving at a better understanding of the underlying physics. 

To be able to optimize description of observables which can be tested experimentally, we will need to determine the critical spin-values corresponding to the onset of tri-axial (Jacobi) and left-right asymmetry (Poincar\'e) shape transitions. Introducing the deformation-dependent congruence energy term will be shown to be one of the most important elements of improvement especially for certain mass regions (cf.~Sect.~\ref{Sect.IV}), in addition to the quantum description of large-amplitude shape-fluctuations driven by flatness of the nuclear energy surfaces, Sect.~\ref{Sect.VIII}. 

Analogous considerations which explicitly include thermal shape-fluctuations,
describe rather satisfactorily the strength function of the Giant Dipole
Resonance (GDR). The latter have been excited and successfully analyses \cite{AMa04} in a few hot rotating compound nuclei in order to study the Jacobi shape transition what encourages the extension of this type of techniques to a more systematic analysis of various nuclear shape fluctuations -- after improving further the performance of the LSD approach. 

In the following Sections we illustrate and discuss our present realization of the Lublin-Strasbourg Drop model in which we include a deformation-dependent
congruence energy contribution in an attempt to improve further the description
of the experimental data on fission barriers. We believe that by doing so we may
achieve an extension of the range of applicability of the macroscopic model at hand to a more refined level of precision. In such a way we could address a delicate balance between Jacobi and Poincar\'e shape transitions at 
high spins especially in the presence of the dynamical effects of the large-amplitude oscillations   and, moreover, introduce a simplified, approximate description of these dynamical effects preceding nuclear scission. 


\section{COLLECTIVE MOTION AND~POTENTIAL-ENERGY HYPER-SURFACES 
         IN~MULTIDIMENSIONAL SPACES}
\label{Sect.III}


We wish to emphasize right at the beginning that employing nuclear potential energy of collective motion in order to provide theory predictions comparable to experiment is ultimately related to the problem of the deformation dependent collective inertia. Indeed the role of the inertia tensor briefly summarized in this Section represents the {\em sine qua non} condition for the theoretical estimates of e.g.~fission life-times as well as those of the shape isomers -- but of course practically all the observables associated with collective motion such as transitions and their probabilities. Consequently, model calculations {\em not including} this element of the theory are in the best case {\em approximations} already on the level of the position of the problem -- not counting further limiting approximations which usually differ from one group of the authors to another (for a brief account of the most typical simplifications and certain of their consequences -- see below).

The role of the collective inertia (tensor) is also one of the most difficult elements for the direct comparison with experiment. Indeed, such comparison can only be performed at a certain significance level after solving collective 
Schr\"odinger equation [Eq.\,(\ref{eqn.13}) in Sect.\,\ref{Sect.III-A}] in the spaces of sufficiently rich dimensionality and for sufficiently many nuclei -- a very complex and challenging task which, to our knowledge, has not been achieved yet except in a rather limited contexts. In the past: Many authors ignored this very basic level of dynamical, multidimensional calculations altogether, or, alternatively, replaced the full multidimensional treatment by a technique involving one-dimensional path integrals.  

One of the important consequences of the quantum and dynamical nature of the  collective problem which, in our opinion is not sufficiently stressed in the literature, is that many apparently important details of the potential hyper-surfaces are naturally `smeared out' when the solutions of the Schr\"odinger equation in the curvilinear spaces are obtained [also with one-dimensional approximations including the inertia tensor]. It then follows that the `very-very exact' numerical properties of the local minima and saddle-points -- whose precise numerical analysis is a non-trivial problem as briefly discussed below -- are partially lost on the way to the final result expressed in terms of probabilities or the collective wave functions. We believe (without discouraging as precise as possible a numerical calculation) that this information may allow for a certain flexibility in finding an optimum between numerical rigor in finding precise positions of minima and saddle points and the computer c.p.u.~time -- 
especially in the large scale calculations.


\subsection{Guidelines Implied by Quantum Theory of~Collective Motion}
\label{Sect.III-A}


Let us begin by a short summary of the general framework of the quantum theory of collective nuclear motion. It introduces principal notions such as the  deformation dependent mass tensor or curvilinear spaces of collective variables and places the potential-energy hyper-surfaces at the right perspective of only one among several factors which combine together in the correct quantum description of nuclear collective motion. Even though for mathematical simplicity reasons we will not use the microscopically calculated mass tensor in this particular article (such project is in progress and the results will be published elsewhere) by beginning with this more general framework we wish to remind the reader about certain general guidelines. 

This part of the discussion will at the same time serve as an introduction to the treatment of the large amplitude motion which in this article will be based on an approximate version of quantum theory of nuclear collective motion (cf.~Sect.~\ref{Sect.VIII} for applications and illustrations).

The number of nuclear collective variables necessary to describe the shape effects realistically needs to be larger than two, what implies that the easy analyses which can immediately be tested using two-dimensional contour plots will not be applicable. Mathematically our problem consists in calculating, analyzing and interpreting the behavior of a scalar function, say $V$ (nuclear potential energy) of a vector argument $\{\alpha_1,\alpha_2,\,\ldots\,\alpha_n\}\equiv \alpha$, as well as a symmetric $n \times n$ inertia tensor, 
$B_{\alpha_{i};\alpha_{j}}(\alpha)$, in an $n$-dimensional vector space, of collective coordinates, $\mathbb{R}^n$. Any function of this type or its projection on a sub-space of smaller number of dimensions is referred to as hyper-surface. 

As it is well known, {\em in principle}, the description of the nuclear motion with the help of the nuclear surface is a fully quantum process described within the collective Hamiltonian, the collective model of Bohr being one of the best known examples. In such an approach one begins with the concept of the inertia tensor entering a classical (to start with) kinetic energy expression  
\begin{equation}
   T_{\rm class.} = \frac{1}{2} \sum_{i,j} 
                    B_{\alpha_{i};\alpha_{j}}(\alpha)
                    \dot{\alpha}_{i}\,\dot{\alpha}_{j} ,
                                                                 \label{eqn.10}
\end{equation}
where $\{\alpha_i\}$ are for the moment unspecified collective coordinates defining the equation of the nuclear surface [for details cf.~e.g.~Ref\,\cite{MBr72}]. The components $B_{\alpha_{i};\alpha_{j}}(\alpha)$ of the mass tensor are usually calculated microscopically using advanced methods of perturbation theory. In relation to, for instance, the usual phenomenological mean-field Hamiltonians such as those based on the deformed Woods-Saxon or Yukawa-folded potentials, the corresponding formulae can be obtained explicitly  following the proposition in Ref.\,\cite{STB59} [though the calculations must be performed numerically; an extended discussion of this problem can be found in \cite{MBr72} whereas for an alternative approach applying the Generator Coordinate Method in the description of the nuclear inertia, the reader is referred e.g.~to Ref.\,\cite{AGo85}]. The quantum version of the Hamiltonian with the classical kinetic energy as in Eq.\,(\ref{eqn.10}) is obtained using the standard by now a 
quantization procedure of Podolsky, Ref.\,\cite{BPo28}. The collective Schr\"odinger equation (whose approximate version will be solved below in Sect.~\ref{Sect.VIII}) is obtained with the Hamiltonian whose form, after Podolsky's quantization, reads:
\begin{equation}
   \hat{H}_{\rm quant.}
   =
   \frac{1}{2m} \sum_{i,j=1}^n
   B^{-1/4} \hat{p}_i \; 
   \left( B^{{\alpha_i} {\alpha_j}}  B^{1/4} \right) 
   \hat{p}_j\, B^{-1/4} + V \,.
                                                                 \label{eqn.11}
\end{equation}
Here, $B=\det[{B^{{\alpha_i} {\alpha_j}}}]$ is the determinant of the mass tensor whereas $\hat{p}_j$ is a canonically conjugate momentum, associated with the generalized coordinate (in our case: deformation coordinate) $\alpha_i$, whereas $V$ is the collective nuclear potential, e.g., the one calculated using Strutinsky method or, as in our case, the Liquid Drop Model.

The role of the tensor of inertia enters not only on the level of the equations of motion with Hamiltonian (\ref{eqn.11}) but very importantly through a `reinterpretation' of the probability of finding the system in the curvilinear space. The latter probability is given by
\begin{equation}
    d\,P_N(\alpha) 
    = 
    \Psi_N^*(\alpha)\Psi_N^{}(\alpha)\,
    \sqrt{B} \,d\alpha \, ,
                                                                 \label{eqn.12}
\end{equation}
where $\Psi_N$ are the solutions of the collective Schr\"odinger equation
\begin{equation}
   \hat{H}_{\rm quant.} \Psi_N = E_N\, \Psi_N \, .
                                                                 \label{eqn.13}
\end{equation} 
Very importantly, in multidimensional spaces it is then sufficient that a few components of the inertia tensor increase in a certain range of the deformation space and the determinant factor in Eq.\,(\ref{eqn.12}) grows very quickly as a functions of the {\em products} of various components of the tensor. This mechanism may turn out occasionally being very important in that the calculated maximum probabilities do not generally follow neither the static minima on the potential hyper-surfaces nor the steepest descent valleys whereas the most probable barrier transmission paths avoid the static saddle points. How importantly the probability expression in Eq.\,(\ref{eqn.12}) may change the simplified interpretations based on the static potential energy hyper-surfaces can be seen e.g.~from the illustrations in Figs.\,(1-3) in Ref.\,\cite{ADo11}.

The collective model schematized above has been applied rather rigorously in the description of the leading, quadrupole nuclear collective motion by several authors and the reader is referred to the rather complete review in Ref.\,\cite{LPr09}. This is a rather exceptional, and indeed rare case in the present context, where the principles of the quantum and microscopic theories are followed down to the final solutions including the calculations of the electromagnetic transition probabilities. 

In contrast, and as already mentioned, with the applications to large amplitude motion in mind such as in nuclear fission, extra simplifying assumptions have been often employed which consist in combining the one-dimensional approximation with path integrals and the Ritz-Rayley method, cf.~e.g.~Ref.\,\cite{ABa81} or using projections (e.g.~on two dimensional subspaces to avoid mathematical complexities of working in the full $n$-dimensional space). 

Not entering into details, a general and rather global conclusion from these studies was that the most probable path through the multidimensional barriers generally {\em does not} pass through the saddle points [more precisely: in the quantum mechanical, dynamical description -- the probability of passing through the saddle (or any other predefined point) is strictly zero.]
Moreover, the idea of considering probabilities for passing through a given area of the deformation space surrounding the static minimum point to another area surrounding another static characteristic point (e.g.~another minimum or scission) along a single path has been implicitly criticized stressing the importance of another mechanism taking the form of `dynamical corrections' -- those originating from the mechanism of the zero-point motion cf.~e.g.~\cite{KPo08}. 

Let us summarize this part of the discussion as follows:
\begin{itemize}
\item The description of the nuclear quantum systems, especially in the regime of the large amplitude motion should employ a quantum formalism such as e.g.~one of the realizations of the collective model;
\item The corresponding description of the motion should take into account the dynamical effects naturally modeled with the help of the inertia tensor; 
\item The nuclear potential energy calculated with the help of either Hartree-Fock, or Strutinsky, or its classical parameterization as e.g.~with the help of the macroscopic models (LDM, LSD etc.) -- plays the role of the potential in the collective Schr\"odinger equation [cf.~Eq.\,(\ref{eqn.13})];
\item Any alternative approach must be treated as an approximation whose applicability and consequences should be separately checked on the case by case basis [for instance drawing experiment-comparable conclusion out of static energy landscapes alone can only be approximate and/or schematic].
\end{itemize}
A couple of other observations should be recalled as certain logical consequences:
\begin{itemize}
\item Agreement between the model results and experiment does {\em not} replace the proof nor discussion of validity and physical adequacy of assumed approximations. Such an agreement could be accidental or a result of compensating errors originating from two or more approximations; 
\item Limitations of each given model should be taken into account first, before attempting the model's optimization or adaptation to avoid contributions from the unphysical and/or uncontrollable regimes (e.g.~macroscopic models, unless proven to the contrary, have no control on the specific quantum processes in the neck region, such as e.g.~cluster formation, especially at extreme deformations and any optimization to this particular geometrical range will most likely deteriorate the model's predictive power, see also below).
\end{itemize}

In the present article, to comply at least partially with the first series of four {\em desiderata} above, we are going to address the question of the shape transitions using potential energies calculated with the macroscopic model and solving an approximate version of the collective nuclear Schr\"odinger equation, 
Sect.\,\ref{Sect.VIII}.

Concerning the second series of remarks we are going to address specifically the problem of the model's performance in the range of the highly developed neck on the way to scission, Sect.\,\ref{Sect.IV}.


\subsection{Guidelines Implied by the Multidimensional Character of the Collective Motion}
\label{Sect.III-B}


The starting point in the construction of macroscopic (and macroscopic-microscopic) nuclear models consists in defining the underlying class of geometrical surfaces and to parametrize nuclear shapes with the collective coordinates within a sub-set of $\mathbb{R}^n$. One of the strategies found in the literature consists in focussing on a very limited number of parameters while preserving the necessary minimum of `really needed' degrees of freedom. For instance, parameterizations employing two spheroids, either overlapping or smoothly joined, were in use as e.g.~the ones joined by another spheroid or a hyperboloid to chose between necked-in and already separated system, Ref.\,\cite{JRN69}; discussion of the latter and some alternative, few-parameter choices, can be found in Ref.~\cite{MNS89}. Some other choices involve elongation, tri-axiality and the left-right asymmetry (see below). 

The issue of parameterization of the nuclear shape is closely related with the mathematical implications on the level of determining the characteristic points of physical interest such as saddle points in multidimensional spaces. This problem is far from trivial and far from being solved as witnessed by contradictory view points that can be found in the literature. Therefore without entering any polemic neither reviewing the issue let us first recall the view points represented by various authors to the extent these contradicting ideas may concern the present study.  

Parameterizations based on any prefixed number of deformation coordinates imply necessarily certain restrictive consequences. At this point it may be instructive to recall a number observations and conclusions of the recent Ref.\,\cite{NDu12} which discusses the uncertainties related to the saddle points in multidimensional spaces. The observations relevant for us are those:
\begin{itemize}
\item Every point which appears as stationary (e.g.~minimum) on the energy surface in an $n$-dimensional space {\em will in general not be stationary} in the $m$-dimensional space for $m>n$;
\item Every point which appears as a saddle in an $n$-dimensional space {\em will in general not be a saddle point} in the $m$-dimensional space with $m>n$.
\end{itemize}
The latter item has natural implications for the present project: Since we wish to study a possible coexistence between the shape transitions of two competing distinct symmetries -- for this type of the project we {\em must not use} any coordinate space of any prefixed dimension. In other words: We must consider a shape representation in terms of an $n$-dimensional subset of the (in principle infinite) basis set of functions, with $n$ playing a role of the control (basis cut off) parameter. 

Yet another set of indications can be drawn from the contradicting conclusions found in the literature about certain presupposed {\em geometrical symmetries} built-in the model:
\begin{itemize}
\item Certain authors working with a prefixed deformation set of an {\em axial symmetry} coordinates within macroscopic-microscopic model conclude obtaining a satisfactory description of the fission barriers in the Actinide nuclei;
\item Other authors, working within the framework of the covariant density functional theory, Ref.\,\cite{HAb10}, (cf.~also Ref.\,\cite{HAb12}) conclude, in relation to the Actinide nuclei, that {\em ``...~they are able to describe fission barriers on a level of accuracy comparable with non-relativistic calculations, even with the best phenomenological macroscopic+microscopic approaches. Tri-axiality in the region of the first saddle plays a crucial role in achieving that...~''.} 
\end{itemize}
Conclusion for the present project: {\em No basis set of any predefined symmetry} can be used in the following discussion; in particular axial and non-axial shapes must be simultaneously taken care of to be able to treat the Jacobi and Poincar\'e shape transitions on the same footing and possibly take position with respect to controversies of the type just quoted.  


\boldmath
\subsection{The Issue of Shape Parameterization and Minimization Scheme in Multipole Space $\{\alpha_{\lambda \mu}\}$} \unboldmath
\label{Sect.III-C}


So far we have arrived at the following indications: In contrast to the choices in certain previous publications we {\em must not} work with neither axially symmetric nor prefixed dimensionality basis when describing the nuclear shapes. Taking this as the guidelines, let us examine the remaining properties, options and conditions which should be considered at this point.

When working with a prefixed dimension of the coordinate space one profits from the implied simplification, while necessarily introducing an uncontrollable rigidity of the model. Indeed, in addition to undesirable features already quoted, it will be very difficult, if not impossible in practice, to verify which class of shapes, that may potentially become important in one or another application, is simply inaccessible within arbitrarily predefined family of shapes. Similarly, it will not be possible to assure that the energies obtained as the result of the minimization correspond indeed to the full variational capacity of the model e.g.~when minimizing nuclear energies.   

In the present realization of the LSD model we wish to make the results of the
calculations independent of the limitations mentioned. This can be done possibly
on the expense of the time of the numerical calculations - and the only way to
achieve such a goal is to expand the nuclear surface in terms of a {\em basis set of functions} such as spherical harmonics $\{Y_{\lambda,\mu}\}$. The latter have been used for a long time in the present context (e.g.~Refs.~\cite{WJS56,CSw63,JRN67}):
\begin{equation}
      R(\vartheta,\varphi)
      = 
      R_{0}c(\alpha)
      \big[ 
           1 + \sum_{\lambda=2}^{\lambda_{max}} 
               \sum_{\mu=-\lambda}^{\lambda} 
               \alpha_{\lambda \mu}^{\star} 
                    Y_{\lambda \mu}^{} (\vartheta,\varphi)
	  \big],	                                                            
                                                                 \label{eqn.14}
\end{equation}
where $\alpha\equiv\{\alpha_{\lambda\mu}\}$ and where function $c(\alpha)$ is
obtained from the nuclear volume conservation condition. The maximum
multipolarity used, $\lambda_{max}$, plays the role of the basis cut-off
parameter and replaces the generic symbol for the space dimension, $n$, used so far. It is an advantage of such an approach that by increasing the cut-off
we can test and decide about achieving (or not) the stability of the final
result with respect to the selected set of the basis functions under the user-chosen stability criteria, cf.~Sects.\,(\ref{Sect.IV}-\ref{Sect.VI}). 

Let us remark at this point that an axially-symmetric analogue of the expansion in terms of the spherical harmonic basis has been considered in Ref.~\cite{Tre80} with the Legendre-polynomial expansion, which can be considered as a particular case of the spherical harmonics series. However, in contrast to 
Eq.\,(\ref{eqn.14}) in which certain exotic forms can not be obtained when the point-position on the nuclear surface can not be expressed as a unique function of $\vartheta$ and $\varphi$, in the quoted reference the expansion variable used was the distance from the nuclear axis to the nuclear surface, thus allowing for a description of shapes going beyond the binary-fission configurations (with e.g.~a possibility of parameterizing three- or four-fragment configurations). The present approach does not include such possibilities, but on the other hand examining the Jacobi and Poincar\'e transitions does not require that.

Minimization of a scalar function depending on vector arguments is a task whose complexity increases with increasing dimension of space and with the {\em degree of non-linearity of the function studied}. In this context the macroscopic model provides simplifications (incomparable to the complexity of the self-consistent Hartree-Fock or relativistic mean-field approaches) which are inherent to the physical nature of the macroscopic energy problem: 
\begin{itemize}
\item The minimized function, the nuclear macroscopic energy, is a very regular
      function of its arguments $\alpha_{\lambda \mu}$ in the physical range of
      their application;
\item With the exception of $\alpha_{20}$ which parametrizes nuclear elongation,
      the majority of the deformation parameters remain not very far from the
      origin $\{\alpha_{\lambda \mu}=0\}$ of the deformation-space 
      reference-frame and seldom approaches 1, except for, occasionally, 
      $\alpha_{30}$ and $\alpha_{40}$;
\item Within the physical range of (i.e.~not too `extravagant') shapes, on the
      average, the larger the multipolarity, the smaller the variation range of
      $\alpha_{\lambda \mu}$, especially within the energy range not very far, 
      say: several MeV, from the absolute minimum;
\item The same observations expressed in other words: With increasing
      multipolarity $\lambda$ the nuclear surface area increases rather fast
      followed by a rapid increase in the (positive) surface energy. Those
      latter areas are avoided by the minimization algorithms.
\end{itemize}

These properties are generally {\em not} satisfied for just any parameterization invented by a physicist -- and, in the present case, they considerably facilitate the use of the minimization algorithm both in terms of the 
stability of the minimization process converging towards the absolute minimum as well as the computer c.p.u.~time. 

Some comments will be in place at this point, related to the strategy of extracting the physics information out of potential energies in the multidimensional shape-coordinate spaces. There have been essentially two techniques applied in the past in this context. One consists in simply tabulating the energies in the $n$-dimensional prefixed {\em \`a priori} a mesh of deformation points with the numbers of the mesh points up to slightly in excess of $10^6$ or less, as reported in the literature. Such an approach suffers from all the potential disadvantages of working with the deformation space of an {\em \`a priori} predefined dimension as discussed above. We must not use such an approach since varying the size of the mesh to be able to study the stability of the final conclusions with respect to increasing the space-size would bring us easily to the mesh of $10^{8} - 10^{9}$ points which would neither be practical nor necessary in the context, see what follows. 

Under these conditions we are left with an alternative, referred to as {\em stochastic technique of projections} as briefly described in the rest of this Section. It consists in calculating the projections of the total energies on the preselected sub-spaces, for instance two-dimensional projections, with the important condition of the repetitive random restarts for any projection point in question. The potential disadvantages may manifest themselves especially when the minimization techniques are not used properly (for instance ignoring the random restarts) in the form of discontinuities which arise especially in the case of self-consistent iterative approaches such as Hartree-Fock methods (cf. Ref.\,\cite{NDu12} for illustrations).

In this article we illustrate the LSD calculation results introducing either one-dimensional energy projections minimizing over various multipole deformations in function of the elongation $\alpha_{20}$, or, alternatively, 
two types of two-dimensional projections. In this second case we project the total energy on the plane of the leading (quadrupole) deformation parameters 
$\{\alpha_{20},\alpha_{22}\}$ using, equivalently, $\{\beta,\gamma\}$-representation -- to address the issue of the Jacobi shape transitions. Alternatively, we employ the $\{\alpha_{20},\alpha_{30}\}$ projections minimized over a number of multipole deformations to illustrate the Poincar\'e type transitions. 

In all the three cases described above we will use the Levenberg-Marquardt non-linear minimization algorithm that is known for its stabilized linear-search properties which increase an over-all stability of the method. In this article, 
minimization algorithm is combined with the standard multi-restart procedure according to which:
\begin{itemize}
\item The initial minimization points are selected at random in a large 
      hyper-cube containing as a rather small subset the physical space of
      interest;
\item We use $N_{\rm rest.}$ random restarts to control the obtained consistency 
      and continuity of the final surface (curve) of interest as well as
      the independence of the final result of the starting point of the
      minimization routine; 
\item Each time the minimization stops when the zero-gradient condition is
      verified within the pre-defined criterion, the consistency of the results 
      obtained from various starting points is analyses and the solution
      satisfying the continuity criteria chosen;
\item This technique allows in particular to detect the mechanism of `crossing
      valleys' as discussed in Ref.\,\cite{NDu12}, otherwise very difficult 
      to treat with the help of automatized search routines.
\end{itemize}
As it is well known, stochastic approaches do not offer mathematical guarantees for satisfying the properties for which no mathematical criteria can be formulated (as in the case of the potential energy hyper-surfaces for which even the adequate dimension of the coordinate space remains unknown and is subject to various `practical' and {\em ad hoc} criteria). However, using these methods one can increase the probability of reaching a success -- {\em the best one can do under the discussed conditions}.

In the calculations presented below we will employ the minimization over up to
12 deformation coordinates simultaneously with $\lambda_{\rm max}\leq 16$ - after having verified that the stability of the final result has been achieved in the context of interest.

Extensive use and tests of the minimization algorithm applied in the variable dimensions of the deformation spaces used in this article convince us of the absence of possible discontinuities and/or other irregularities in the present context which {\em may} arise in the case of alternative shape parameterizations and/or very exotic shapes as well as less careful handling of the non-linear minimization algorithms. 


\section{EXTENDED LSD-MODEL FORMULATION}
\label{Sect.IV}


In this Section we are going to discuss first the properties of stability of the final results with respect to the cut off in terms of axial symmetry sub-set of the spherical harmonic basis, Sect.\,\ref{Sect.IV-A}, and the axial asymmetry, in Sect.\,\ref{Sect.IV-B}. Having stabilized the convergences properties of the algorithm with respect to the basis symmetries and cut-off, we will obtain a parameterization of the deformation-dependent congruence energy term with certain criteria specified below, Sect.\,\ref{Sect.IV-C}, finishing by formulating a series of comments about necking properties in Sect.\,\ref{Sect.IV-D}. 


\subsection{Basis Stability Conditions: Axial Symmetry}
\label{Sect.IV-A}


To begin with, let us recall that the extended LSD energy expression developed in this article contains the deformation-dependent congruence-energy term whose presence strongly influences the quality of comparison of the nuclear fission barriers (for illustration cf.~Table \ref{tab.02} below) with experiments and increases the impact of the higher-$\lambda$ multipoles at the 
large-elongation limits. When discussing the stability with respect to the basis cut-off in this Section we will use the full, extended LSD expression, including deformation dependent congruence energy term, despite of the fact that the parameterization of the congruence contribution will be {\em presented and illustrated in the following Section only}.

We start by examining the conditions of the basis cut-off in relation to the order $\lambda$ of the spherical-harmonic basis $\{Y_{\lambda\mu}\}$, for the axial-symmetry deformations $\alpha_{\lambda \mu=0}$. Since we will be interested in particular in the Jacobi shape transitions involving non-axial (to the first order tri-axial shapes) we will specifically illustrate also the role of the higher-order tri-axiality degrees of freedom, $\alpha_{42}$ and 
$\alpha_{62}$, the natural `partner' deformations possibly coupling with the quadrupole-triaxial deformation $\alpha_{22}$, in the next Section as the next step.

An undesirable feature of a description of the nuclear surfaces in
terms of any set of basis functions is an appearance of {\em local surface fluctuations} which are physically meaningless in the context - yet an unavoidable consequence of expanding a given function in terms of the basis functions which one way or another correspond to polynomials of an increasing order. This mechanism can be present in particular at a large elongation, especially in the presence of the nuclear neck. More generally, an attempt to describe surfaces with locally strong curvature may always be accompanied by such fluctuations. 

We are confronted here with contradicting tendencies well known in the discussed context: On the one hand-side, the tendency to increase the basis size in order to improve the variational minimization conditions and lower the energy of the final solution, but on the other hand increasing the presence of (to an extent meaningless) small amplitude fluctuations on the nuclear surface. In order to examine the behavior of the LSD energy expression in function of increasing basis cut-off parameter, $\lambda_{\rm max}$, we have performed two types of tests. 

First, we have tested the total energy behavior in function of the increasing quadrupole deformation $\alpha_{20}$ (also referred to as elongation parameter) at spin zero by minimizing the total energy over the axial-symmetry deformation-parameters $\alpha_{\lambda\mu=0}$, for increasing even $\lambda$. 
\begin{figure}[htbp!]
   \begin{center}
     \includegraphics[width=8.0cm]{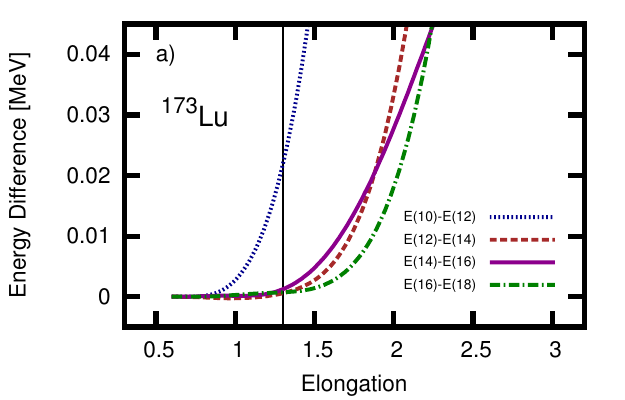}\vspace{-0.1cm}
     \includegraphics[width=8.0cm]{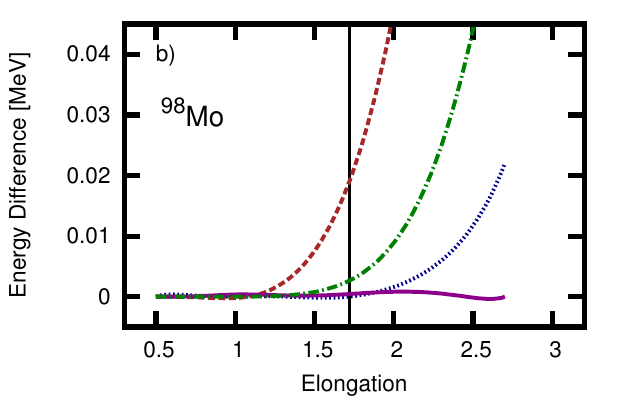}\vspace{-0.1cm}
     \includegraphics[width=8.0cm]{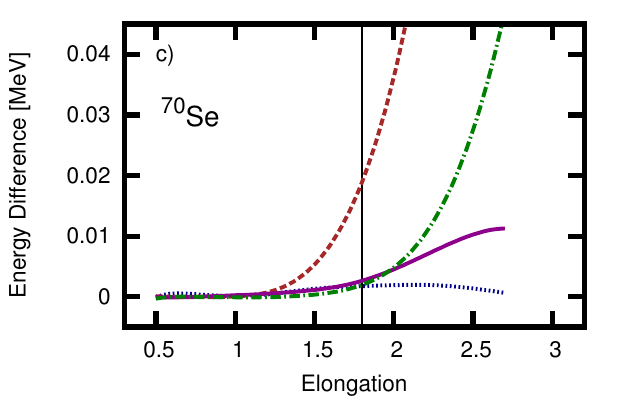}
     \caption{Axial-basis cut-off stability-test for three nuclei
              representative for the mass ranges illustrated in this article.
              We select as a measure of stability the energy differences of
              Eq.\,(\ref{eqn.15}). By definition vertical lines illustrate the
              elongation at which the strongest contribution equals 20 keV.
              Observe that, characteristically, the convergence in traditional
              sense (`the higher the $\lambda$-value the smaller the 
              discrepancy') does not apply here. [For comments, see 
              Figs.\,\ref{fig.02}-\ref{fig.03} and discussion in the text.]
              }
                                                                 \label{fig.01}
   \end{center}
\end{figure}
Results in Fig.~\ref{fig.01} show the energy differences: 
\begin{equation}
       \delta E_{\lambda_{\rm max.}}(\alpha)
       \equiv
       E_{\lambda_{\rm max.}-2}(\alpha)
       -
       E_{\lambda_{\rm max.}}(\alpha) \, ,
                                                                 \label{eqn.15}
\end{equation}  
for $\lambda_{\rm max.}=10,12,14,...$, for three nuclei representing the mass ranges on which we focus in this article. Since above we subtract the result with the richer basis size from the result with the poorer basis size, the corresponding difference is necessarily non-negative and we keep this convention for the graphical convenience.

As it can be seen from Fig.\,\ref{fig.01}, up to certain value of the elongation the nuclear energy is stabilized with respect to adding new deformation parameters at the level of a few keV in terms of discrepancies between various multipolarity contributions -- already for $\lambda_{\rm max}=12$. This characteristic limiting value of the elongation relates to building up of the necking (as illustrated in the following figures). In Fig.\,\ref{fig.01} we have somewhat arbitrarily placed vertical reference lines to mark the points further on  referred to as $\alpha^{\rm stab.}_{20}$; by definition at these points the discrepancy caused by the first-contributing pair of multipolarities is equal to 20 keV. The latter symbol and the reference lines will facilitate the discussion below.  Observe that the position of the `stability limit',  
$\alpha^{\rm stab.}_{20}$, decreases with increasing nuclear mass as shown in Fig.\,\ref{fig.01}. 

\begin{figure*}[htbp!]
   \begin{center}
   \vspace{-1cm}
     \includegraphics[width=8.6cm]{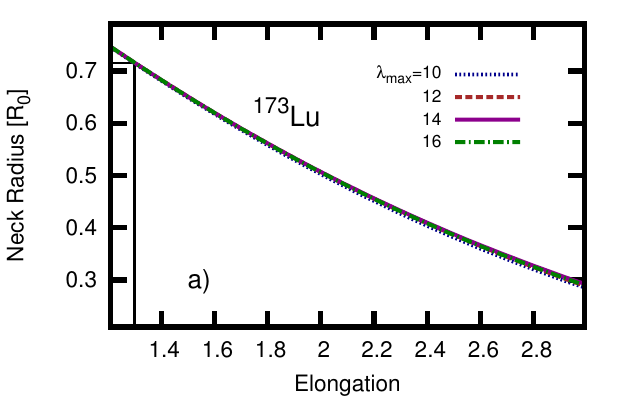}
     \includegraphics[width=8.6cm]{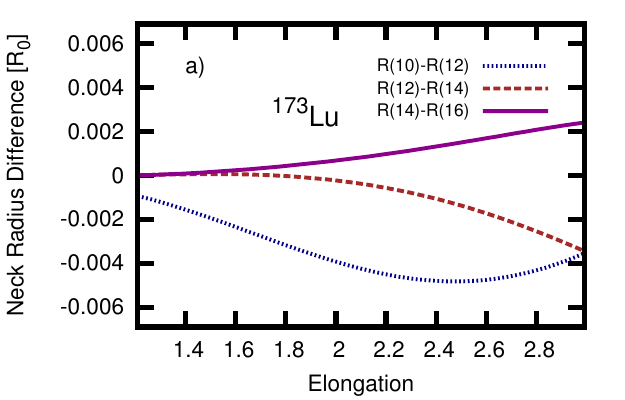}\vspace{-0.0cm}
     \includegraphics[width=8.6cm]{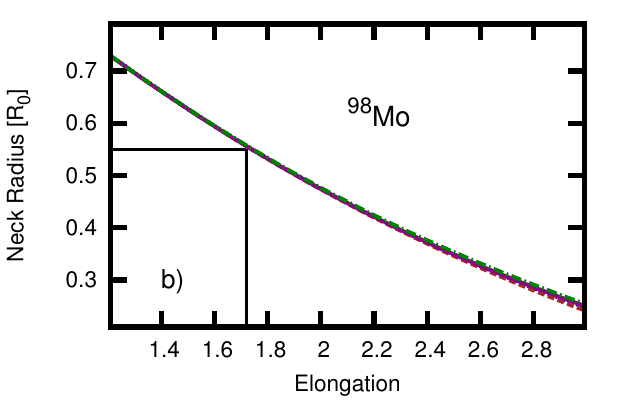}
     \includegraphics[width=8.6cm]{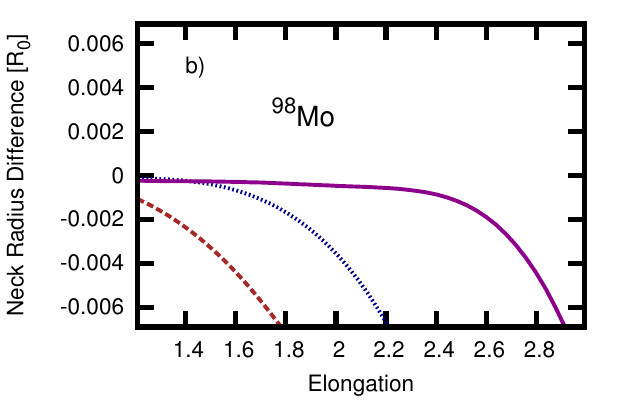}\vspace{-0.0cm}
     \includegraphics[width=8.6cm]{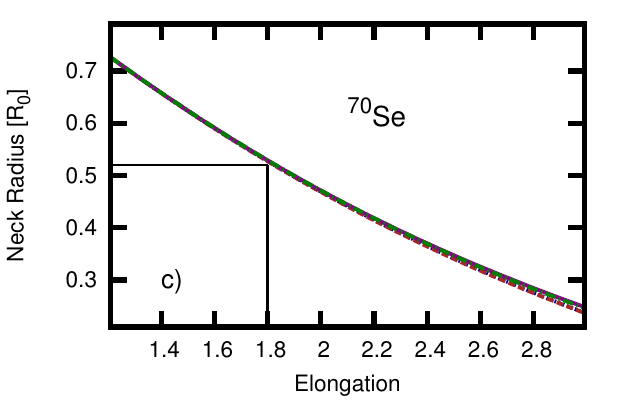}
     \includegraphics[width=8.6cm]{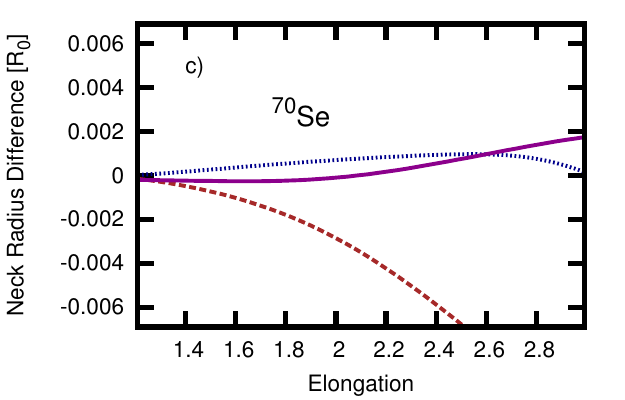}
     \caption{Left: Neck radii in function of nuclear elongation for increasing
              multipolarities in the nuclear surface expansion, in units of the
              corresponding spherical radius $R_0$. Positions of the vertical
              lines are the same as those in Fig.\,\ref{fig.01}, for more
              details see text. [Observe a perfect stabilization of the nuclear
              neck-radius curves in terms of the multipole expansion.] Right:
              Decomposition of the nuclear neck radii in terms of contributions
              from various multipolarities as indicated [Observe
              characteristically `erratic' behavior of various contributions
              in terms of differential quantities,
              which does {\em not} resemble the usual multipole-expansion 
              convergence properties (the bigger the $\lambda$ the smaller the
              contribution).]
              }
                                                                 \label{fig.02}
   \end{center}
\end{figure*}

Figure \ref{fig.02}, left-hand side, shows the neck radii for the three nuclei discussed, in function of their elongations. The crossing points between the curves and the vertical lines whose positions are copied from Fig.\,\ref{fig.01}, define the `degree' of the neck formation. Recall that at neck radii in the range, typically (0.3-to-0.4)\,$R_0$, the nuclei which are still described with the help of one, common surface, should rather be considered as effectively two separate nuclear objects with vanishing probability of returning to the original compact configuration.

When the system reaches such a configuration by bypassing the no-return point, the probability of fission becomes equal to the probability of arriving at such a configuration. On the one hand side, as it often happens, the nuclear macroscopic energy may {\em keep increasing} with increasing elongation, reaching the saddle point only later (i.e.~for even larger quadrupole deformations). On the other hand these are the saddle points which are traditionally thought of as classical no-return points -- what brings us to an apparent interpretation conflict. 

In our opinion, as just suggested, such a conflict is only apparent. Indeed, no element in any macroscopic model is capable of describing the quantum physics of the diluted nuclear matter in the neck area. Neither there exists any reliable way of fitting the model predictions at this point to any neck-related observable and unless explicitly proven to the contrary, the macroscopic models {\em become partially inadequate at this point} for further describing realistically the nuclear energy. By the same token, the estimating nuclear fission life-times by including the information about the saddle points under these conditions can be seen as at the limit of validity\footnote{We are dealing at this point with a not an uncommon situation, where one is forced to push the use of a phenomenological model over the limits of its applicability range. Unfortunately, the predictions of such a model may still `look good' whereas its physics context deteriorates gradually together with the model's predictive power.}.

Positions of the vertical lines indicate that the basis instability areas measured in terms of the nuclear elongation in Fig.\,\ref{fig.01} clearly correspond to the advanced stages of the neck formation -- and thus further analysis of these instabilities must focus on the behavior of the nuclear necks. This is illustrated in the right-hand side of Fig.\,\ref{fig.02} by showing the neck-radius contributions coming from various multipolarities: Here, following the convention in Fig.\,\ref{fig.01}, we take as a measure of those contributions the differences between the neck radii obtained with a given $\lambda$ and $\lambda-2$, denoted symbolically $R(10)-R(12)$, $R(12)-R(14)$, etc. Despite the fact that the neck-radii themselves are pretty robust in terms of the stability of the multipole basis expansion, the detailed contributions of various multipolarities to the neck radius are not regular (`erratic').

Let us emphasize at this point that from the microscopic point of view, the neck zones are in fact the scenes of complex quantum few body processes in the dilute (decreasing density of) nuclear matter, governed by two-body and, possibly, three-body correlations, occasionally influenced by the cluster formation etc., -- in short: the mechanisms whose description with the help of classical concepts and a single two-dimensional surface is clearly impossible. Therefore the `erratic' fluctuations in question are not so much {\em the issue of stabilizing the algorithm with respect to the basis cut off} but rather a manifestation of the limit of applicability of the macroscopic energy formula. This being said we find it instructive to pursuit a short discussion of the neck geometry properties within the macroscopic algorithm despite the fact of an inadequacy of the macroscopic model to the microscopic description of the neck formation mechanism just mentioned. 

In other words, results in Figs.\,\ref{fig.01} and \ref{fig.02} reveal a characteristic feature: Depending on the nucleus -- and thus the details of the neck formation within the model -- the magnitude of the discrepancies expressed in terms of differential quantities does not depend in any regular manner on the increasing $\lambda$. This supports the interpretation of the convergence failure caused by the {\em model in-adaptation}.    

\begin{figure}[htbp!]
   \begin{center}
     \includegraphics[width=8.9cm]{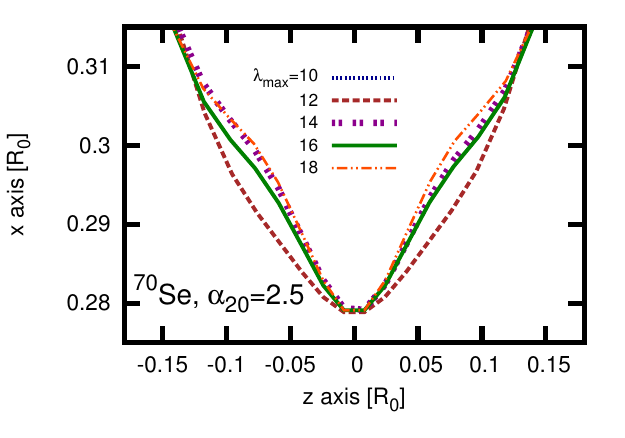}\\[-0mm]
     \caption{Illustration of the build-up of the shape fluctuations in the 
              neck region with increasing elongation and thus progressing neck
              formation, for $^{70}$Se as an example at $\alpha_{20}=2.5$\,. }
                                                                 \label{fig.03}
   \end{center}
\end{figure}

The mechanism is further illustrated in Fig.\,\ref{fig.03} in which the distance $x$ from the nuclear axis $\mathcal{O}_z$ is shown as a function of $z$ for a large value of the elongation. As seen from the Figure, clearly distinguishable fluctuations occur with no `evident convergence scheme' behind the order of the curves. Similar neck profiles can be drawn for smaller elongation with the conclusion that although the amplitude of the fluctuations decreases with decreasing elongation, as expected, the order of the curves on the diagrams remains `erratic'.

From the above illustration and from the results of similar calculations for nuclei in the mass-range considered we conclude that the energy contributions of the multipole deformations in excess of $\lambda=12$ for 
$\alpha_{20}>\alpha_{20}^{\rm stab.}$ have a character of irregular but otherwise {\em small fluctuations whose energy impact also remains small}. Let us remark again that no aspect of the model can pretend being an adequate tool to describe the dynamics of the neck formation, in particular it does not take into account e.g.~alpha-particle formation in the neck area. Therefore we believe that the energy fluctuations of this order represent generic inadequacy of the macroscopic model in the description of inevitably microscopic mechanism of creation of light clusters and/or the motion of the nucleons in the strongly `diluted' nuclear matter in the neck range -- rather than the basis cut-off uncertainties.


\subsection{Basis Stability Conditions: Tri-Axial Shapes}
\label{Sect.IV-B}


Examining the axially symmetric deformations alone will not be sufficient to convince oneself about stability properties of the basis cut off when studying the Jacobi shape transitions. 
\begin{figure}[htbp!]
   \begin{center}
   \vspace{1cm}
     \includegraphics[width=8.0cm]
     {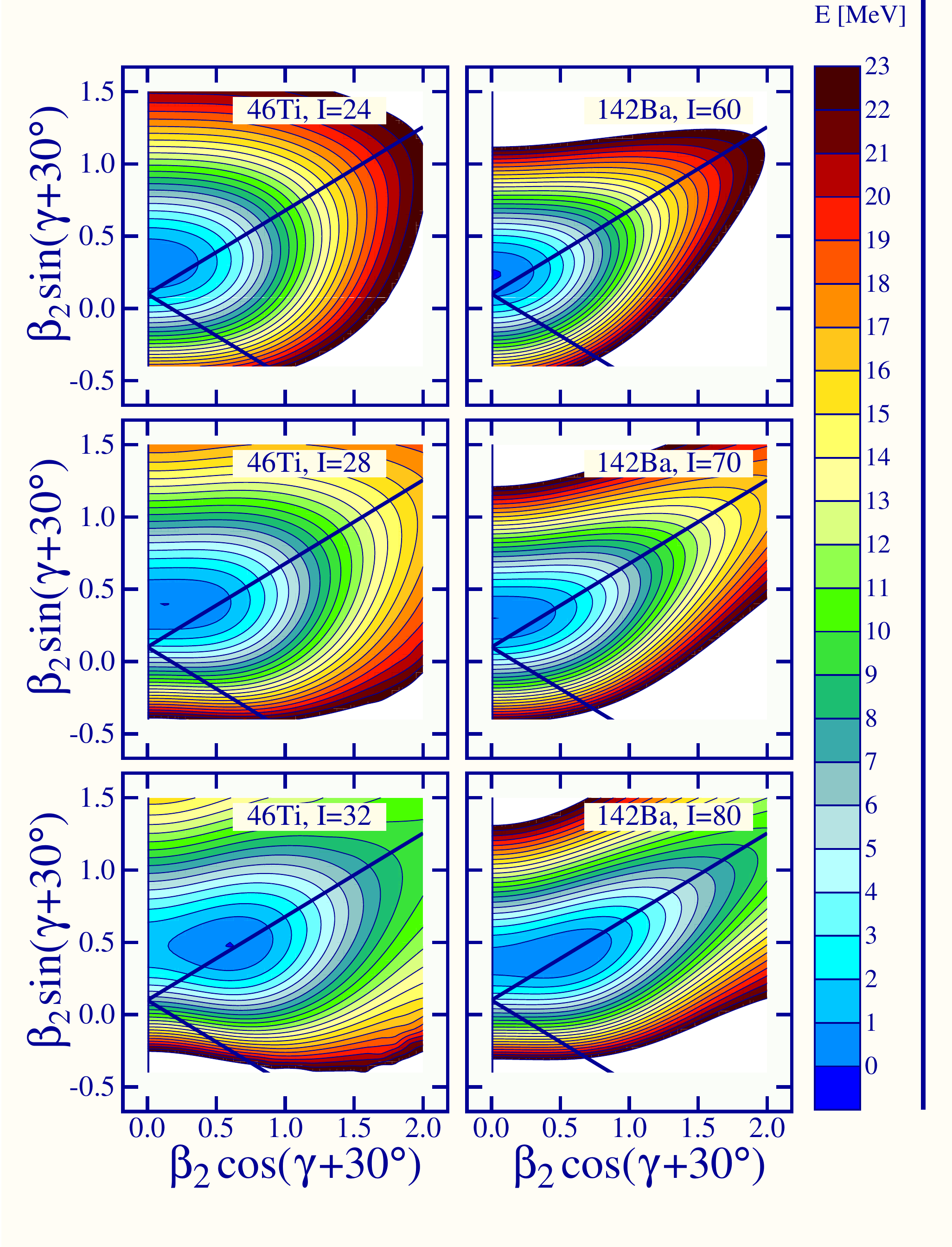}
     \caption{Total energy $\{\beta,\gamma\}$-plane projections for the nuclei
              indicated at spins along the Jacobi transitions (see text).
              At each $\{\beta,\gamma\}$-point the energy was
              minimized over the even-$\lambda$ deformations 
              $\alpha_{\lambda 0}$ for $\lambda \leq 16$.
              }
                                                                 \label{fig.04}
   \end{center}
\end{figure}
\begin{figure}[htbp!]
\vspace{1cm}
   \begin{center}
     \includegraphics[width=8.5cm]
     {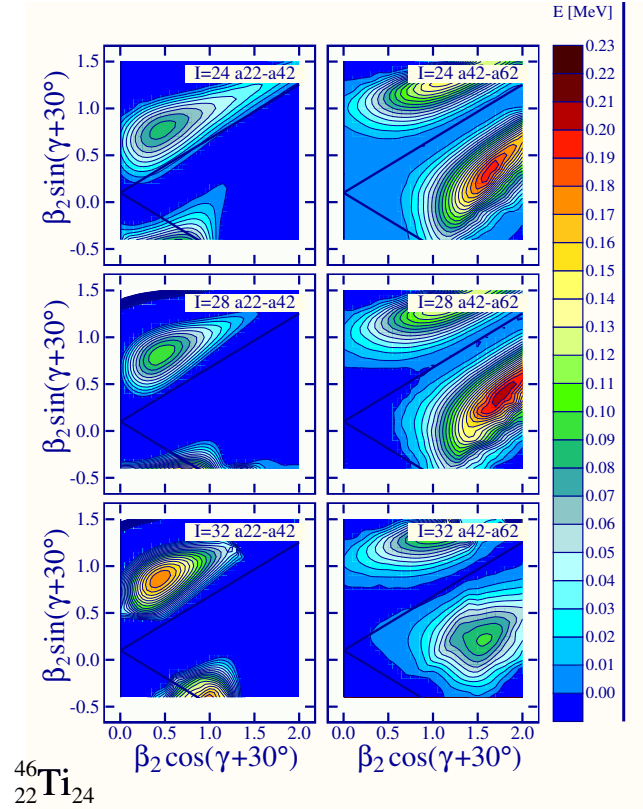}
     \caption{Energy differences for $^{46}$Ti at spins indicated, representing
              the energies in the preceding Figure minus the similar projections
              minimized in addition with respect to $\alpha_{42}$, according to
              the notation introduced in the text: $E_{22}$-$E_{22,42}$ (left
              column). Three maps in the right column were constructed in
              analogy to the previous ones, except that the energy difference
              corresponds this time to the results obtained including the
              minimization over $\alpha_{42}$ and $\alpha_{62}$ minus the
              energies corresponding to the minimization over $\alpha_{42}$
              i.e.~$E_{22,42}$ - $E_{22,42,62}$. Other minimization conditions
              are same here as in Fig.\,\ref{fig.04}.}
                                                                 \label{fig.05}
   \end{center}
\end{figure}
\begin{figure}[htbp!]
   \begin{center}
     \includegraphics[width=8.5cm]{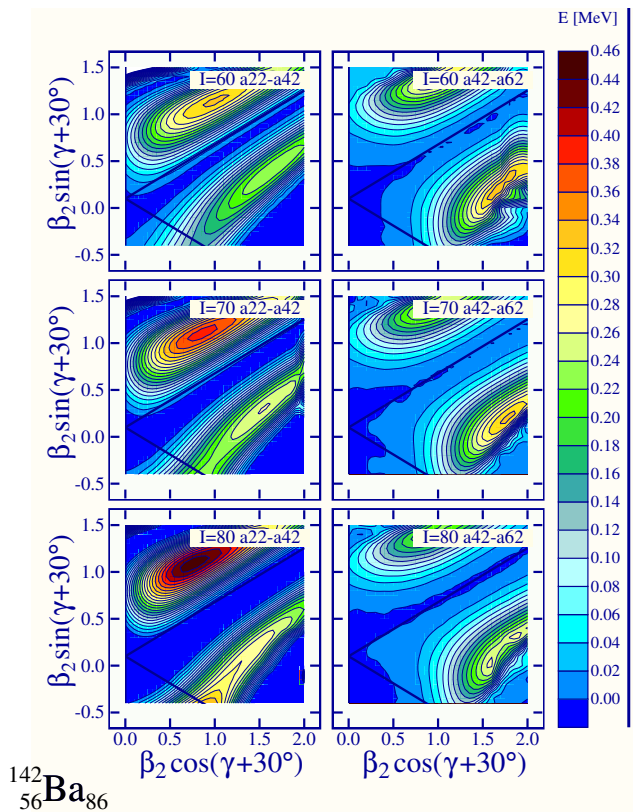}
     \caption{Analogous to the preceding Figure, here for $^{142}$Ba,
              representative for the heavier nuclei studied in this paper.}
                                                                 \label{fig.06}
   \end{center}
\end{figure}
Therefore we have also performed the
calculations for spins characteristic of the Jacobi shape transitions using at first the usual $[\{\beta,\gamma\}\leftrightarrow\{\alpha_{20},\alpha_{22}\}]$-plane representation, but next allowing for extra minimization over the similar tri-planar geometry deformations, $\alpha_{42}$ and $\alpha_{62}$, at each 
$\{\alpha_{20},\alpha_{22}\}$ point. 

The results corresponding to the usual $\{\beta,\gamma\}$-plane -- i.e.~with neither $\alpha_{42}$ nor $\alpha_{62}$ extra minimization -- are given in Fig.\,\ref{fig.04}, for two representative nuclei in the mass range studied. These energy maps are meant to define the energy scale within which the basis cut-off instability tests will be analyses. The potential energies of Fig.\,\ref{fig.04} will be labelled $E_{22}$ for further reference, despite the fact that minimization over several axially-symmetric deformations is performed at each $\{\beta,\gamma\}$-point. 

Next we define two extra energy $\{\beta,\gamma\}$-plane projections. The first one differs from the one illustrated in Fig.\,\ref{fig.04} in that in addition to the minimization over all even-$\lambda$ deformations 
$\alpha_{\lambda \mu=0}$ with $\lambda\leq 16$, the minimization over 
$\alpha_{42}$ is performed at each $\{\beta,\gamma\}$-point. The corresponding energies are denoted by $E_{22,42}$ for short. In the second projection, an extra simultaneous minimization over $\alpha_{42}$ and $\alpha_{62}$ is performed at each $\{\beta,\gamma\}$-point, with the resulting energies abbreviated to $E_{22,42,62}$.

To estimate the impact of the tri-axial deformations $\alpha_{42}$ and 
$\alpha_{62}$ we constructed the differences of the type $E_{22}-E_{22,42}$ and $E_{22,42} - E_{22,42,62}$ representing the impact of each single tri-axial deformation parameter mentioned. The results of the two energy differences for the lightest among the nuclei considered, $^{46}$Ti, are illustrated in Fig.\,\ref{fig.05}, showing no impact whatsoever in the deformation areas surrounding the energy minima. The differences of the order of 200 keV at most, representing the energy-gain when minimizing in addition over the $\alpha_{42}$ deformation superpose with the energies at the range of 5-to-10 MeV above the minimum thus having no impact on one of the central issues of focus in this article: The dynamical (most probable) deformations that accompany large amplitude shape fluctuations associated with the Jacobi shape transitions at the lowest vibration energy limits. Similar conclusion applies for the effect of the $\alpha_{62}$-deformation as seen from the three maps on the right-hand side 
column in Fig.\,\ref{fig.05}. 

For $^{142}$Ba representing heavier-mass nuclei considered in this article, the relative impact of the higher order triaxial deformations is even weaker, since the small maxima in Fig.\,\ref{fig.06} should be compared with the total energies of the order of 15-to-20 MeV above the minimum. This can be seen directly from the three maps on the left-hand side of Fig.\,\ref{fig.06} representing the effect of the $\alpha_{42}$ deformation and, similarly from the three maps on the right hand-side of the figure for $\alpha_{62}$.

Analogous results apply to other nuclei in the mass range considered in this article and we conclude that the effect of the higher-order tri-axial deformations $\alpha_{42}$ and $\alpha_{62}$ is negligible for the Jacobi shape transitions in the nuclei studied.  


\subsection{Parameterization of the Deformation-Dependent Congruence Energy}
\label{Sect.IV-C}


As mentioned in Sect.\,\ref{Sect.II}, there exist in the literature a few types of phenomenological assumptions about the form of the deformation-dependent congruence energy term and, to an extent, they give similar results as far as improvements of the description of the fission-barriers are concerned [cf.~Refs.\,\cite{WDM97}, \cite{PMo04} or \cite{KPo04}; in the latter the deformation-dependent congruence energy term of \cite{WDM97} has been used in conjunction with the LSD approach]. One of the tendencies in the past was to focus on the parameterizations with possibly small number of adjustable parameters. This may have certain aesthetic and occasionally practical advantages even though with the present day computing power these arguments are of much less importance as compared to the past.  

To our knowledge there are no rigorous arguments behind the phenomenological expressions used in the literature other than those based on certain rather vague intuition. This does not allow to construct the clear-cut physics criteria discriminating one approach against the other. Consequently we do not have at our disposal, neither have the other authors, any precise guideline so as to what should be the functional relation between evolving nuclear neck (or any of its parameterizations) representing the process of the {\em creation} of two fission fragments {\em and} the nuclear macroscopic energy. Under these circumstances we propose a functional form, in which the only leading idea is to have a certain parametric flexibility allowing to test the reaction of the model with respect to {\em `accelerated or slowed-down turning on the congruence term with increasing elongation'}, while improving the description of the fission barrier heights -- otherwise the functional form used below remains an ad hoc 
postulate -- similarly to the other forms discussed in the literature. 

Below we will use the experimental information about the fission barrier
heights in nuclei in which this information is available, in order to optimize
parameters of the following simple analytical expression below referred to as {\em neck factor}:
\begin{equation}
      F_{\rm neck}(\alpha_{20})
      = 
      1+\textstyle\frac{1}{2} 
      \left\{
           1 
           +
           \tanh\left[({\alpha_{20}^{}-\alpha_{20}^0})/{a_{\rm neck}^{}}\right]
      \right\}.
                                                                 \label{eqn.16} 
\end{equation}
Above, $\alpha^0_{20}$ and $a_{\rm neck}^{}$ are two, at this time yet unknown
adjustable parameters. With the above assumption the deformation-dependent
congruence energy contribution will be defined as:
\begin{equation}
      E_{\rm  cong.}(N,Z;\alpha)
      \stackrel{df}{=}
      W_0(Z,N) \cdot F_{\rm neck}(\alpha_{20}), 
                                                                 \label{eqn.17}
\end{equation}
where the so-called Wigner energy term, cf. Refs.\,\cite{EW37a,EW37b}, denoted
$W_0$, is still parametrized as in \cite{WDM96}, i.e.:
\begin{equation}
       W_0(Z,N)
       =
       -C_0 \exp(-W\vert I \vert/C_0),
                                                                 \label{eqn.18} 
\end{equation}
with $I\equiv(N-Z)/A$, $C_0=10$ MeV and $W=42$ MeV. The rest of this Section is devoted to the description of the determination of the phenomenological parameters $\alpha_{20}^0$ and $a^{}_{neck}$ of the nuclear neck-formation.  

For the present applications it will be convenient to introduce an
$A$-dependence through a simple linear form, i.e.,~$\alpha^0_{20}\to\alpha^0_{20}(A)$ in which we `arbitrarily parametrize the $\alpha_{20}^0$-parameter' as a function of the mass number :
\begin{equation}
      \alpha^0_{20}(A)
      =
      \alpha^{\rm min.}_{20}
      +
      \frac{(\alpha^{\rm max.}_{20}-\alpha^{\rm min.}_{20})}
           {(A^{\rm max.}-A^{\rm min.})}
      \cdot
      (A-A^{min.})
                                                                 \label{eqn.19} 
\end{equation}
where $\alpha^{\rm min.}_{20}$, $\alpha^{\rm max.}_{20}$, $A^{\rm min.}$ and 
$A^{\rm max.}$ are predefined a priori as:
\begin{eqnarray}
   & \alpha^{\rm min.}_{20}=1.5, \quad \, A^{\rm min.}\,=\;\;70;
                                                               \label{eqn.20}\\
   & \alpha^{\rm max.}_{20}=3.5, \quad A^{\rm max.}=220 ,
                                                               \label{eqn.21}
\end{eqnarray}
so that effectively only the $a_{\rm neck}$-parameter can be seen as an adjustable constant. 

\begin{table}[htbp!]{
\caption{Comparison of the barrier heights for nuclei listed. Columns 2-5
         contain: Experimental values [Exp], reference of origin [Ref], 
         LSD model results with congruence ignored, and congruence contribution
         from Myers and \'Swi\c{a}tecki, [denoted C. M.-S.], Ref.\,\cite{WDM97}.
         The last three columns represent the results obtained using
         hypotheses: $a_{\rm neck}=0.5, 1.0$ and~1.5\,.}
  \begin{center}
      \begin{tabular}{ c r c c c c c c }
                                                                         \hline
Nucleus & Exp Ref & No C. & C. M.-S.& \mc{3}{c}{$\alpha_{20}^0$ $A$-dependent}\\
        &         &       &$a_{neck}\to$&0.5& 1.0&1.5                         \\
                                                                         \hline
$^{70}$Se  & 39.4 \cite{fan00}& 50.618  & 43.337& 38.973&40.393& 41.825\\
$^{76}$Se  & 44.5 \cite{fan00}& 54.323  & 49.624& 43.944&45.084& 46.068\\
                                                                         \hline
$^{75}$Br  & 41.0 \cite{del91}& 52.603  & 47.062&  42.169&43.410&44.599\\
                                                                         \hline
$^{90}$Mo  & 42.0 \cite{jin99} & 50.890  & 45.519& 40.995&42.308&43.359\\
$^{98}$Mo  & 46.0 \cite{jin99} & 54.571  & 50.651& 46.495&47.443&48.132\\
                                                                         \hline
$^{173}$Lu & 29.0 \cite{mor72} & 28.707  & 25.635& 27.433&26.797&26.616\\
$^{228}$Ra & 6.3  \cite{mor72} &  6.204  &  6.013&  6.204&6.186 & 6.120\\
                                                                         \hline
       \end{tabular}
  \end{center}
                                                                  \label{tab.02}
   }
\end{table}
By a repeated minimization of the nuclear energy over the multipole deformations $\alpha_{\lambda 0}$ with $\lambda\in[3,16]$ in function of the elongation 
$\alpha_{20}$ for various parameter values of  $a_{\rm neck}$ we have verified that the $a_{\rm neck}=0.5, 1.0, 1.5$ values presented in the Table can be considered as an acceptable solution of the minimization of discrepancies between the model and experimental fission barriers\footnote{The experimental values in the Table have been obtained similarly as in Ref.\,\cite{WDM97}, i.e.~subtracting the shell energy contribution at spherical shapes and using the fact that the shell energies at the saddle points can be considered negligible according to \'Swi\c{a}tecki's ``topological property''.} as given in 
Table \ref{tab.02}.

Lublin-Strasbourg Drop model expression of Eq.\,(\ref{eqn.01}), together with
the modifications, which aim at including the deformation-dependent congruence
energy term, Eqs.~(\ref{eqn.16}) and (\ref{eqn.17}-\ref{eqn.21}) will be
referred to as LSD-C, `C' standing for `congruence'.


\subsection{More Comments about Nuclear Necking}
\label{Sect.IV-D}


The very notion of the nuclear neck is of course a fully classical, geometrical
concept in the present approach. As it is well known the nuclear mean-field
interaction can be parametrized phenomenologically using, e.g.,~the Woods-Saxon
potentials with the diffusivity parameter, $a\approx 0.6$ fm, what implies, that
the nuclear skin thickness, defined as the distance for which the potential
decreases from 10\% to 90\% of its minimal value, corresponds to $4\,a\approx
2.5$\,fm: In other words, the radius value at which the potential falls to its 90\% of the minimum value is $R_{90\%}=R - 2a = R - 1.2$\,fm. 

Let us consider a nucleus with $A=125$ nucleons for which the radius estimated as usual as $R = r_0 A^{1/3}$, with $r_0=1.2$\,fm gives $R=6$\,fm. At the neck-value of the order of $0.4\times R= 2.4$\,fm, the Woods-Saxon potential is equal to its 90\% depth at $R_{90\%}^{0.4} \approx (2.4 - 1.2)$\,fm\,$\approx\,$1.2\,fm which is {\em the half of the nucleon size!} We conclude that there is no way, a macroscopic model with the concept of classical surfaces can approach any realistic description of this part (neck) of the nucleus.

When the neck radius approaches this range one may say that the spatial
nucleonic content of this part of the nucleus is ``merely composed of the
nuclear skin'' and that the corresponding configuration nears the scission
configuration. As already mentioned, for heavy and moderately heavy nuclei the corresponding typical neck-size for deformations neighboring the scission configuration can be expressed, rather roughly, as 
$R_{\rm scission}\approx (0.3-0.4)\,R_0$ where $R_0=r_0\, A^{1/3}$ and where $r_0\approx 1.2$\,fm. 

Despite doubtful capacity of describing physics of the nucleonic behavior in the nuclear neck-content within the macroscopic models, it is necessary to discuss the associated geometrical elements for reasons of imposed continuity/regularity of the equations of nuclear surfaces. Formally, the neck radius in an axially-symmetric fissioning nucleus can be defined in terms of sections which are perpendicular to the symmetry axis, say $\mathcal{O}_z$, of the elongating nuclear surface. Indeed, for the nuclear elongation sufficiently high, one can find the perpendicular section with the minimal surface corresponding to a separation between the two nascent fragments (usually at the $z$-values not too far from the origin of the reference frame). 

In a similar manner, for non-axial shapes one can still define the minimum
surface and the associated minimum and maximum radii with the possibility of
defining the ``effective'' neck radius e.g.~as an arithmetical or weighted
average of the two. 

\begin{figure}[htbp!]
   \begin{center}\vspace{-0cm}
     \includegraphics[angle=-90,scale=0.62]{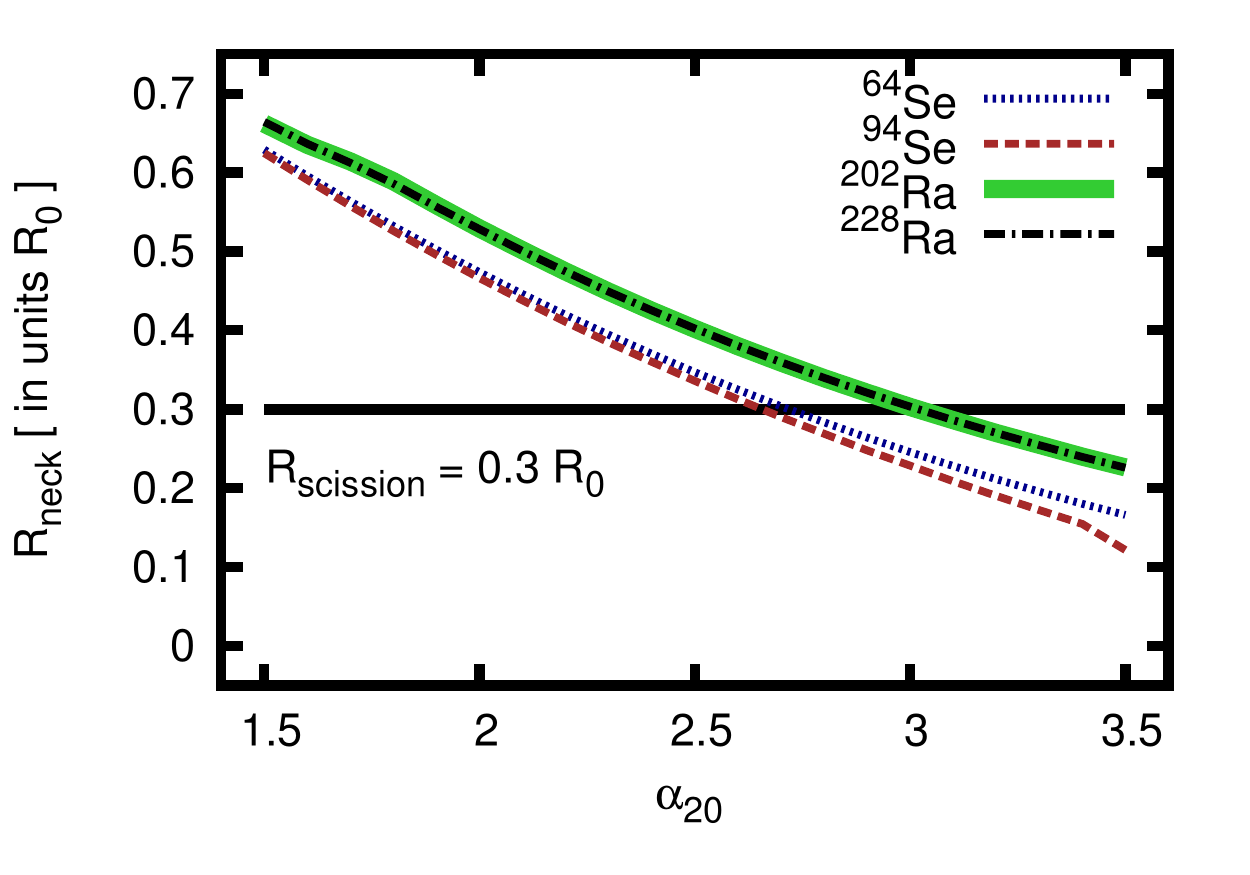}
     \vspace{-0cm}
      \caption{Neck radii of the nuclei indicated, in function of the
              quadrupole deformation. Plotted values are normalized to the
              radii of the corresponding spherical nuclei, the former defined by
              $R_0=r_0\,A^{1/3}$ with $r_0=1.2$ fm. Horizontal line shows
              the geometrical scission reference defined by
              $R_{\rm scission}{=}0.3 R_0$. Let us emphasize that the latter
              parameter can be seen as a somewhat schematic reference value
              whose exact definition is neither possible nor important in a
              qualitative discussion.}
                                                                 \label{fig.07}
   \end{center}
\end{figure}

In what follows it will be instructive to obtain a global information about the
neck evolution with increasing quadrupole deformation $\alpha_{20}$ which,
within the multipole parameterization of the nuclear surface used in this
article, is a leading component in describing the nuclear elongation. To obtain
such an illustration we have calculated, as before, the total nuclear energies
for increasing $\alpha_{20}$ for a few nuclei for which the experimental data on
the fission barriers exist in the literature. Figure \ref{fig.07} shows the neck
radii relative to the radius of the equivalent spherical nucleus, in function of
nuclear elongation obtained by using the same calculations as the ones used to
obtain Fig.\,\ref{fig.01}. In this case, $R_{\rm neck} = x$ at $z=0$.

\begin{figure}[ht]
   \begin{center}
     \vspace{-0cm}
     \includegraphics[angle=-90,scale=0.620]{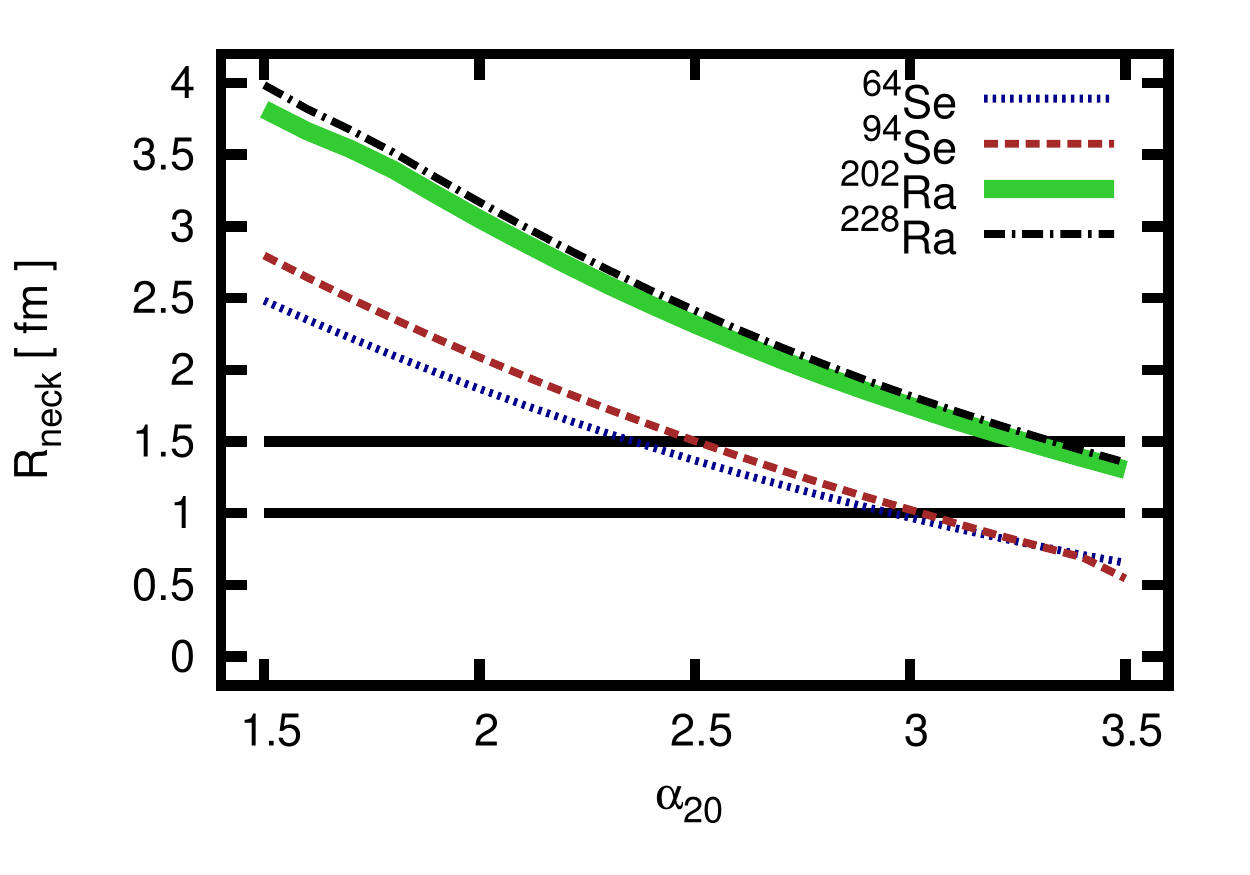} 
     \vspace{-0cm}
     \caption{Illustration similar to the one in Figure \ref{fig.07},
              but with the neck radius expressed in [fm]. At the level of the
              neck-size of the order of 1.5\,fm there is hardly any space left
              for any single nucleon to move orthogonally with respect to the
              elongation axis. Since it is difficult to believe in an adequacy
              of the classical model to describe this geometrical range, by
              approaching the illustrated elongation we approach at the same
              time the range of decreasing adequacy of macroscopic models.} 
                                                                 \label{fig.08}
   \end{center}
\end{figure}

Curves obtained without taking into account the deformation-dependent congruence
energy coincide with the ones with the deformation-dependent congruence-energy
included and therefore in the Figure we place only one set of them. Let us 
emphasize here that the process of the creation of the nuclear neck on the way to fission (referred to as ``necking'') depends very little on the mass number,
$A$, and even less on the nucleon excess $|N-Z|$ as results in 
Figs.\,\ref{fig.07}-\ref{fig.08} indicate.
 
The neck-radii of the corresponding selected nuclei, illustrated in the Figures, decrease almost linearly with the quadruple deformation, the
negative slope depending only slightly on the mass of the nucleus at least in
the cases examined. To verify that these results do not depend very much on the
isospin we have included in the comparison the nuclei with relatively large
differences in the nucleon excess, $N - Z$.

As it can be seen from Fig.\,\ref{fig.07}, for the light nuclei such as
$^{64,94}$Se, the neck radius approaches $R_{\rm scission}$ for the elongation
$\alpha_{20}\approx 2.5$ at the most, probably markedly earlier. For heavy nuclei, represented by Radium isotopes, the neck radius approaches the discussed limit at slightly higher quadrupole deformation of $\alpha_{20}\approx 3$ or earlier. 

In the Selenium case the isospin dependence is practically non-existent - one
cannot distinguish among the positions of the illustrated curves for the span in
the neutron number of $\Delta N =30$. Let us notice that the results
concerning the neck description but obtained using the Myers-\'Swi\c{a}tecki
prescription are indistinguishable from the ones presented within the scale of
the plot.


\section{Deformation Dependence of the Fission Barriers}
\label{Sect-V}


We will use the experimental fission barrier heights to compare with the model
results of the optimized here LSD-C approach for $^{70,76}$Se \cite{fan00}, $^{90,98}$Mo \cite{jin99} and $^{173}$Lu \cite{mor72}. Our results are compared with those of Ref.\,\cite{WDM96} showing an alternative parameterization of the deformation-dependent congruence energy term\footnote{Let us remark in passing that the experimental macroscopic barriers for the lightest nuclei cited here are deduced from the corresponding excitation functions and are dependent on the level density parameter used in this type of analysis. Uncertainties in the level density parameters may influence deduced fission barrier heights by few MeV.}. 

\begin{figure}[htbp!]
   \begin{center}\vspace{-0mm}
     \includegraphics[angle=-90,scale=0.62]{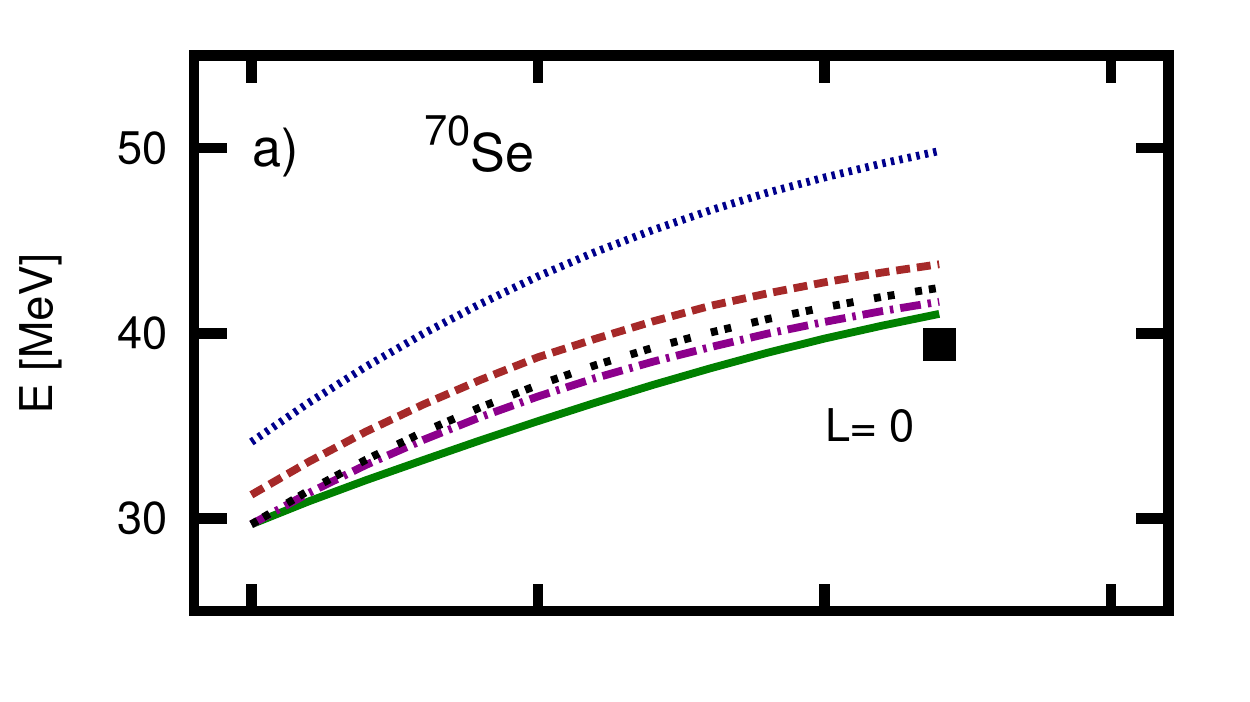}\\[-6mm]
     \includegraphics[angle=-90,scale=0.62]{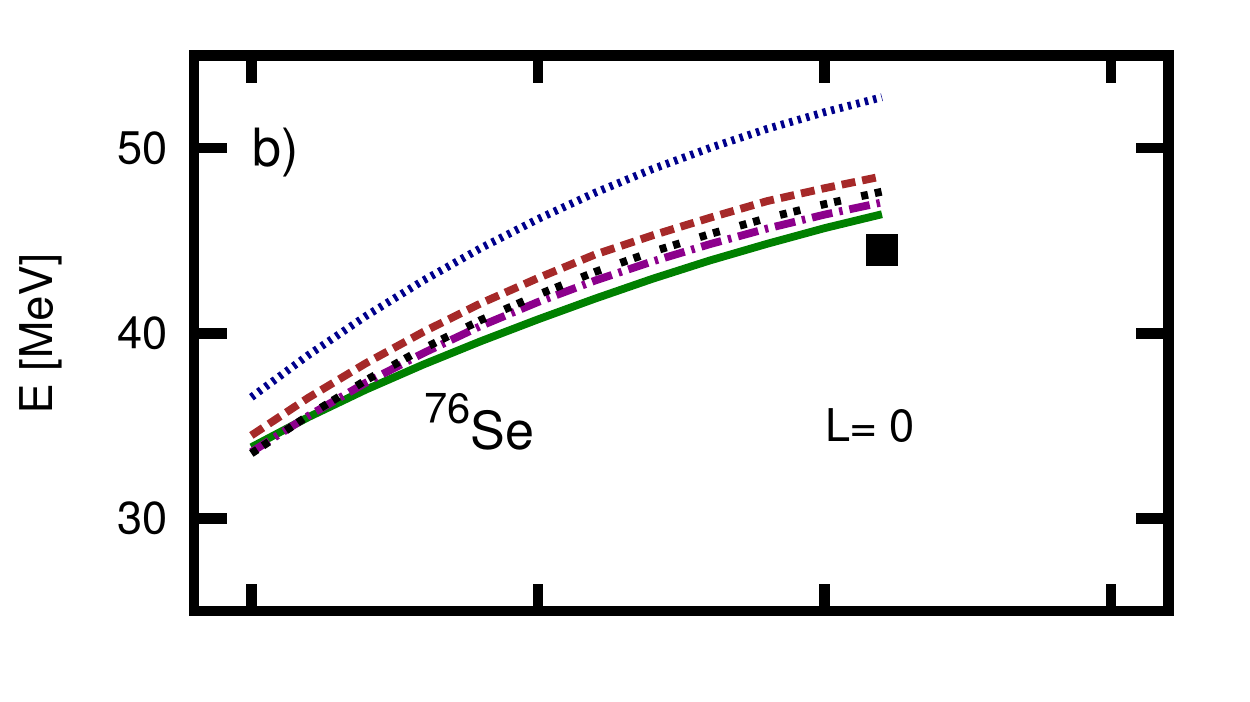}\\[-6mm]
     \includegraphics[angle=-90,scale=0.62]{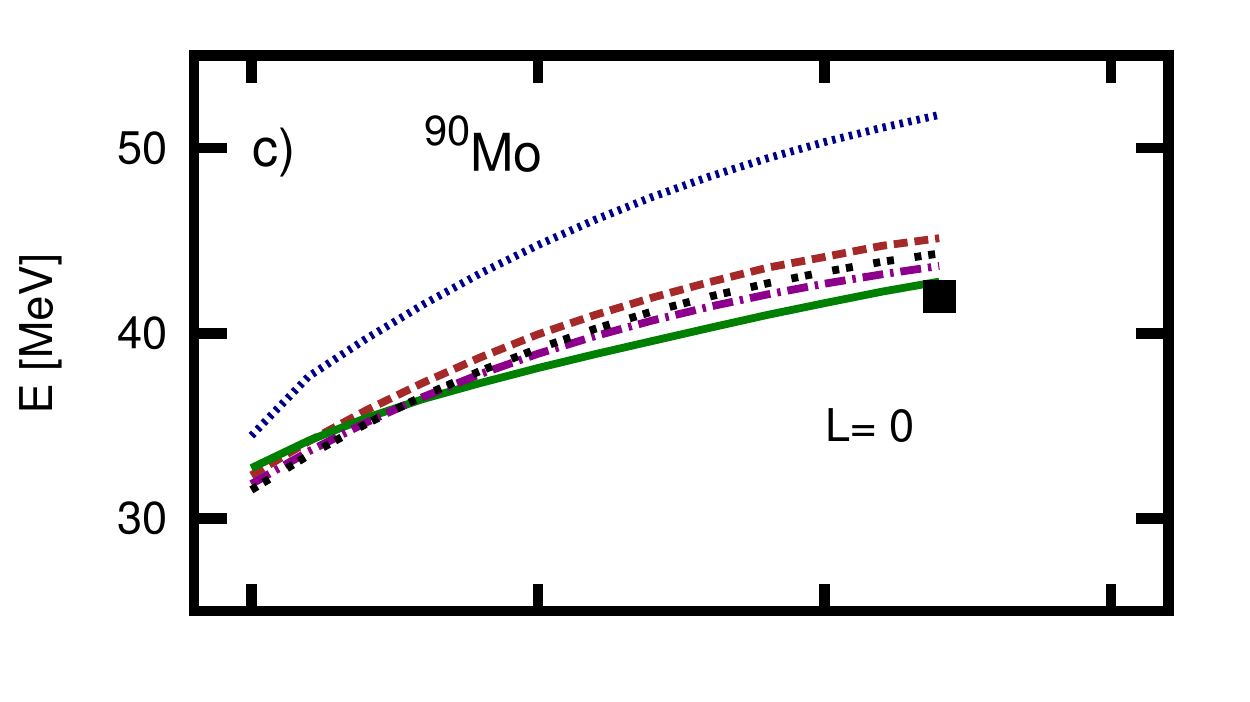}\\[-6mm]
     \includegraphics[angle=-90,scale=0.62]{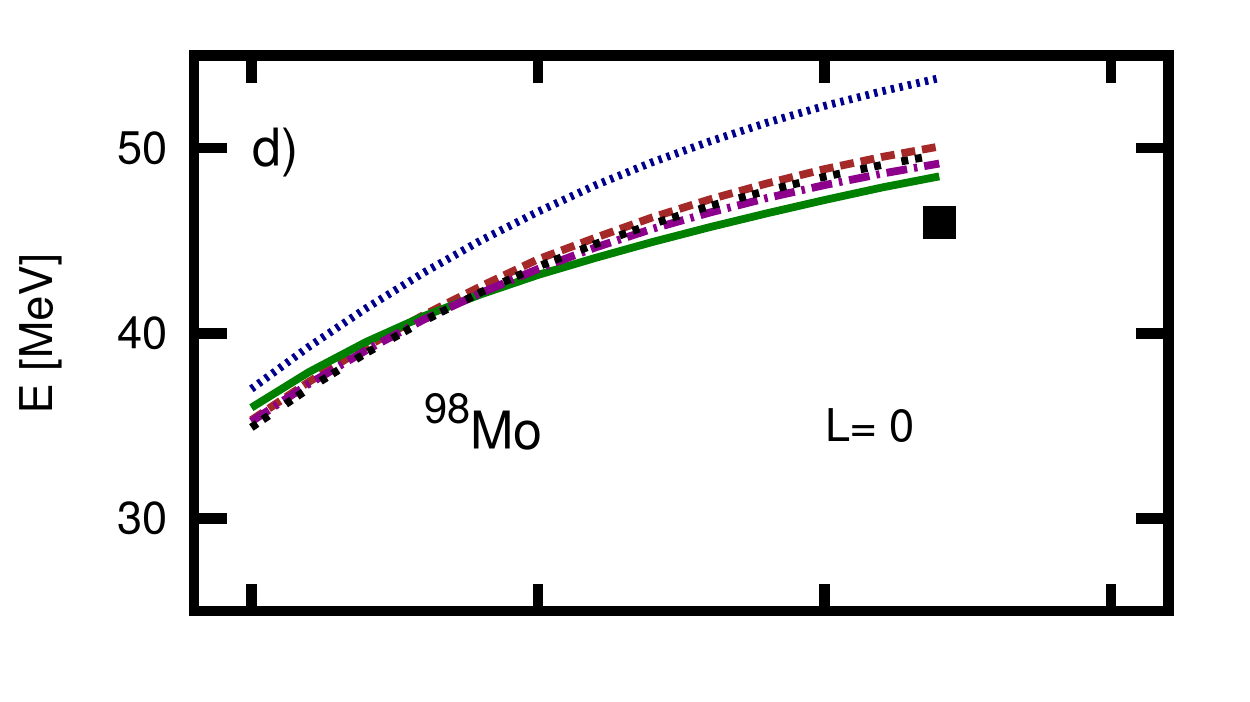}\\[-6mm]
     \includegraphics[angle=-90,scale=0.62]{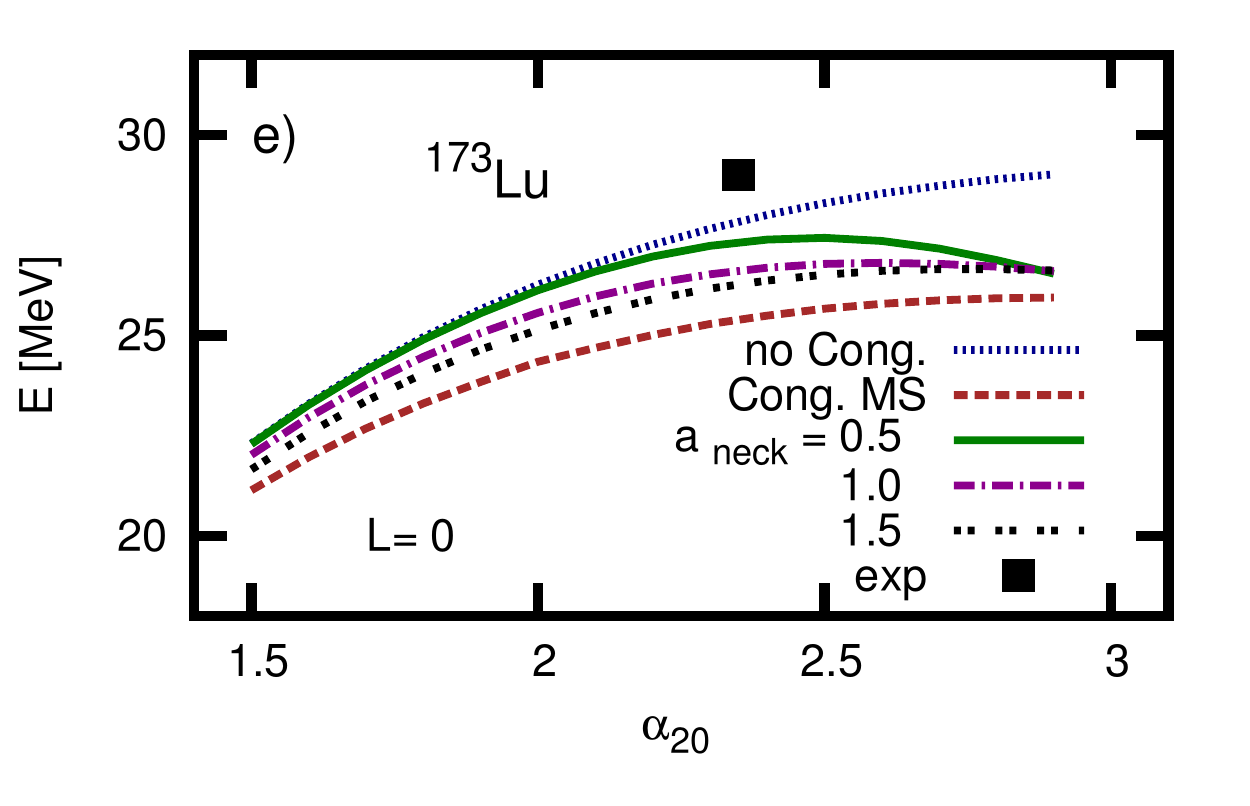}  \vspace{-0cm} 
     \caption{
              LSD-C model energies in function of $\alpha_{20}$, minimized over 
              $\{\alpha_{\lambda 0}\}$ for $\lambda\in[3,12]$ for the 
              $a_{\rm neck}$ values indicated - compared with the energies from
              of Myers and \'Swi\c{a}tecki, Ref.\,\cite{WDM96}, and labeled
              (Cong.\,MS). Experimental values, squares, from the references
              given in Table \ref{tab.02} are placed at either scission
              elongations defined by condition $R_{\rm neck}=0.3 \times R_0$ 
              (all but one), or the saddle-point deformation in the case of
              $^{173}$Lu cf. the text for more details).
}
                                                                 \label{fig.09}
   \end{center}
\end{figure}

Let us begin by illustrating the obtained parametric dependence of the LSD-C realization of the model in terms of the parameter $a_{\rm neck}$, the latter controlling the way (smoother vs.~more abrupt) congruence energy contribution lowers the barrier when elongation increases. The corresponding results are given in Fig.\,\ref{fig.09}, where the fission barriers at $L=0$ (no rotation) are shown. These results were obtained at each given elongation $\alpha_{20}$ by minimizing the nuclear energy over 10 deformation parameters\footnote{Although the odd-$\lambda$ multipoles have been formally allowed in the minimization, the final results depend only on the even-$\lambda$ deformations.}  $\alpha_{\lambda 0}$ with $\lambda\in[3,12]$. In what follows we compare the results obtained for three values of the `neck-parameter', $a_{\rm neck}=0.5$, 1.0 and 1.5. 

Adopted parameterization of the congruence-energy contribution lowers the nuclear fission barriers in a way that is rather independent of the neutron excess down to the scission point. The improvement brought by the congruence term is $\approx-10$ MeV for light nuclei. Let us notice that our results
with $a_{\rm neck}=0.5$ are the closest to the experiment, with the typical
discrepancy of the order of 1 MeV, the model overestimating the experimental
values for all the nuclei except for $^{173}$Lu. In the latter case the LSD-C
result underestimates the experimental value but is close to it.  

The LSD-C energy curves extend {\em formally} to the very high elongation values in terms of $\alpha_{20}$. However, the scission point defined conventionally by the condition that $R_{\rm neck}\approx 0.3\,R_0$, quite often sets in for markedly smaller $\alpha_{20}$-values, cf.~discussion in 
Sect.\,\ref{Sect.IV-D}. The corresponding effect is illustrated in Fig.\,(\ref{fig.09}) in which the experimental values of the fission barriers [solid squares] are placed, by convention, at the scission-point elongation. We consider the nuclei at this stage of their shape evolution as effectively composed already of two fragments. Further increase in the model energy represents more the shortcoming of the model rather than physical mechanisms. Indeed, for elongation exceeding scission point deformations, the LSD-C energy expression -- like all other macroscopic energy expressions -- nears the range of model's limited applicability. Indeed, not only no element of the model could be physically adequately optimized for this deformation range but, what is even more important, there are conceptual difficulties in associating the behavior of the dilute nuclear matter in the neck range with the purely geometrical features of the model. To illustrate this aspect we plot most of the curves (all but the one for $^{
173}$Lu) in such a way that they terminate, by definition, at the conventionally defined scission points.

Interestingly, the results of Ref.\,\cite{WDM96} overestimate (underestimate)
the experiment in the same nuclei in which LSD-C overestimates (underestimates)
the data. In absolute terms, the LSD-C provides an improvement by at least a factor of 2 (4 on average), as compared to the above cited reference.


\section{Basis Cut-off, Nuclear Potential Energies and Energy Minima}
\label{Sect.VI}


We will illustrate the reaction of the LSD-C energy formula induced by adding more and more terms in the spherical harmonic basis. Let us observe in passing that in certain applications enlarging the basis may be considered of strong disadvantage and/or even a prohibitive step as in the description of the motion in terms of e.g.~Langevin equations. Indeed in this case the increasing number of differential equations which need to be solved could represent a prohibitive aspect of such methods. Although in such situations alternative parameterizations of nuclear shapes may be preferable -- yet in any case, testing those alternative parameterizations must pass by the basis cut-off ultimate verification to guarantee that such an alternative (i.e.~not using a basis expansion but rather an arbitrary presupposed parameterization) is indeed acceptable.

Since the evolution of the nuclear shapes with spin is one of the interest here we included the angular momentum dependence in the tests.
\begin{figure}[htbp!]
   \begin{center}
     \vspace{-0cm}
     \includegraphics[angle=-90,scale=0.620]
                     {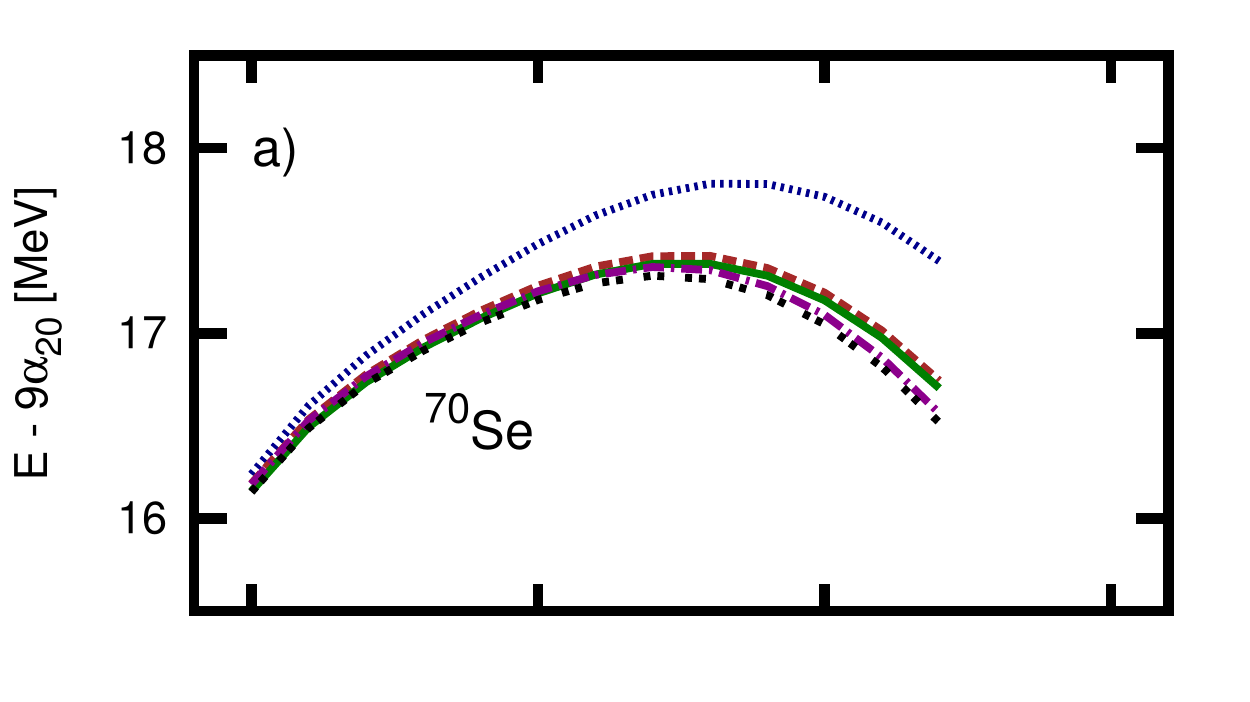}\\[-6mm]
     \includegraphics[angle=-90,scale=0.620]
                     {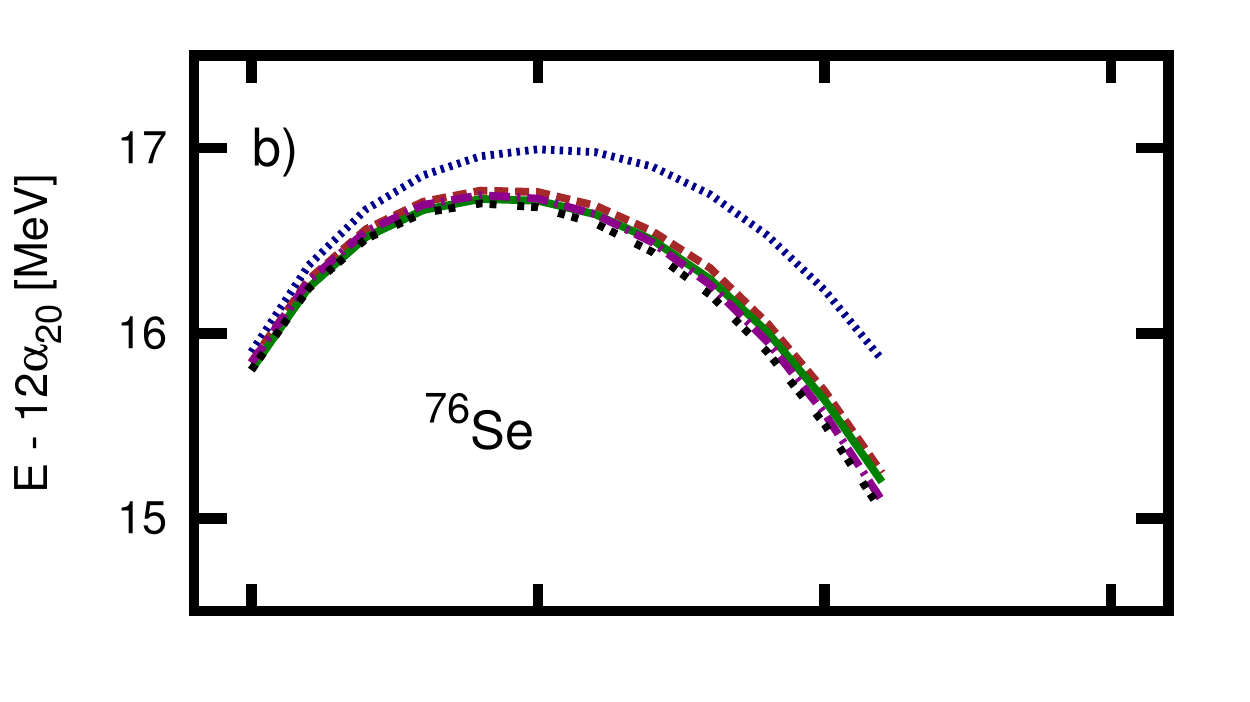}\\[-6mm] 
     \includegraphics[angle=-90,scale=0.620]
                     {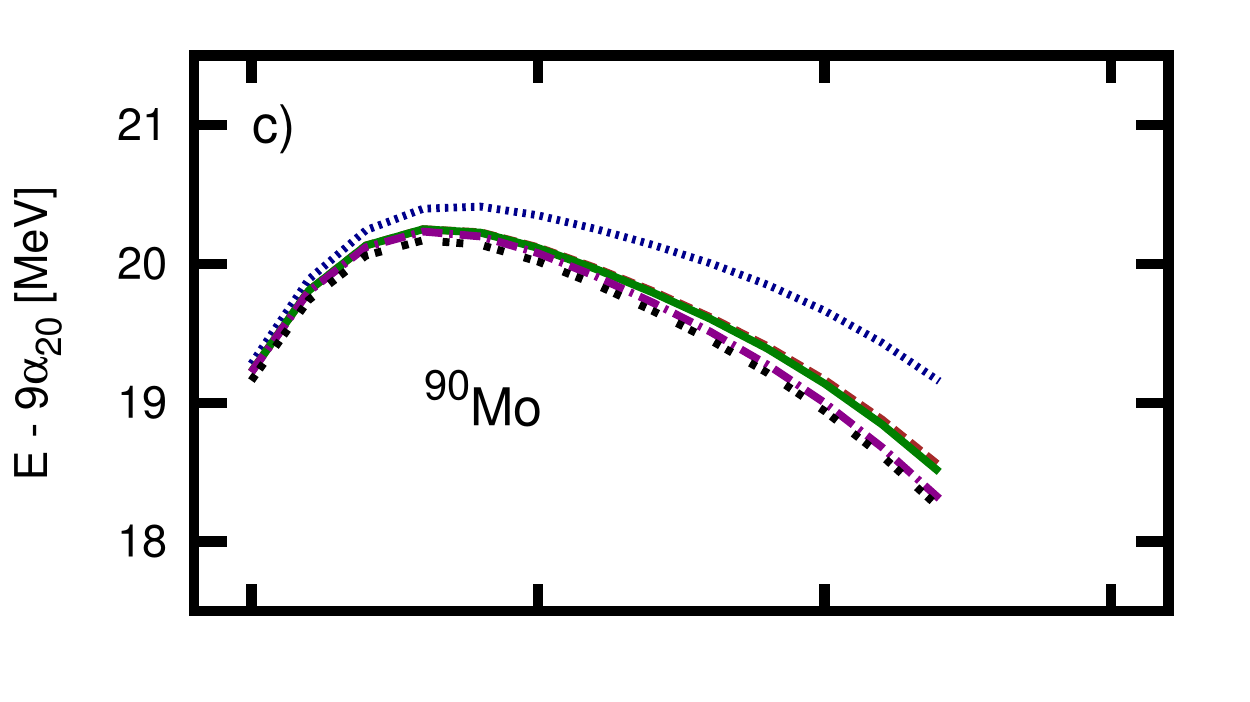}\\[-6mm] 
     \includegraphics[angle=-90,scale=0.620]
                     {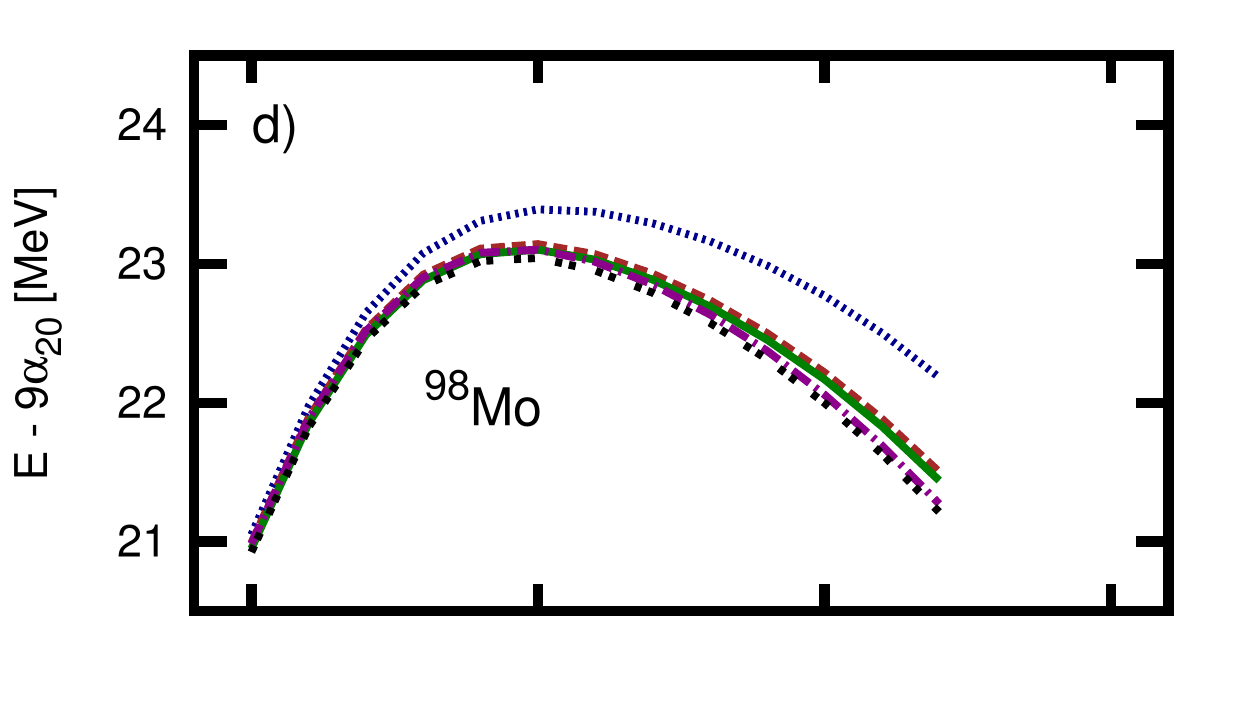}\\[-6mm]
     \includegraphics[angle=-90,scale=0.620]
                     {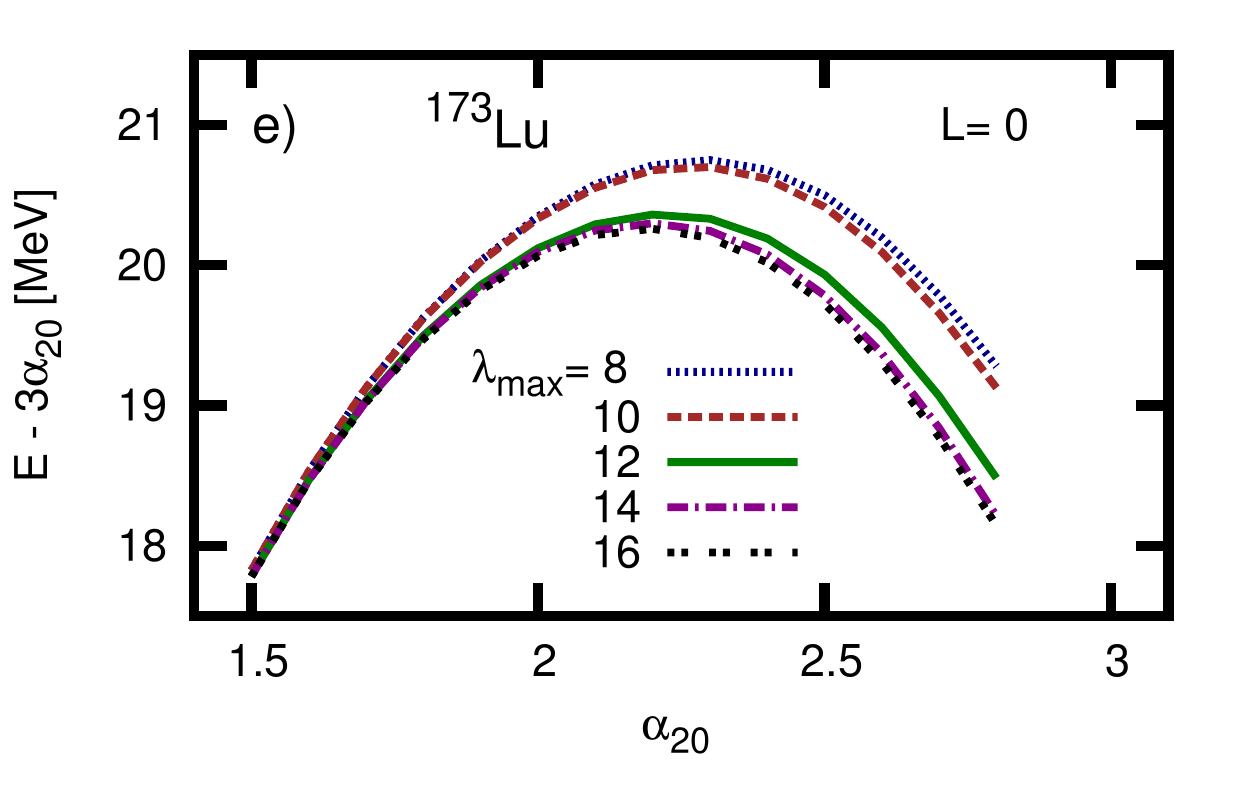}  
     \vspace{-0cm} 
     \caption{Dependence of the nuclear energy calculated using the LSD-C
              expression with $a_{\rm neck}=0.5$ at the large elongation regime
              and at spin $L=0$~$\hbar$ -- in function of the basis cut-off. 
              The energies have been minimized over deformations
              $\alpha_{\lambda 0}$ for $\lambda \leq \lambda_{\rm max}$
              indicated. To increase the legibility of the present illustration
              a linear reference has been subtracted as indicated in the
              description of the vertical axis. By convention the curves stop 
              at the deformation at which $R_{\rm neck}<0.3 \times R_0$ for all
              but $^{173}$Lu case in which the saddle point deformation comes
              first.
              }
                                                                 \label{fig.10}
   \end{center}
\end{figure}

\begin{figure}[htbp!]
   \begin{center}
     \vspace{-0cm}
     \includegraphics[angle=-90,scale=0.620]
                     {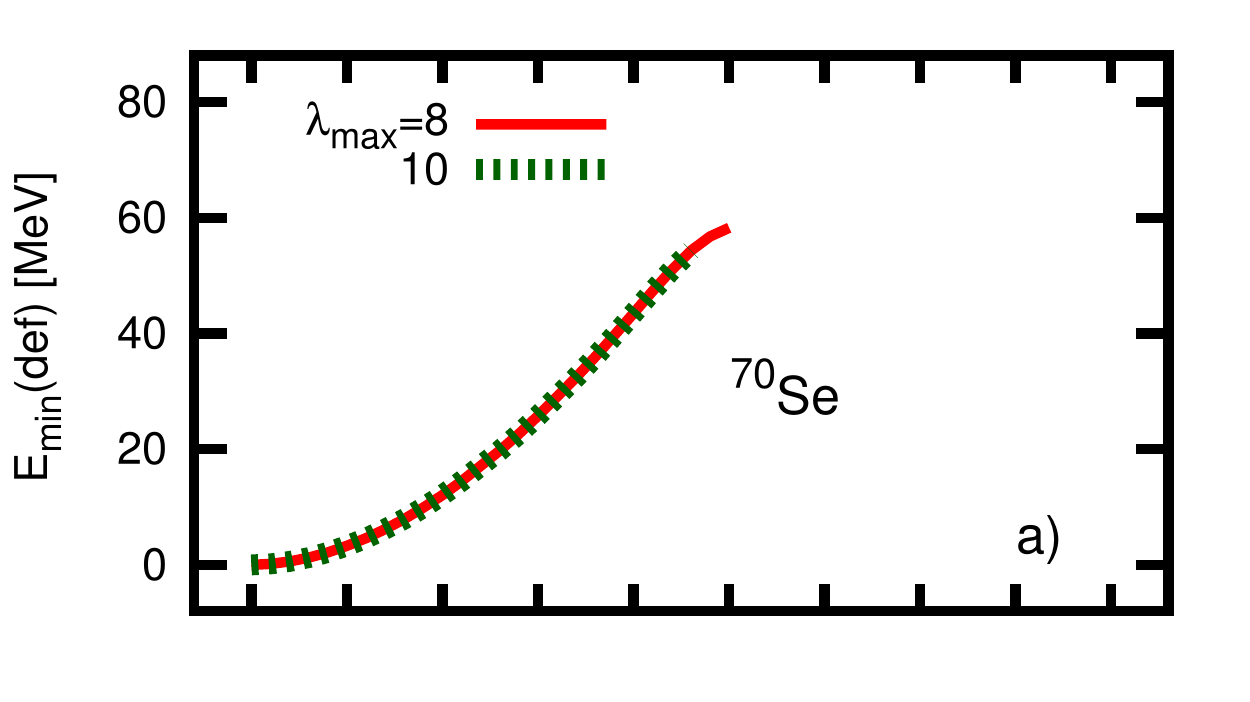}\\[-6mm]
     \includegraphics[angle=-90,scale=0.620]
                     {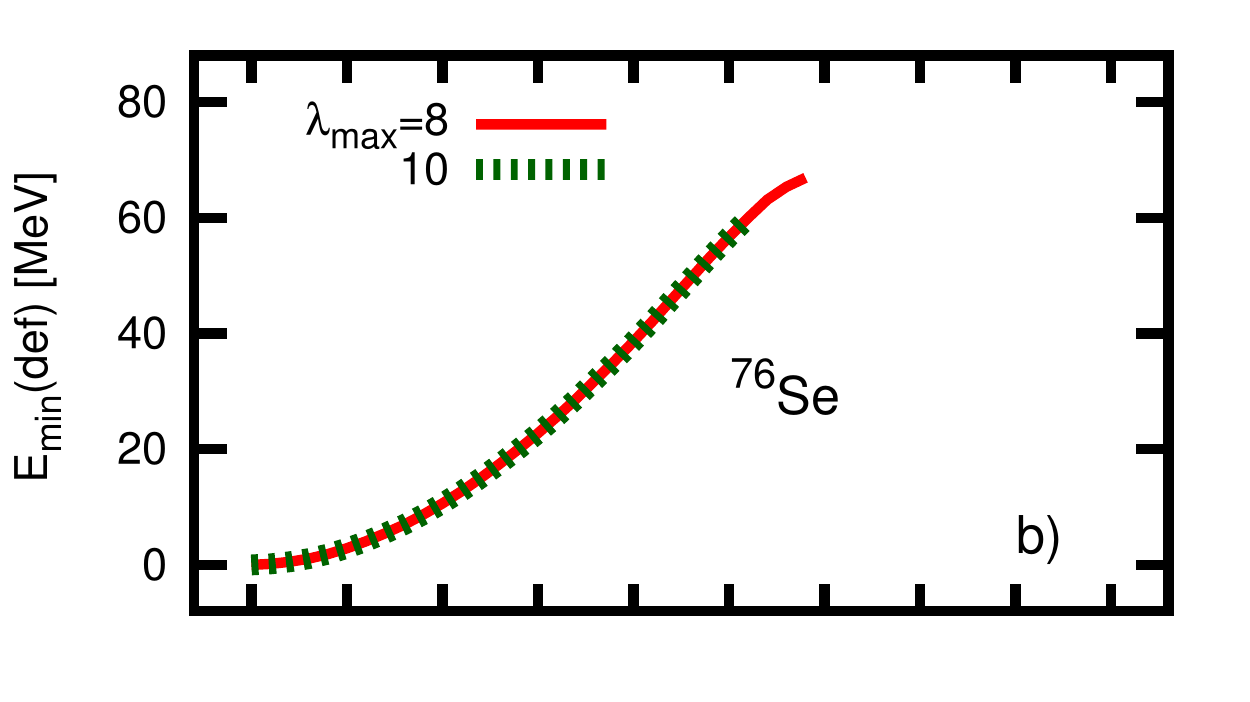}\\[-6mm]
     \includegraphics[angle=-90,scale=0.620]
                     {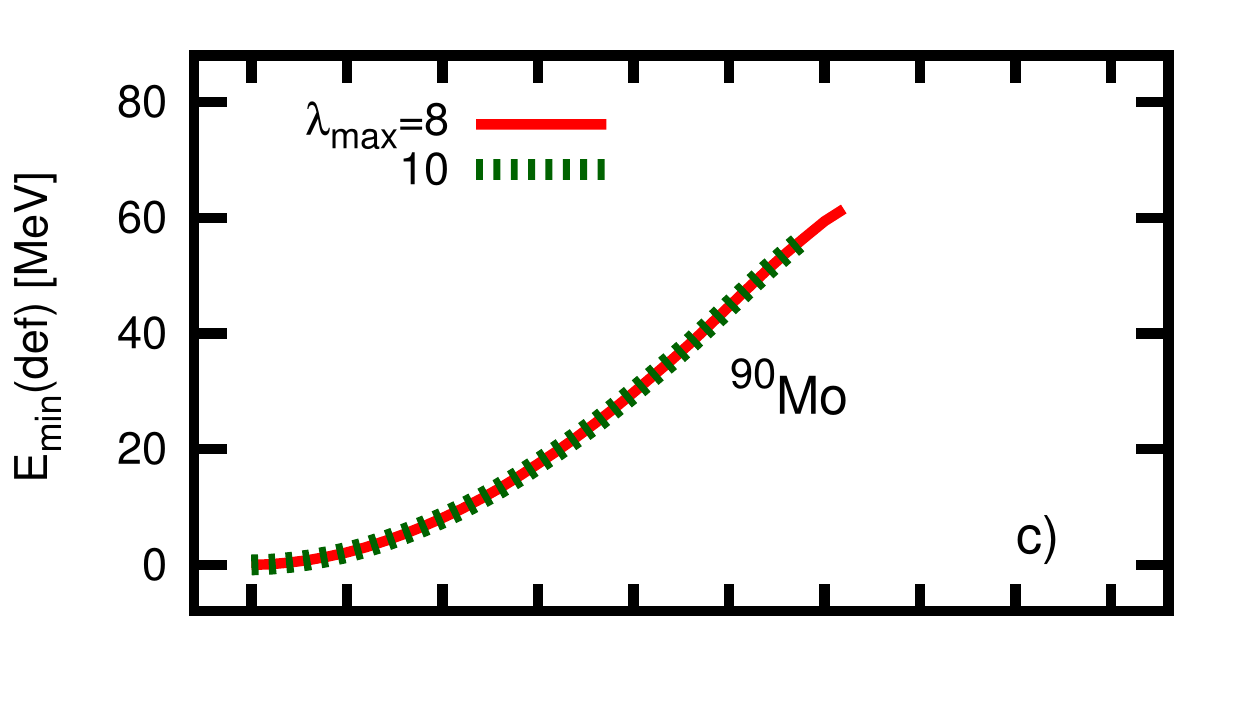}\\[-6mm] 
     \includegraphics[angle=-90,scale=0.620]
                     {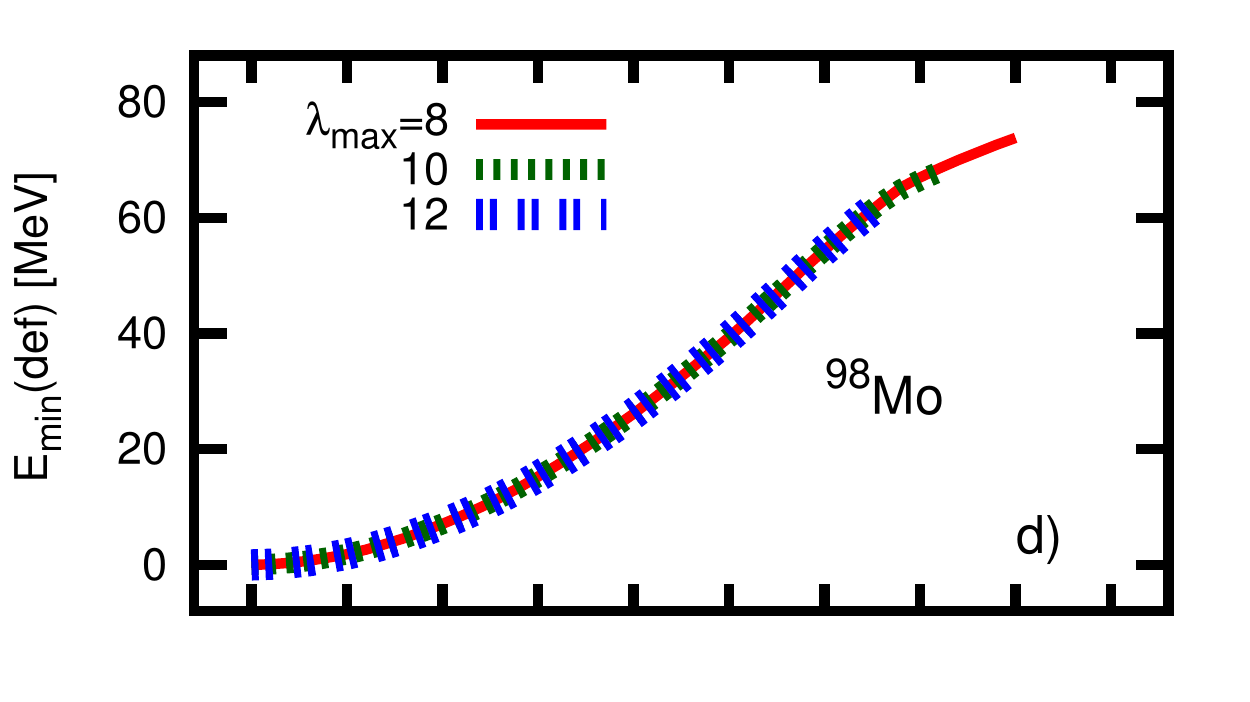}\\[-6mm]
     \includegraphics[angle=-90,scale=0.620]
                     {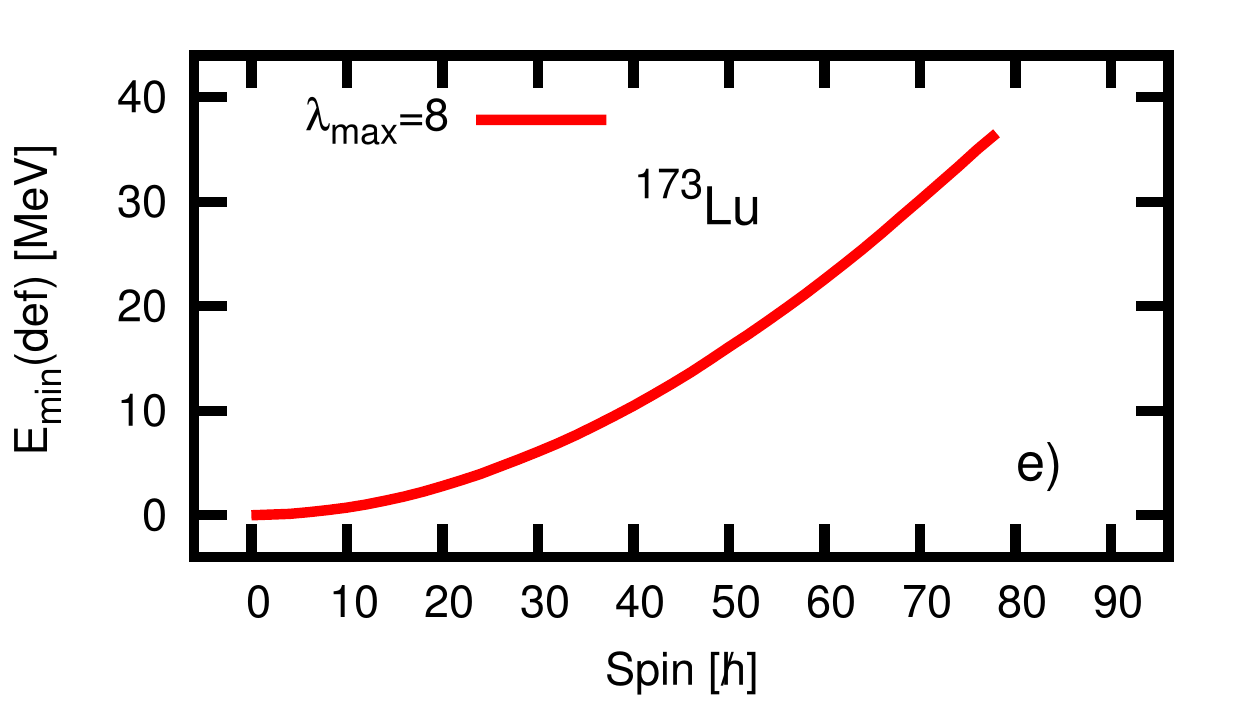}
     \vspace{-0cm}
     \caption{Dependence of the nuclear minimum energy on the basis cut-off
              parameter $\lambda_{\rm max}$ as indicated -- in function of
              increasing spin. Minimization performed over $\alpha_{\lambda,0}$
              for $\lambda\leq \lambda_{\rm max}$. The macroscopic energies
              include
              the shape-dependent congruence energy with $a_{\rm neck}=0.5$.
              Curves end if the scission condition is met for spins lower than
              the spins at which the barrier vanishes -- otherwise at spins of
              the vanishing barriers. Notice that the stability is obtained
              already at $\lambda_{\rm max}=8$.}
                                                                 \label{fig.11}
   \end{center}
\end{figure}

Figure \ref{fig.10} in which the static ($L=0\,\hbar$) nuclear energies are plotted for various cut-off parameters $\lambda_{\rm max}$, illustrates the impact of the basis cut-off on the total nuclear energy along the path to fission. As it can be seen from the Figure the strongest impact is expected at the largest elongation, with $\alpha_{20}\sim 2.5$: There, increasing $\lambda_{\rm max}$ from 12 to 14 may lower the energy by not more than about 100 keV. It should be emphasized at this point that for better legibility of the Figure, the total energies are plotted after having subtracted a smooth linear reference curve as indicated and consequently the maxima in those curves {\em do not} represent the positions of the saddle-points (fission barrier heights).

Results of the analogous tests are shown in Figure \ref{fig.11} for the absolute
{\em minima} of the potential energies. There the main effect of the basis
cut-off manifests itself in lowering the critical (scission or saddle) spin values with increasing $\lambda_{\rm max}$ [recall that the saddle-point elongation $\alpha_{20}$ is in all but $^{173}$Lu case larger than the elongation corresponding to the scission-condition introduced and discussed above.] In other words -- the formal minimum points loose their meaning as the physical equilibrium deformations at spins at which the fission barrier disappears or at which the nucleus has reached the scission configuration since for yet larger elongation the model cannot represent/control the fissioning systems which it is supposed to describe (model's limited adequacy described earlier). Other than that, the stability of the final results for the numerical values of the {\em energy minima} is achieved already at $\lambda_{\rm max}=8$ for all the cases studied.  


\section{Nuclear Yrast Energies at Increasing Spin}
\label{Sect.VII}


Within classical nuclear models the nuclear rotation is usually accounted for by
adding to the spin-independent macroscopic energy the classical rigid-body rotational-energy term, in the form of Eq.\,(\ref{eqn.08}) with the rigid-body moment of inertia, $\mathcal{J}(N,Z;\alpha)$, calculated using the uniform nuclear density distribution corresponding to nucleons contained within the surface given by Eq.\,(\ref{eqn.14}). In order to calculate the rotational energy, one uses the rotation axis associated with the largest moment of inertia thus providing the lowest energy contribution. 
\begin{figure}[htbp!]
   \begin{center}
     \includegraphics[angle=-00,scale=1.32]{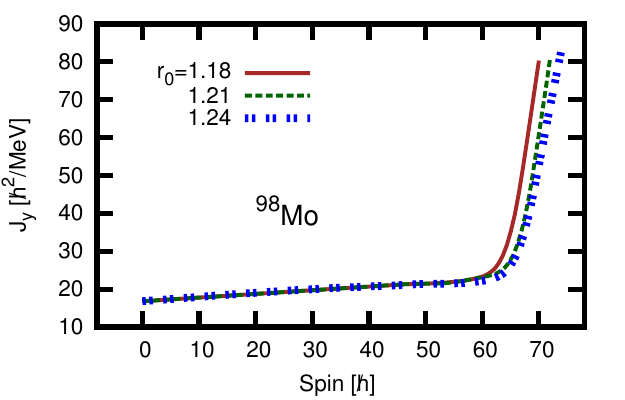}
     \caption{Illustration of the typical behavior of the classical moment of
              inertia at the deformations corresponding to the minimum of the
              potential energy for increasing spin at three typical values of
              the nuclear radius parameters $r_0$; for details see text and 
              Fig.\,\ref{fig.11}. Results for other nuclei have very similar
              form.
              }
                                                                 \label{fig.12}
   \end{center}
\end{figure}

Some authors calculate the moments of inertia using the diffused-surface
assumption, cf.~e.g.~Ref.\,\cite{KTR76}, what may be considered slightly more physical whereas, on the other hand, the relatively small differences which result can be at least partially accounted for by possibly readjusting the nuclear radius constant~$r_0$. In order to illustrate the order of magnitude of variations/uncertainties possibly caused by the freedom in choosing the radius parameter when reproducing the values of the classical moment of inertia (and to convince ourselves about the possible sizes of these uncertainties) we compare in Fig.\,\ref{fig.12} the results obtained for three characteristic values of the $r_0$ parameter, showing that only at the deformations very close to the fission/scission some `slightly visible impact' of the radius uncertainty can be expected.
\begin{figure}[htbp!]
   \begin{center}
     \vspace{-0cm}
     \includegraphics[angle=-90,scale=0.62]
                     {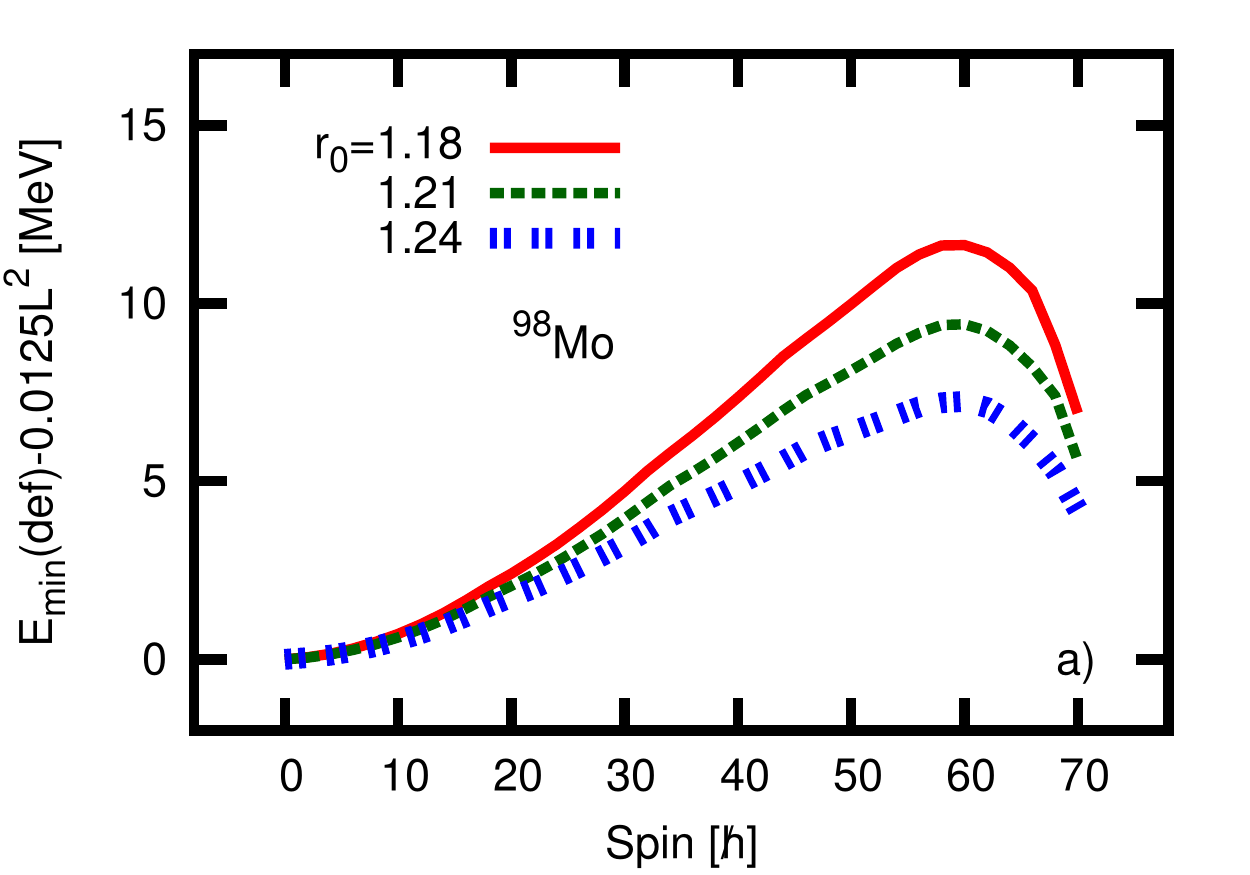}
     \vspace{-0cm}
      \caption{Illustration of the typical impact of the uncertainties in the
              classical nuclear moments of inertia on the total energy
              minimization result - here: in terms of the total energies at the
              nuclear energy minima for the three radius-parameter values
              indicated -- relative to a parabolic reference to increase the
              legibility of the Figure. 
             }
                                                                 \label{fig.13}
   \end{center}
\end{figure}

Since the moment of inertia has a direct impact on the total energy description, and thus the description of the shape transitions, we illustrate in Fig.\,\ref{fig.13} the results analogous to the ones in Fig.\,\ref{fig.11}. In this article we are using the radius parameter value $r_0=1.21$~fm. The variation of $r_0$ within $\pm 0.03$~fm which can be considered already very large in the context, leads to about $\pm 2$~MeV absolute shifts in terms of the total minimum energies at the highest spins. However, the latter dependence influences the total energy surfaces in a very smooth, regular and easily foreseeable manner, its impact being mainly to lower or increase the slope of the yrast lines.  

\begin{figure}[htbp!]
   \begin{center}
     \includegraphics[scale=1.130]{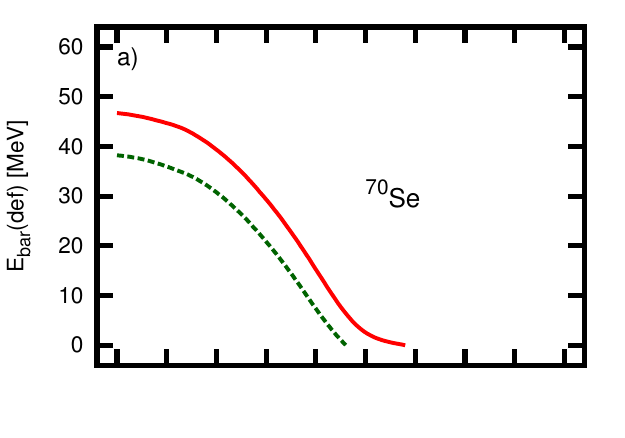}\\[-9mm]
     \includegraphics[scale=1.130]{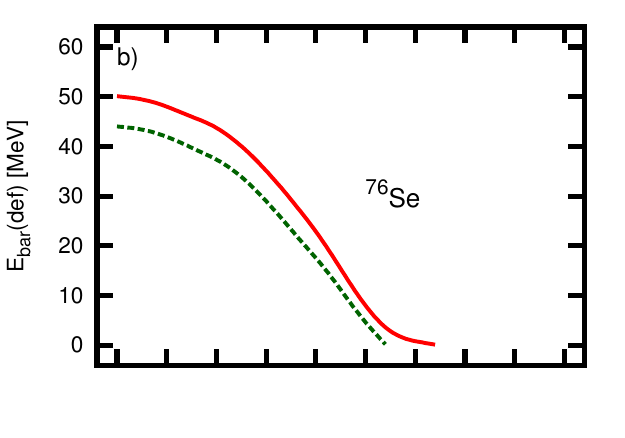}\\[-9mm]
     \includegraphics[scale=1.130]{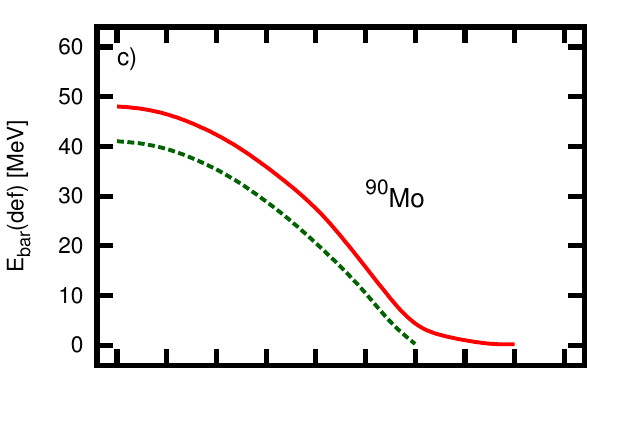}\\[-9mm]
     \includegraphics[scale=1.130]{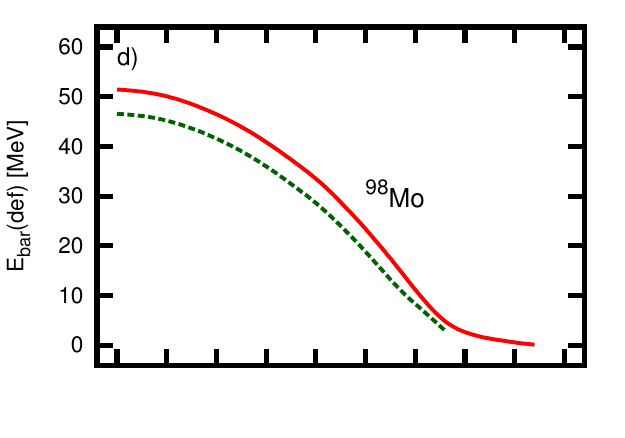}\\[-9mm]
     \includegraphics[scale=1.130]{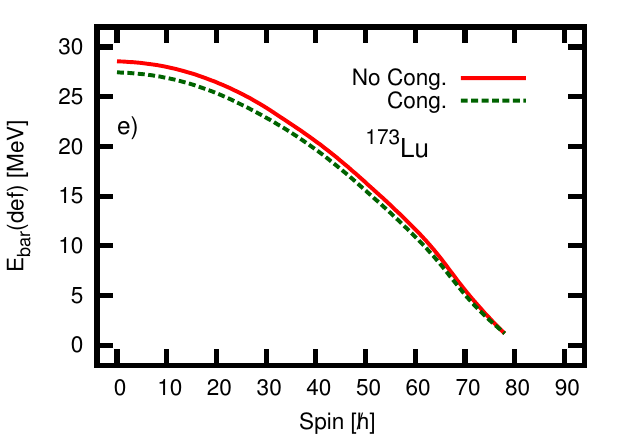}
      \caption{   
               Fission barrier heights obtained with (dashed line) and without  
               (full line) congruence energy contributions by using, in the
               former case, $\textrm a_{\rm neck}=0.5$. The barrier heights are 
               defined, either as the nuclear energy at the scission point - if
               the saddle point corresponds to stronger elongation (compared to
               the scission point) - or else to the saddle point energy.
              }
                                                                 \label{fig.14}
   \end{center}
\end{figure}

Figure \ref{fig.14} presents the dependence on spin of the scission point energies or the saddle-point energies, calculated
with (dashed lines) and without (full lines) congruence energy contribution. 
Since the congruence energy contribution is negative and its value decreases
(increases in absolute terms) when the nuclear deformation approaches scission,
the barrier heights calculated with the congruence energy contribution are
systematically lower. 


\section{Large Amplitude Effects in Nuclear Shape Transitions}
\label{Sect.VIII}


In this Section we will discuss the problem of the nuclear large amplitude motion for spins in the vicinity of the critical shape-transition spins, beginning with the Jacobi-type transitions first. All other shape transitions can be treated similarly, and we will present the Poincar\'e shape transitions next.

Critical shape-transition spins are defined as follows. In the framework of the static description -- the Jacobi critical-spin value, $L^{\rm crit.}_{\rm J}$, is given by the first spin at which the absolute energy minimum corresponds to a non-axial deformation. Similarly, within the static description of the Poincar\'e shape transitions -- the Poincar\'e critical-spin value, $L^{\rm crit.}_{\rm P}$, is defined as the first spin at which the absolute energy minimum corresponds to a left-right asymmetric shape. Calculations show that the congruence energy term, in addition to lowering the calculated energies systematically, also lowers the critical (Jacobi and/or Poincar\'e) transition-spins. [Let us mention in passing that a more physical sense should be associated with the analogues of those critical spins obtained by taking into account the dynamical affects discussed below.]

\begin{figure}[ht!]
\vspace{1cm}
   \begin{center}
     \includegraphics[scale=0.39]{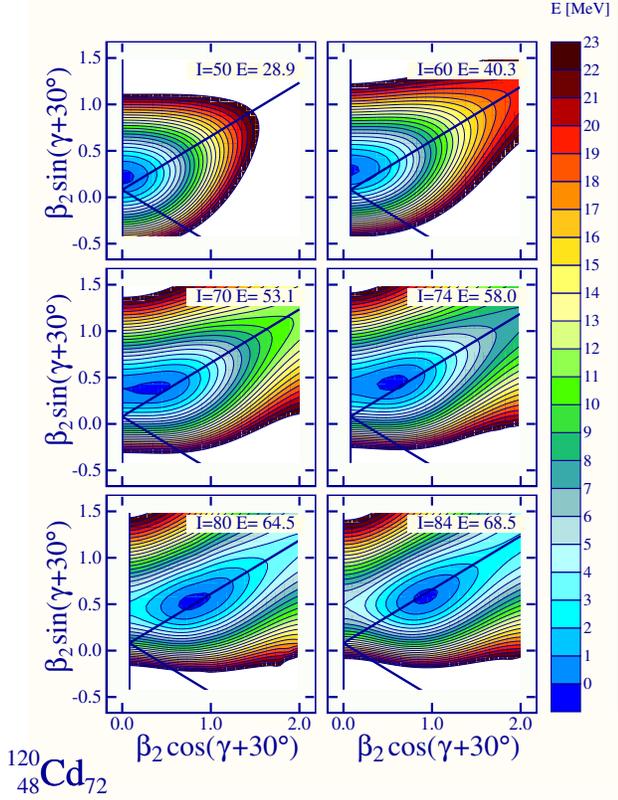}
     \caption{Total energy surfaces for increasing spin in $^{120}$Cd, 
              using a `traditional'
              shape-coordinate representation with the quadrupole deformation
              parameters $(\beta,\gamma)$ of Bohr as often found the literature.  
              Vertical axis corresponds to $\gamma=60^{\circ}$ (oblate) whereas
              the path to fission ($\gamma=0^{\circ}$-axis) has $30^{\circ}$
              inclination with respect to $\mathcal{O}_x$-axis. Down-sloping
              lines correspond to $\gamma=-60^{\circ}$, oblate shapes with
              nucleus turning about an axis perpendicular to the symmetry axis. 
              [Minimization over axial deformation coordinates 
              $\alpha_{\lambda 0}$ with $\lambda \leq 12$.]}
                                                                 \label{fig.15}
   \end{center}
\end{figure}

\begin{figure}[ht]
\vspace{1cm}
   \begin{center}
     \includegraphics[scale=0.39]{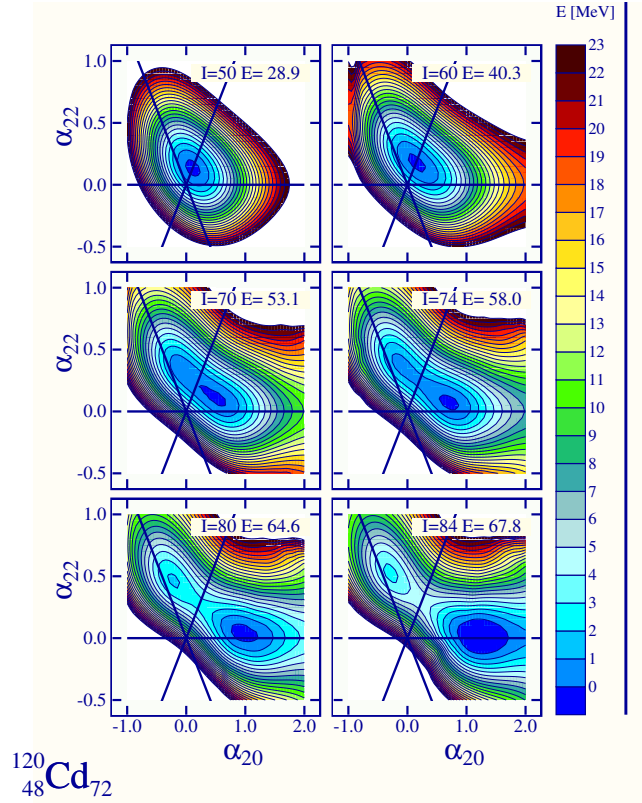}
     \caption{Surfaces of potential energy minimized over axial
              deformation parameters $\alpha_{\lambda 0}$ with $\lambda\leq 12$,
              for spins around the critical spin value for the Jacobi
              shape transition in $^{120}$Cd. Here we use the $\alpha_{20}$ and
              $\alpha_{22}$ co-ordinates well suited for constructing the
              Hamiltonians describing the collective motion with large amplitude
              fluctuations (see text). Compared to Fig.\,\ref{fig.15},
              horizontal lines correspond to $\gamma=0^{\circ}$ (axial)
              deformation, 
              the straight lines with the positive inclination correspond to 
              the $\gamma=60^{\circ}$-axis, whereas the one with the negative
              inclination corresponds to the $\gamma=-60^{\circ}$.  
              }
                                                                 \label{fig.16}
   \end{center}
\end{figure}


\subsection{The Case of Jacobi Transitions}
\label{Sect.VIII-A}


When approaching the critical spin value for a given type of the shape transitions, the corresponding energy landscapes flatten, often forming characteristic `valleys' in subspaces of two or more dimensions in the multidimensional deformation space. Typical results of the LSD-C calculations for the Jacobi shape transitions with the energy minimized in multi-dimensional deformation spaces are illustrated using two-dimensional projections in Figs.\,\ref{fig.15}-\ref{fig.16}. The purpose of comparing these two Figures is to provide a translation from the often applied $\{\beta,\gamma\}$-representation of Bohr in which the first of the two variables has the interpretation of the radial distance from the origin of the reference frame whereas the other one has the interpretation of the polar angle -- and an alternative, $\{\alpha_{20},\alpha_{22}\}$-representation, in which the two quadrupole shape-coordinates appear at the same footing. This latter representation will be of advantage when solving the Schr\"
odinger equation for the collective motion - the main subject of this Section. 

In reference to the results in Figs.~\ref{fig.15}-\ref{fig.16}, observe a gradual displacement of the absolute minimum along the oblate-shape axis ($\gamma=60^{\circ}$ vertical axis according to the convention of Fig.\,\ref{fig.15} and positive-inclined axis according to convention of Fig.\,\ref{fig.16}) at $L=50-60$~$\hbar$, followed by the transition towards increasing $\alpha_{20}$ at increasing tri-axiality (in other words: for non-zero $\alpha_{22}$, or, equivalently, angle $\gamma$ decreasing from $60^{\circ}$ towards 
$\gamma \approx 0^{\circ}$) when spin increases. Arriving at spins $L=80-84$~$\hbar$, we note that $\alpha_{20}\approx 1.0$ with $\alpha_{22}\approx 0$, i.e.~a strongly 
elongated, almost axially-symmetric shape with $\gamma \approx 0^{\circ}$, marking a gradual termination of the Jacobi transition in this nucleus.

Despite the fact that the {\em model} used here to calculate the nuclear macroscopic energy is classical, the physical system is not -- and therefore the motion of the latter in the deformation space should be described using the nuclear collective model, Sect.\,\ref{Sect.III-A}, whose Schr\"odinger equation can be written down in the usual form 
\begin{equation}
    [\hat{T} + \hat{V}(\alpha)]\Psi_{N}(\alpha) = E_N\,\Psi_N(\alpha) ,
                                                                 \label{eqn.22}
\end{equation}
in which the kinetic energy term, $\hat{T}$, depends in principle on the inertia (mass) tensor, cf.~Eq.\,(\ref{eqn.11}), the latter being in general a complicated object depending non-trivially on the deformation coordinates, whereas the potential, $\hat{V}(\alpha)$, with the single symbol $\alpha$ standing for {\em all} the deformation coordinates used, represents the same nuclear energy whose two-dimensional projections are illustrated in the Figures.

\begin{figure}[ht!]
   \begin{center}
     \includegraphics[angle=-90,scale=0.30]{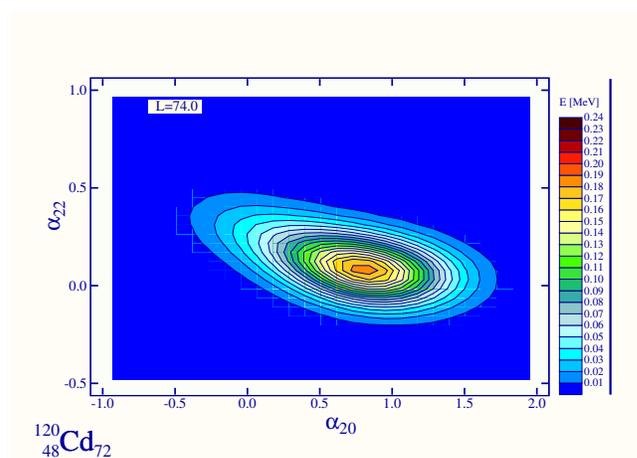}
     \caption{Contour-plot representation of the absolute value of the 
              wave-function -- solution to the two-dimensional collective 
              Schr\"odinger equation, Eq.\,(\ref{eqn.22}), with the constant
              mass-parameter approximation as discussed in the text.}
                                                                 \label{fig.17}
   \end{center}
\end{figure}

In our approach involving the collective solutions of the Schr\"odinger equation, Jacobi transitions will be described using two-dimensional projections on the plane of the quadrupole variables $(\alpha_{20},\alpha_{22})$. Using the collective wave-functions will allow to distinguish between the most probable (`dynamic') quadrupole deformations, 
$(\bar{\alpha}_{20},\bar{\alpha}_{22})_{\rm dyn.}$ and the static ones, 
$(\alpha_{20},\alpha_{22})_{\rm stat.}$, the latter corresponding by definition to the minimum on the potential energy landscape. An example of a typical behavior of the collective wave-function in the shape-transition range close to the critical transition-spin is given in Fig.\,\ref{fig.17}, for illustration. As it can be seen from the Figure, the wave-function varies very slowly along a huge deformation stretch ranging from $\alpha_{20}\sim 0.25$ to 
$\alpha_{20}\sim 1.25$, the variation of the wave-function corresponding to merely a factor of two.

The absurd of possibly continuing to use static description for discussed transitions can be seen clearly in the case of flat valleys with nearly constant
energy along the bottom and very slowly varying associated collective wave-function. Given the flatness of the potential valley, the energy changes only a little whereas the nuclear deformation usually varies considerably. Under these conditions the system oscillates at the energy level of the order of (0.5 - 1) MeV above the formal minimum. When this happens neither the energy value nor the corresponding deformation carry any meaningful quantum interpretation and the need of explicit use of the quantum description becomes evident\footnote{Let us remark in passing that here we have used the constant mass tensor approximation. In the microscopic treatment of the inertia tensor with an explicit presence of temperature to allow for the variation in the nuclear excitation, cf.~e.g.~Ref.\,\cite{ABa94}, the variation of probability density including the factor 
$\sqrt{B}$, as in Eq.\,(\ref{eqn.12}) will add to a less regular structure of the wave-functions in question.}.

The explicit knowledge of the wave-functions will allow to calculate explicitly the dynamical (most likely) deformations and thus provide a quantitative distinction between the static and the dynamic description of the transition quadrupole-deformations, $(\bar{\alpha}_{20},\bar{\alpha}_{22})_{\rm dyn.}$ and  
$(\alpha_{20},\alpha_{22})_{\rm stat.}$. Generally, we may select as a measure of the most probable value of a given shape coordinate say, 
$\alpha_{\lambda\mu}$, the associated r.m.s.~values, 
$\bar{\alpha}_{\lambda\mu}$, defined by
\begin{equation}
   \langle \alpha_{\lambda\mu}^2 \rangle
   =
   \int \mathrm{d}\alpha 
   \Psi^*_{n}(\alpha) \alpha_{\lambda\mu}^2 \Psi^{}_n(\alpha)
   \to
   \bar{\alpha}_{\lambda\mu} = \sqrt{\langle \alpha_{\lambda\mu}^2 \rangle}.
                                                                 \label{eqn.23}
\end{equation}

Calculations show that, although the total energy is flat in the two-dimensional
projection of the quadrupole variables $\alpha_{20}$ and $\alpha_{22}$, it is
generally a much steeper function in terms of most of the other 
$\alpha_{\lambda\mu}$ variables, especially those with relatively high 
$\lambda$. In the present context we are interested in the low-energy
large-amplitude motion in terms of the two quadrupole variables which describe,
to the first order, the transition from axially symmetric oblate to tri-axial
symmetry configurations. Under these conditions, to obtain the first order estimate of the mechanism in question using the nuclear collective wave functions we will introduce two approximations which will now be briefly discussed. 

The first approximation concerns kinetic energy term in Eqs.\,(\ref{eqn.11})-(\ref{eqn.12}) together with the nuclear inertia tensor,
$B_{\alpha_{\lambda\mu};\alpha_{\lambda'\mu'}}(\alpha)$. Although the issue of the deformation-dependent nuclear-inertia tensor is an important and interesting  problem in itself -- it is also relatively complex and it is not our intention to enter this problem at this point. Instead, since we are working with hot nuclear systems approximated by using the liquid drop analogies, we
will rather use the simplest estimation of the inertia parameter according to
the irrotational flow model which, following Ref.\,\cite{sob69}, can be
approximated as $B_{\rm irr.}=(2/15)MAR_0^2$, where $M$ denotes the nuclear mass  [cf.~also Table \ref{tab.03} for comparison of the estimated values]. 
\begin{table}[h!]
\caption{The irrotational flow mass parameter $B_{\rm irr.}$ and the vibration 
         energy for the zero spin calculated with the stiffness parameters
         obtained from LSD-C model. We use:
         $B_{\rm irr.}=2/15MAR_0^2$, and $R_0=r_0A^{1/3}$ with $r_0=1.2$~fm.
                                                                 \label{tab.03}      
        }
        \begin{tabular}{cccc}
	         \hline\hline \\[-4mm]
             Nucleus         &$B_{\rm irr}$ & $E_{\rm vib,x}$ & $E_{\rm vib,y}$   
             \\[1mm] \hline
             $^{46}$Ti       &2.54      &3.34       &5.20  \\
             $^{88}$Mo       &8.06      &1.87       &3.55  \\
             $^{120}$Cd      &13.5      &1.53       &2.90  \\
             $^{128}$Ba      &15.0      &1.39       &2.68  \\
             $^{142}$Ba      &17.9      &1.30       &2.52  \\                 
             $^{147}$Eu      &18.9      &1.19       &2.35  \\                 
            \hline\hline
        \end{tabular}
\end{table}

\begin{figure}[h!]
   \begin{center}
     \includegraphics[scale=1.15]{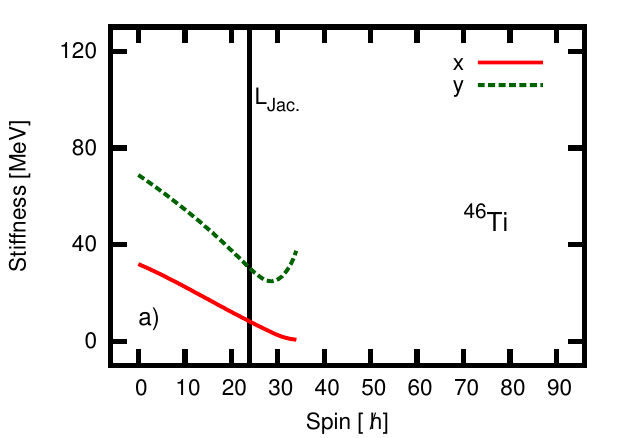}
     \includegraphics[scale=1.15]{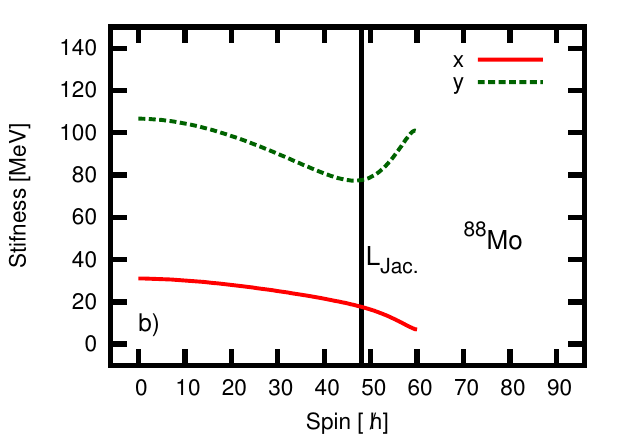}
     \includegraphics[scale=1.15]{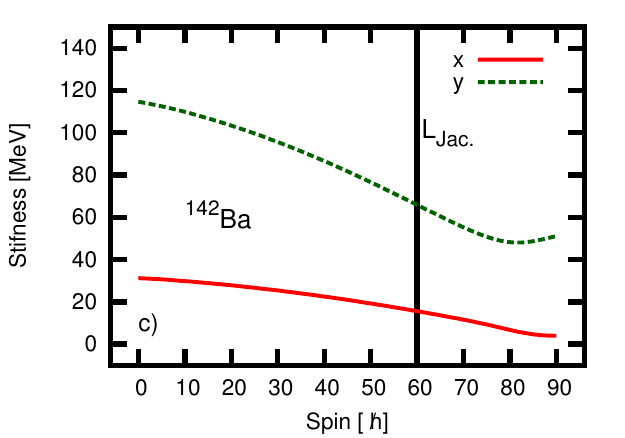}
      \caption{Stiffness coefficients, $C_{x}\equiv C_{\alpha_{20}}$
              and $C_{y}\equiv C_{\alpha_{22}}$ obtained from the 
              two-dimensional projection of the total energy surfaces like the
              ones in Figs.\,\ref{fig.15}-\ref{fig.16} and Eq.\,(\ref{eqn.25}).
              }
                                                                 \label{fig.18}
   \end{center}
\end{figure}

Our second approximation consists in using the two-dimensional projection obtained through minimization over several extra variables treated as depending exclusively on $\{\alpha_{20}, \alpha_{22}\}$. According to this scheme the potential for the collective Schr\"odinger equation in quadrupole coordinates will be constructed as:
\begin{equation}
    V_2(\alpha_{20},\alpha_{22})
    =
    \min_{\alpha_{\lambda \mu}: \lambda>2}V(\alpha),
                                                                 \label{eqn.24}
\end{equation}
i.e.~as if the other deformation coordinates were frozen when considering the quadrupole motion. In what follows we will consider Schr\"odinger equation in the general form of Eq.\,(\ref{eqn.22}) with the approximate form of the potential of Eq.\,(\ref{eqn.24}) and with the constant and diagonal mass tensor represented as an approximation by the irrotational-flow mass-parameter introduced above.

\begin{figure}[h!]
   \begin{center}
     \includegraphics[scale=1.15]{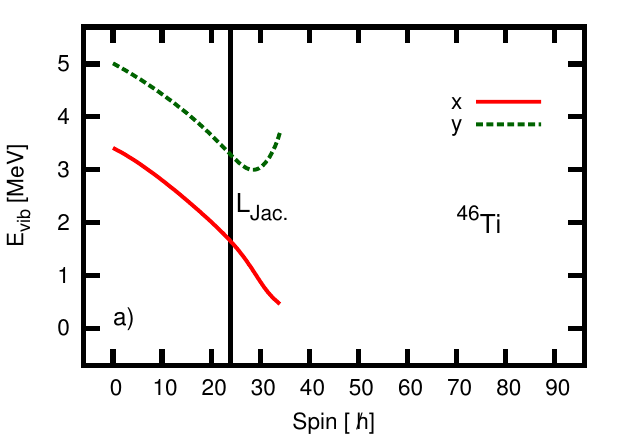}
     \includegraphics[scale=1.15]{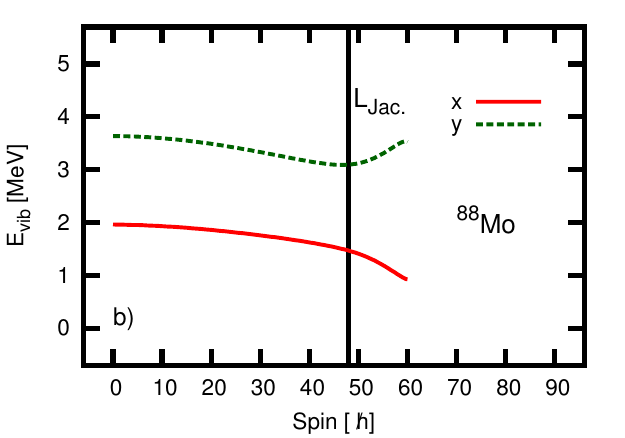}
     \includegraphics[scale=1.15]{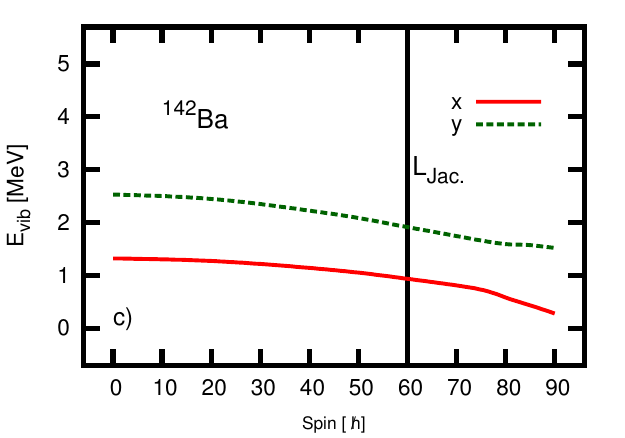}
     \caption{Estimated auxiliary vibration-energies obtained from the harmonic
              approximation expression with the help of formula: 
              $E_{{\rm vib.;}x,y}=\sqrt{C_{x,y}/B_{\rm irr.}}$,
              cf.~Eqs.\,(\ref{eqn.25})-(\ref{eqn.26}).
             }
                                                                 \label{fig.19}
   \end{center}
\end{figure}
We will introduce, for the test purposes, the stiffness coefficients 
$C_{x}\equiv C_{\alpha_{20}}$ and $C_{y}\equiv C_{\alpha_{22}}$ defined with the
help of the potential energies $V_2(\alpha_{20},\alpha_{22})$ as 
\begin{eqnarray}
   V_2(\alpha_{20},\alpha_{22})
   &\approx&
   V_2(\alpha_{20}^{\rm min.},\alpha_{22}^{\rm min.})
                                                                 \nonumber \\
   &+&
   \textstyle\frac{1}{2} C^{}_{\alpha_{20}} 
            (\alpha^{}_{20} - \alpha_{20}^{\rm min.})^2
                                                                 \nonumber \\
   &+&
   \textstyle\frac{1}{2} C^{}_{\alpha_{22}} 
            (\alpha^{}_{22} - \alpha_{22}^{\rm min.})^2 .
                                                                 \label{eqn.25}
\end{eqnarray}
With the help of Eq.\,(\ref{eqn.25}) we further introduce (also exclusively for the purposes of the order of magnitude estimate) the usual harmonic
approximation of the energies of the corresponding collective motion and express the associated vibration energies as:
\begin{equation}
\left.
\begin{array}{ccc}
   E_{\rm vib.;\alpha_{20}} &=& \sqrt{C_{\alpha_{20}}/B_{\rm irr.}}
                                                                \nonumber\\[2mm]
   E_{\rm vib.;\alpha_{22}} &=& \sqrt{C_{\alpha_{22}}/B_{\rm irr.}}
                                                                 \label{eqn.26}
\end{array}
\right\}.
\end{equation} 

Let us emphasize here that to perform the calculations of the most probable
deformation-values we solve numerically the corresponding Schr\"odinger equation
without harmonic approximation. However, considerations of the harmonic
approximation are useful, among others, for `academic' purposes, e.g., they will
allow verifying whether the irrotational flow approximation for the inertia
parameters gives right order of magnitude for the vibrational energies expected
to be in the range of one- to a few MeV, depending on the mass of considered
nuclei.

It is instructive to illustrate the behavior of the stiffness coefficients obtained according to Eq.\,(\ref{eqn.25}) in function of the angular momentum in order to follow their possible spin evolution. The corresponding typical illustrations are presented in Fig.\,\ref{fig.18} for the three nuclei selected. Recall: The larger the stiffens coefficient -- the `stiffer', i.e.~the steeper, the potential. Comparison shows that, on the average, the stiffens coefficients evolve by decreasing, typically, by about 30\%-to-50\% when spin approaches the fission limit what implies that the total energy landscapes get accordingly flatter and flatter and the associated effects of the large amplitude motion more and more important.

The results just discussed are translated, in Fig.\,\ref{fig.19}, into vibration
energies -- using the irrotational-flow inertia parameters. According to
Eq.\,(\ref{eqn.26}), the corresponding curves represent the same trends as the ones in Fig.\,\ref{fig.18}, since the vibration energies  are proportional to 
$\sqrt{C_\alpha}$, however it is instructive to observe that the obtained results expressed in MeV correspond to a `reasonable' order of magnitude expected for the discussed nuclei. 

Let us notice that the quantum character of the collective motion implies that
each nuclear deformation should be associated with the probability density
function $P(\alpha)$ so that the considered probabilities of finding the nucleus
in a given `shape interval', $[\alpha,\alpha+\mathrm{d}\alpha]$, are given by
\begin{equation}
   \mathrm{d}P(\alpha) = |\Psi(\alpha)|^2 \mathrm{d}\alpha ,
                                                                 \label{eqn.27}
\end{equation} 
where $d\alpha$ denotes the associated volume element in the deformation space.
It then follows that the dynamical character of the described shape phenomena
can be, to a first approximation, described in terms of the expected values of the deformation involved, but also by the spreading of the probability distribution whose measure is often selected as the dispersion coefficients
\begin{equation}
   \sigma_{20}
   \equiv
   \sqrt{\langle \alpha_{20}^{2} \rangle-\langle \alpha_{20}^{} \rangle^{2}}
   \;\textrm{and}\;
   \sigma_{22}
   \equiv
   \sqrt{\langle \alpha_{22}^{2} \rangle-\langle \alpha_{22}^{} \rangle^{2}} ,
                                                                 \label{eqn.28}
\end{equation}
which allow to approach the problem of varying flatness of the energy landscapes
in terms of four (spin dependent) quantities of the type: $\bar{\alpha}_{20}$ and $\sigma_{20}$ as well as $\bar{\alpha}_{22}$ and $\sigma_{22}$. 
\begin{figure}[t]
   \begin{center}
       \includegraphics[scale=1.19]{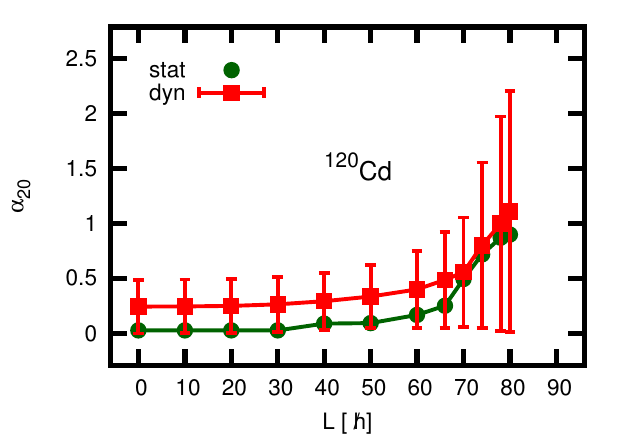}
      \caption{Static values of the axial-symmetry quadrupole deformation 
               $\alpha_{20}$, circles, taken at the minimum of the total energy
               landscape and the dynamical average quadrupole deformation, 
               $\sqrt{\langle \alpha_{20}^{2} \rangle}$, (squares). The vertical
               bars give the $\pm\sigma$ deviations around the centroids
               defined by the latter, Eq.\,(\ref{eqn.28}). 
               }
                                                                 \label{fig.20}
   \end{center}
\end{figure}
\begin{figure}[h]
   \begin{center}
      \includegraphics[scale=1.19]{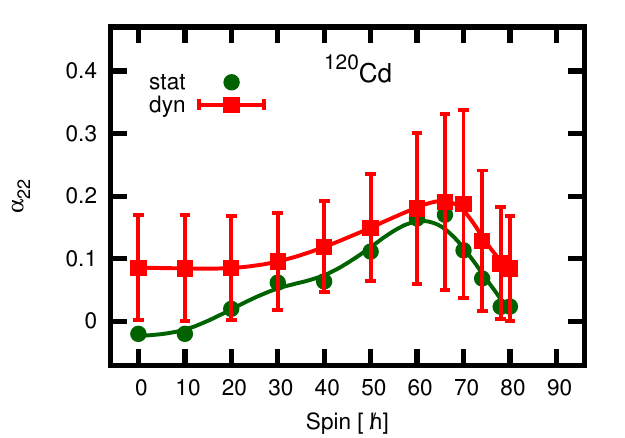}
      \caption{Illustration similar to the one in Fig.~\ref{fig.19} but for the
               non-axial quadrupole deformation parameter $\alpha_{22}$.
               }
                                                                 \label{fig.21}
   \end{center}
\end{figure}

The corresponding differences between the centroid positions of the static and dynamical quantities (circles and squares) are given in Fig.\,\ref{fig.20} for $^{120}$Cd nucleus. These differences illustrate the impact of the shape-fluctuation (dynamical as opposed to static) effects. The Figure illustrates at the same time the spreading of the associated probability distribution (`shape uncertainty') with the help of the vertical bars -- for the axial quadrupole deformation.  

Analogous illustration for the non-axiality associated with the quadrupole
motion is presented in Fig.\,\ref{fig.21} constructed following the same pattern
as the preceding one. 

Let us observe a much larger spreading in terms of the $\alpha_{20}$-fluctuations
expressed in terms of the $\bar{\alpha}_{20}$ deformation together with the
associated dispersion, $\sigma_{20}$, systematically bigger than $\sigma_{22}$,
cf.\,Fig.\,\ref{fig.21} and compare with Fig.\,\ref{fig.20}. This result underlies the fact that in the case of the Jacobi shape transitions the flattening of the energy landscape in the `direction of elongation' plays a leading role, although the effect of fluctuations in terms of tri-axiality is quite important as well. This result has also significant consequences for the future nuclear structure calculations which will need to take explicitly into account the mechanism of the strong, so-called $K$-mixing both in the rotational-band description in the continuum excitation regime as well as in the calculations of the corresponding electromagnetic transition probabilities. 


\subsection{The Case of Poincar\'e Transitions}
\label{Sect.VIII-B}

All what has been said so far about Jacobi shape transitions expressed to the
leading order with the help of two quadrupole variables, $\alpha_{20}$ and 
$\alpha_{22}$ simultaneously, can be formulated as well for the Poincar\'e shape
transitions. The latter involve the so-called left-right asymmetry (sometimes: mass-asymmetry) i.e.~the transition from the inversion-symmetric to the inversion asymmetric shapes. They can be expressed, to the leading order, by the single multipole, $\alpha_{30}$, the octupole coordinate, often also referred to as `pear-shape' deformation.  

\begin{figure}[htbp!]
   \begin{center}\vspace{1cm}
      \includegraphics[scale=0.38]{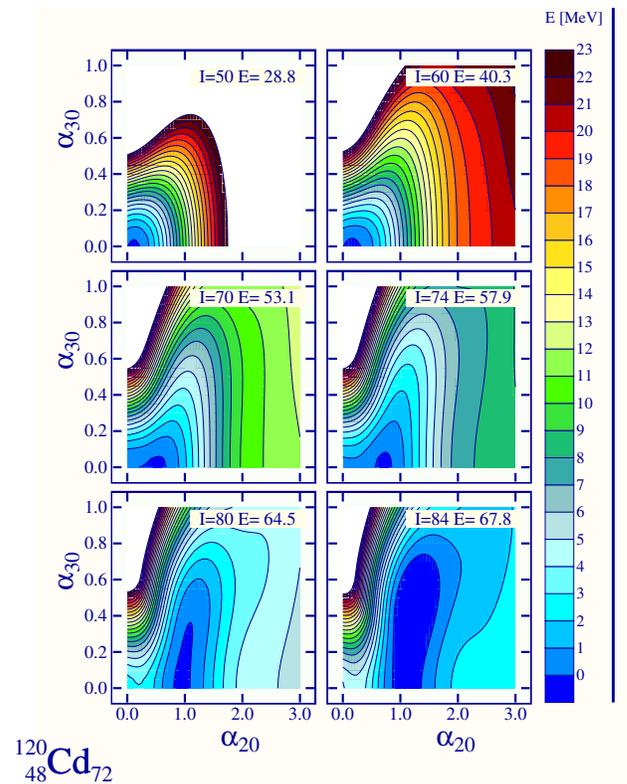}
      \caption{Example of the Poincar\'e-type shape evolution with
               spin using the two-dimensional projections 
               $(\alpha_{20},\alpha_{30})$ - analogous to the one in 
               Fig.\,\ref{fig.16} for the Jacobi-type transition. Two
               dimensional projections corresponding to these two Figures were
               obtained using the same full space of collective coordinates over 
               which minimization has been performed. Observe that, strictly
               speaking, the static Poincar\'e transition takes place at the
               very highest spins only, and this, closely to the fission
               critical spin (disappearance of the fission barrier). However, 
               the evolution of the dynamic effects with lowering of the octuple 
               valley extending to the `north' is clearly visible.
               }
                                                                 \label{fig.22}
   \end{center}
\end{figure}

A typical illustration of the Poincar\'e type shape transition using the $(\alpha_{20},\alpha_{30})$ projection is shown in Fig.\,\ref{fig.22}. Observe the characteristic evolution of the octupole susceptibility which naturally generates fission-fragment mass-asymmetry through the most probable octupole deformations whose exact values increase (see below). This growth is accompanied by a gradual decrease in the fission barrier height.

\begin{figure}[htbp!]
   \begin{center}
      \includegraphics[scale=1.39]{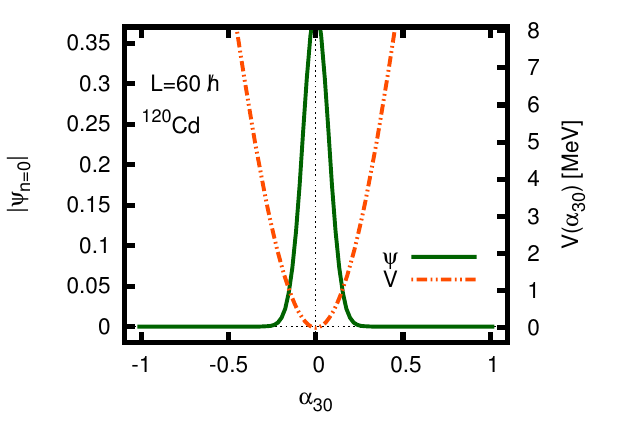}\\[-4mm]
      \includegraphics[scale=1.39]{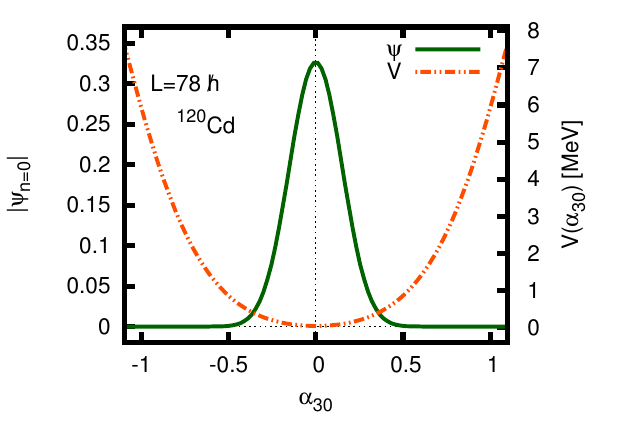}\\[-4mm]
      \includegraphics[scale=1.39]{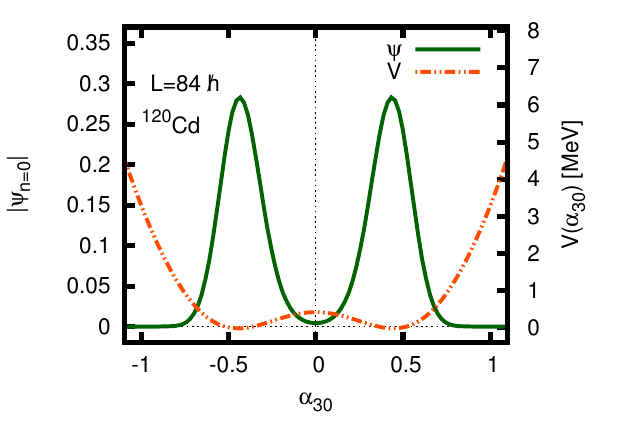}
      \caption{Example of the pear-shape octupole $\alpha_{30}$ evolution with
               spin. As before the static deformations are taken
               at the equilibrium, whereas the dynamic ones, defined as 
               $\bar{\alpha}_{30}$, are given by Eq.\,(\ref{eqn.23}).
               The wave functions (full lines) correspond to the left-hand 
               side scale, whereas the potentials (dot-dashed lines) -- to the 
               right-hand scale.}
                                                                 \label{fig.23}
   \end{center}
\end{figure}

To calculate the most probable (r.m.s.)~$\alpha_{30}$ deformations, the one-dimensional approximation of the potential energies in the direction of $\alpha_{30}$ has been obtained first, through projections on the `octupole valleys' illustrated in Fig.\,\ref{fig.22}. These projections have been used to solve the Schr\"odinger equation in Eq.\,(\ref{eqn.22}) and the corresponding results representing simultaneously the lowest energy wave-function (left scale) and the potential energy curve (right scale) are illustrated in Fig.\,\ref{fig.23}. Observe a characteristic flattening of the potential energies with increasing spin eventually evolving into a double-minimum landscape (here plotted for $L=84\,\hbar$) symmetric with respect to transformation 
$\alpha_{30}\to-\alpha_{30}$. Observe a rather tiny barrier between the two minima at spin $L=84\,\hbar$, however, with the octupole mass parameter chosen here for the semi-quantitative illustration at $B_{30}=100\,\hbar^2/$MeV, sufficient to generate the wave function with a double hump structure.

The forms of the corresponding potential energy curves translate directly into the characteristic evolution of the corresponding collective wave-functions which increasing spreading (cf.~spins $L=60$ and $78\,\hbar$) finishing with the  double hump form at the highest spin illustrated.

\begin{figure}[htbp!]
   \begin{center}
      \includegraphics[scale=1.40]{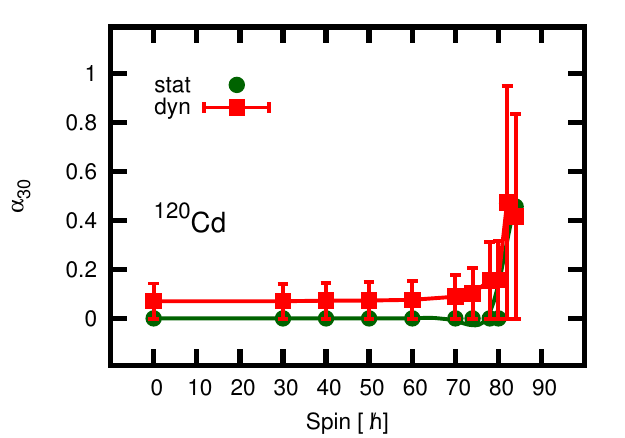}
      \caption{Example of the pear-shape octupole $\alpha_{30}$ evolution with
               spin, in terms of the static and dynamic deformations. As before
               the static deformations are taken at the equilibrium (minima),
               whereas the dynamic ones, defined as $\bar{\alpha}_{30}$, are
               given by Eqs.\,(\ref{eqn.23},\ref{eqn.28}).
               }
                                                                 \label{fig.24}
   \end{center}
\end{figure}
The characteristic evolution of the pear-shape forms is illustrated in
Fig.\,\ref{fig.24} in function of increasing spin in terms of the static and dynamic representations of the shape evolution. Observe that the left-right symmetry breaking obviously takes place in the dynamical description already at the lowest spins. This follows from the fact that at those low spins the zero-phonon collective wave functions have approximately the behavior of a Gaussian, so that at vanishing static equilibrium deformation, $\alpha_{\rm stat.}$, we necessarily find:
\begin{equation}
   0=\alpha_{\rm stat.}^2 <
   \langle {\alpha}^2 \rangle_{\rm dyn.}
   \sim
   \int \alpha^2 \exp\left(-\frac{\alpha^2}{\sigma^2}\right) \mathrm{d}\alpha
   \neq 0\,,
                                                                 \label{eqn.29}
\end{equation} 
what has an immediate impact on the mass-asymmetry. 

Information presented in Fig.\,(\ref{fig.24}) can be completed with the one
showing explicitly the mass asymmetry of the fission fragments translating the r.m.s.~deformation parameter $\alpha_{30}$ into the mass asymmetry. To obtain the experiment-comparable fission-fragment mass-asymmetry one may construct an auxiliary surface composed of two touching ellipsoids and minimized the volume between such an auxiliary object and the actual nuclear surface. The ratio of the volumes of the two ellipsoids allows to obtain an approximate mass ratio of the fission fragments. Of course alternative of extracting the fission-fragment mass-asymmetry can be considered, for instance integrating directly the volumes of the fragments starting from the point where the nuclear neck can be introduced -- the results remain close.

Combining the information from the last two illustrations allows to transform
the results of the dynamical deformation estimates directly into the
observables: The fission-fragment mass-asymmetry in function of spin. A systematic analysis of the corresponding predictions will be published elsewhere.


\section{Summary and Conclusions}
\label{Sect.IX}


In this article we develop a new algorithm based on the macroscopic
Lublin-Strasbourg Drop (LSD) Model. This approach allows to calculate, with an
improved precision, the mechanism of the nuclear shape transitions with varying spin in hot rotating nuclei and the accompanying discussion, we believe, offers a more realistic physics insight as compared to previous discussions of the subject. 

We focus on two families of the shape transitions known already from the `historical' astrophysics works: The so-called Jacobi and Poincar\'e shape transitions, Refs.\,\cite{Jac84,Poi85}, respectively. Jacobi transitions lead from axially symmetric nuclear configurations to the tri-axially symmetric ones. Poincar\'e transitions lead from the inversion-symmetric forms to the ones that break the inversion symmetry. Both can be viewed upon as symmetry-breaking phenomena, but unlike their astrophysical realizations, their nuclear realizations treated by us involve explicitly the description of the quantum critical shape-fluctuations.

The present article can be seen as a contribution to increasing the performance of the Macroscopic-Microscopic approaches which combine the powerful nuclear mean-field theory with the approach based on the Liquid Drop Model. Indeed by focussing on the studies of the nuclear states at high-temperature where the quantum shell effects can be neglected we profit from the unique opportunity of optimizing the macroscopic LSD-C model alone, independently of the nuclear mean-field theory aspects. The Jacobi and Poincar\'e nuclear shape transitions offer valuable experimental test grounds in this context which are discussed in this article.

To be able to formulate the necessary and sufficient criteria which would allow
for (if possible unambiguous) an identification of those transitions in nuclei
one must take into account that both of them may compete when angular momentum
increases. To this end it is important to be able to calculate, in a realistic
manner, the probabilities of signals from nuclei which are axially symmetric,
tri-axial, left-right asymmetric and/or which combine these features -- when
spin increases. Such a competition depends on the large-amplitude fluctuations
in terms of both of these modes (Jacobi, Poincar\'e). We obtain these
probabilities approximately by solving numerically the corresponding collective-model Schr\"odinger-equations and construct the most probable families of shapes from the corresponding collective wave-functions.

Taking explicitly into account the presence of either zero-point or large-amplitude motion in the direction of the mass asymmetry coordinate (here: 
$\alpha_{30}$ octupole deformation) introduces the possibility of estimating  the fission-fragment mass asymmetry with the help of the quantum, collective model technique right from the beginning. This technique reflects in a sensitive manner the octupole-deformation {\em susceptibility} (e.g.~increasing flatness of the nuclear energy landscape without necessarily producing static left-right asymmetric total energy minima) as opposed to the traditional analyses based on the static minima in terms of octupole-type coordinates. [Using this dynamical description also introduces a qualitative difference with respect to estimates of the early Businaro-Gallone approach, Ref.\,\cite{ULB55}.]

Transitions in question have been traditionally characterized by the critical
spin values associated with the static total energy minima: The last spin value
at which the preceding symmetry occurs and the first spin value at which the new
symmetry arises define the critical spin for the transitions considered. To
obtain more realistically these critical spin values, as well as the description
of the fission barriers, we have modified the original LSD model expression of
Refs.\,\cite{KPD03,JDu04}, by introducing a deformation dependent congruence
energy term with certain phenomenological parameters whose values have been
optimized to the known experimental values of the fission barriers.  

This new, LSD-C energy expression, with the shape dependent congruence energy
term, lowers the previous discrepancies for the fission barriers for the lighter
nuclei considered ($^{70,76}$Se, $^{75}$Br, $^{90,98}$Mo) by about 10~MeV
whereas it modifies only slightly the fission barriers for the heavier nuclei
for which the agreement has already been good. The remaining discrepancies
are of the order of 1 MeV for the mass range between A$\sim$80 and 
A$\sim$230.

The shape-dependent congruence energy term is expected to imply a better
description of the fission barrier heights for increasing spin as well as for
fast rotating nuclei with the better prospects for the description of the
fission cross-sections and the shapes of the charge and mass distribution of the
fission fragments. The implied additional binding energy lowers also the fission
critical spin by 5-10~$\hbar$.

Finally, the stability of the newly obtained LSD-C expression with respect to
the basis cut off, $\lambda_{\rm max}$ in the nuclear shape parameterization as well as the parametric uncertainties involved by introducing the new parameters to the energy expression have been studied and the results are discussed. The results in this article have been principally limited to a few illustrative 
cases only; the results of the systematic calculations using these methods will be presented elsewhere.

\section*{Acknowledgements}
This work has been supported by the COPIN-IN2P3 Polish-French Collaboration under Contract No. 05-119 and LEA COPIGAL project: ``Search for the high-rank symmetries in subatomic physics`` and the Polish Ministry of Science and Higher Education (Grant No.~2011/03/B/ST2/01894). We would like to thank A.~G\'o\'zd\'z and M. Kmiecik for fruitful discussions.



\end{document}